\documentclass[a4paper, 11pt, reqno]{amsart}
\usepackage[utf8]{inputenc}

\usepackage{amssymb, amsmath, amsfonts, amsthm, graphicx, booktabs, xcolor,comment}
\usepackage{bm} 
\usepackage{mathrsfs} 
\usepackage{commath} 
\usepackage[top = 0.7in, bottom = 0.5in, left = 0.5in, right = 0.5in]{geometry}
\usepackage{enumitem}
\usepackage{mathtools}
\usepackage{ulem} 
\usepackage{caption}
\usepackage{subcaption}
\usepackage[normal, flushleft]{threeparttable}
\usepackage[scr=boondoxo , scrscaled=1.1]{mathalpha}
\usepackage{pdflscape}
\usepackage[foot]{amsaddr}

\usepackage{tabstackengine}
\stackMath

\usepackage{setspace}
\usepackage{natbib}
\bibpunct[, ]{(}{)}{,}{a}{}{,}

\usepackage[many]{tcolorbox}    	
\newtcolorbox{boxk}{
    sharpish corners, 
    boxrule = 0pt,
    leftrule = 4.5pt, 
    enhanced,
    fuzzy shadow = {0pt}{-2pt}{-0.5pt}{0.5pt}{black!35} 
}

\usepackage{hyperref}
\hypersetup{colorlinks = true, linkcolor = blue, citecolor = gray, filecolor = magenta, urlcolor = olive}

\allowdisplaybreaks[4]
\theoremstyle{plain}
\newtheorem{thm}{Theorem}[section]

\theoremstyle{definition}
\newtheorem{dfn}[thm]{Definition}

\newtheorem{rem}[thm]{Remark}

\newcommand{\pder}[2][]{\frac{\partial#1}{\partial#2}}

\DeclareMathOperator*{\argmax}{arg\,max}

\newcommand{\calF}{\mathcal{F}}

\newcommand{\calW}{\mathcal{W}}

\newcommand{\frakf}{\mathfrak{f}}
\newcommand{\frakg}{\mathfrak{g}}

\newcommand{\E}{\mathbb{E}}				
\newcommand{\Q}{\mathbb{Q}}

\newcommand{\vm}[1]{\mathbf{#1}}

\let\oldFootnote\footnote
\newcommand\nextToken\relax
\renewcommand\footnote[1]{%
    \oldFootnote{#1}\futurelet\nextToken\isFootnote}
\newcommand\isFootnote{%
    \ifx\footnote\nextToken\textsuperscript{,}\fi}

\usepackage[linecolor=black,textwidth=30mm,textsize= footnotesize]{todonotes}

\begin{document}

\normalem 

\title[Hybrid Variable Annuities]{Variable annuities: A closer look at ratchet guarantees, hybrid contract designs, and taxation}

\author[]{Jennifer Alonso-Garc\'ia}
\address[JAG]{Department of Mathematics, Universit\'e Libre de Bruxelles, Belgium; Centre of Excellence in Population Ageing Research (CEPAR), Australia; Netspar, The Netherlands}
\email[JAG]{jennifer.alonso.garcia@ulb.be}

\author[]{Len Patrick Dominic M. Garces}
\address[LPDMG]{School of Risk and Actuarial Studies, University of New South Wales, Australia}
\email[LPDMG]{l.garces@unsw.edu.au \textrm{(Corresponding Author)}}

\author[]{Jonathan Ziveyi}
\address[JZ]{School of Risk and Actuarial Studies, University of New South Wales, Australia; Centre of Excellence in Population Ageing Research (CEPAR), University of New South Wales, Australia}
\email[JZ]{j.ziveyi@unsw.edu.au}


\thanks{\textit{Funding}: LPDMG acknowledges the financial support provided by the Australian Research Council Centre of Excellence in Population Ageing Research (Project Number CE170100005), the CEPAR HDR and ECR Fellow Travel Scheme, and the University of Technology Sydney Faculty of Science start-up grant. JZ acknowledges the financial support from the Australian Research Council Discovery Project DP210101195.}

\date{\today}

\begin{abstract}
This paper investigates optimal withdrawal strategies and behavior of policyholders in a variable annuity (VA) contract with a guaranteed minimum withdrawal benefit (GMWB) rider incorporating taxation and a ratchet mechanism for enhancing the benefit base during the life of the contract. Mathematically, this is accomplished by solving a backward dynamic programming problem associated with optimizing the discounted risk-neutral expectation of cash flows from the contract {\color{black} from the policyholder's perspective}. Furthermore, inspired by traded VA contracts in the market, we consider hybrid products providing policyholders access to a cash fund which functions as an intermediate repository of earnings from the VA and earns interest at a contractually specified cash rate. We contribute to the literature by revealing several significant interactions among taxation, the cash fund, and the benefit base update mechanism. {\color{black} The policyholder's valuation of the contract increases when a cash fund is included in the contract or when the benefit base updates via a ratchet mechanism. When the tax rate is sufficiently high, including both features is necessary to stimulate demand for the contract. Furthermore, the ratchet mechanism tends to discourage early surrender as it provides enhanced downside market risk protection. Similarly, the cash fund discourages active withdrawals, with policyholders preferring to transfer the guaranteed withdrawal amount to the cash fund to leverage the cash fund rate and the varying taxation rules around interest earnings.} 

\end{abstract}


\maketitle

\section{Introduction}

\subsection{Background and Review of Related Literature}

A variable annuity (VA) is a long-term insurance contract where the policyholder agrees to pay a single premium or make a stream of periodic payments in exchange for guaranteed minimum periodic {\color{black} payments} from the insurer. Designed to enhance the policyholder's income (usually after retirement) or to provide some financial protection in the event of death, VAs may have benefits that are underwritten before the policyholder's retirement or may solely consist of post-retirement benefits. Guarantees embedded in VAs are collectively known as ``GMxBs'', where ``x'' may stand for death (D), accumulation (A), income (I), or withdrawal (W). The latter three riders are also referred to as guaranteed minimum living benefits as these provide income protection while the policyholder is still alive. In general, the benefits embedded in VAs are contingent on the performance of investment funds (hence VAs are equity-linked or equity-indexed annuities), but guarantee a particular cash flow architecture despite variable fund growth. In other words, VAs provide policyholders the opportunity to participate in the equity market in conjunction with downside protection via the contractual guarantees. For a detailed discussion of VAs and the guarantee riders, see for example \citet{Bacinello-2011}, \citet{Bauer-2008}, and the references therein.\footnote{\citet{Bacinello-2011} and \citet{Bauer-2008} provide general unified valuation frameworks for variable annuities with various riders (including the minimum withdrawal benefit) using a risk-neutral framework in the absence of market frictions such as taxation. Similar studies of variable annuities have been conducted under more complex financial model dynamics, including stochastic interest rates, stochastic volatility, and regime-switching dynamics \citep[see, for example,][]{Ai-2022, AlonsoGarcia-2018, Forsyth-2014, Gudkov-2019} or path-dependent guarantees \citep[see, for example,][]{Ai-2022, DeelstraHieber-2023}. Methods in continuous-time stochastic optimal control have also been used to investigate optimal policyholder withdrawal strategies for VA with GMWB or guaranteed lifelong withdrawal benefit (GLWB) riders; see, for example, \citet{AzimzadehForsyth-2015, ChenVetzalForsyth-2008, Dai-2008, Forsyth-2014, Huang-2014}. Due to the analytical intractability of the stochastic optimal control associated to the analysis of VAs with withdrawal benefits (GMWB or GLWB), various numerical methods have been proposed towards its resolution under a variety of modelling assumptions \citep[see, for example,][]{AlonsoGarcia-2018, ChenVetzalForsyth-2008, Forsyth-2014, Gudkov-2019, Shevchenko-2017}.} In this study, we focus on VAs with a GMWB rider. Under this rider, the policyholder is entitled to withdraw a cash amount periodically (usually at the policy anniversary dates) and is guaranteed a minimum withdrawal amount regardless of the performance of the investment account associated to the VA.\footnote{For VAs with a GMWB rider, the minimum guaranteed withdrawal amount is typically set to be a contractually specified proportion (typically $100\times\frac{1}{N} \%$ for an $N$-year contract) of the prevailing benefit base.} For VAs with withdrawal benefits, policyholder withdrawal strategies can either be described as \textit{static}, where policyholders only withdraw the guaranteed withdrawal amount, or \textit{dynamic}, where policyholders (optimally) decide how much they withdraw at each opportunity.


The guarantees embedded in a VA contract are calculated based on a benefit base whose initial value is equal to the premium paid by the policyholder at the conception of the contract.  Oftentimes, VA contracts allow the benefit base to grow over the life of the contract. In this paper, we focus on VAs with a GMWB rider in which the benefit base grows through a \textit{ratchet} mechanism. This implies that at pre-determined dates throughout the life of the contract the benefit base is set to the maximum between the prevailing investment account value and the benefit base. 
However, since the analysis of the ratchet mechanism requires including the benefit base as a state variable (in addition to the value of the VA investment account), the ratchet mechanism in the context of the GMWB rider is relatively understudied in the literature.\footnote{Earlier work on unified approaches to variable annuities only consider ratcheting for GMWBs under static policyholder behavior \citep{Bacinello-2011} or in contracts with other types of riders, such as death, accumulation, or income benefits \citep{Bauer-2008}.} 


For rational policyholders, it has been shown that surrendering the contract is optimal when the value of the investment account exceeds a certain threshold (or when the guaranteed benefit becomes out-of-the-money). In this case, it may be more reasonable for the policyholder to invest directly into the underlying fund or to lapse on the current contract and purchase a new policy, possibly at the same price as the original, in which the guarantee is at-the-money \citep{Moenig-2018}. From the insurer's point of view, the expected profit and the risk the insurer faces are heavily impacted by the misalignment between expected and actual policyholder behavior around contract lapsation. To discourage policyholder surrender in VA contracts, particularly in the earlier years of the contract, an exponentially-decaying surrender penalty structure has been proposed \citep[see for example][]{BernardMacKayMuehlbeyer-2014}. Strategies to disincentivize policyholder surrender have also been incorporated in the design of the VA benefit base; see, for example, \citet{Harcourt-2024, Moenig-2018}. 
However, despite having surrender mitigation strategies in place, surrender is still optimal in some situations \citep{AlonsoGarcia-2024, BernardMacKayMuehlbeyer-2014, Moenig-2016}. 
 
Policyholder surrender behavior is also linked to the guarantee fee structure that is effective for the policy. While the constant proportional fee structure is simple to implement and convenient for further analysis, it has been shown that this fee structure incentivizes policyholders to surrender when the guarantee is deep out-of-the-money and induce a misalignment in the value of the guarantee and the fee income \citep{BernardHardyMacKay-2014}. Furthermore, under a constant proportional fee structure, charging fees when the investment account balance is low tends to accelerate the ruin of the investment account and trigger guaranteed payments by the provider \citep{FengJingNg-2025}. In addition, proportional guarantee fees also tend to exacerbate liquidity and basis risks faced by providers of variable annuities \citep{Ankirchner-2014}. To this end, alternative fee structures have been proposed, including state-dependent fees \citep{BernardHardyMacKay-2014, Wang-2021}, time-dependent fees \citep{Bernard-2019}, or fees indexed to a financial index, such as the VIX \citep{KouritzinMacKay-2018}. More recently, contract specifications in which the guarantee fee is levied on a cash account separate from the VA investment account have been analyzed as an alternative to the more traditional fee structure \citep{AlonsoGarcia-2025}. Ultimately, these alternative fee structures have been shown, through extensive numerical experiments, to be effective in preventing policyholder surrender. 


There is also mounting evidence suggesting that taxation plays a substantial role in explaining policyholder behavior in VAs and life insurance products by inducing an inconsistency between the policyholder's valuation and the insurer's valuation of the contract \citep[see, for example,][]{AlonsoGarcia-2024, Horneff-2015, Moenig-2016, Molent-2020, Ulm-2020}.\footnote{Indeed, majority of studies on the interaction between variable annuities and taxation \citep[see for example][]{Horneff-2015, Moenig-2016} are set in a US setting, where VA contracts are tax deferred. In contrast, the work of \citet{Ulm-2020} and \citet{AlonsoGarcia-2024} more closely reflect the taxation environment in New Zealand and Australia, respectively. Without explicitly dealing with taxes, \citet{GaoUlm-2015} establish a similar gap for a VA with a GMDB rider between the policyholder's willingness-to-pay (established using an optimal lifecycle utility approach) and the insurer's risk-neutral valuation given the policyholder's optimal strategy.} This is primarily due to the fact that VA issuers are focused on the total payments (and the cost of covering guarantees) made to policyholders throughout the contact, whereas the policyholders are concerned about after-tax cash flows from the contract \citep{Bauer-2023}. Within a risk-neutral valuation framework, \citet{Moenig-2016} find that incorporating taxation leads to VA contract prices which closely match empirically observed values. Using a life-cycle utility approach, \citet{Horneff-2015} find that taxation typically lowers the demand for VAs with GMWB riders, with policyholders finding direct investment in equity more advantageous. \citet{Moenig-2021} shows that, through an extension of the subjective risk-neutral valuation framework of \citet{Moenig-2016}, a higher tax rate on investments made outside the VA contract (similar to the external investments considered in \citet{Horneff-2015}) makes the VA contract more attractive to the policyholder. Furthermore, \citet{Moenig-2021} shows that optimal policyholder withdrawal behavior is substantially different if the policyholder does not have the option to invest their withdrawals in a separate investment account.\footnote{Another key finding of \citet{Moenig-2021} is that the optimal withdrawal behavior of the policyholder obtained through the subjective risk-neutral valuation framework, accounting for taxation and allowing the policyholder to invest the proceeds from their VA in alternative investments (separate from the VA contract), is consistent with that arising from life-cycle utility approaches \citep[for example]{Horneff-2015}. Given the difficulty in ascertaining the utility function of policyholder, this result justifies the use of the risk-neutral valuation approach in assessing optimal policyholder withdrawal behavior. \label{fn-RiskNeutralUtility}}


The bulk of the studies mentioned above focus on the impact of taxation on VA contracts with a single rider, typically a GMAB or a GMWB rider. In practice however, VA contracts are complex instruments which bundle several baseline riders and features and are sold to policyholders as a \textit{hybrid} product.\footnote{\citet{Bauer-2023} investigate several complex VA contracts with GMWB riders issued in the USA. These contracts include a return-of-premium death benefit and differ in various aspects such as surrender fee schedule, fee-free withdrawals, and benefit base update scheme. Examples of similarly complex VA-type contracts offered in Australia include the \href{https://www.mlc.com.au/adviser/retirement/investment-protection/mlc-masterkey-investment-protection}{MLC MasterKey Investment Protection}, \href{https://www.northonline.com.au/adviser/products/guarantees}{MyNorth Super and Pension Guarantee}, and the \href{https://www.challenger.com.au/individual/what-we-offer/lifetime-annuities}{Challenger Lifetime Annuity} (URLs are correct as of the time of writing).} Analyzing VA contracts with a GMWB rider offered in the USA, \citet{Bauer-2023} find that, in the presence of market frictions such as taxation, insurers incur a negative marginal cost to bundling an additional GMDB rider while increasing the policyholder's valuation of the contract. 
\citet{AlonsoGarcia-2025} also analyze VA contracts with bundled and stand-alone accumulation and death benefits in the presence of taxation under an alternative fee structure in which investment and management fees are charged on the VA investment account, but the guarantee fee is levied on a separate cash account. 
The authors find that policyholders perceive the bundled contract more valuable compared to separate, stand-alone contracts, and that a bundled contract has a lower optimal surrender boundary compared to stand-alone contracts. However, the marginal cost to the insurer of offering the additional rider was not discussed, in contrast to \citet{Bauer-2023}.

\subsection{Key Research Questions and Main Contributions}

As the aforementioned studies suggest, contract features, market frictions, and fee structures all have a quantifiable and significant effect on policyholder behavior, including optimal withdrawal and/or surrender strategies. In this paper, we analyze the interaction between contract features, taxation, and optimal policyholder behavior in the context of a VA contract with a GMWB rider. In terms of contract features, we investigate the impact of adding a secondary investment option, separate from the VA investment account {\color{black} but embedded in the same insurance wrapper}, which is still administered by the VA issuer. The secondary investment earns the cash rate and is accessible solely through the purchase of the VA contract. We shall refer to this secondary investment as the \textit{cash fund}.
This type of hybrid VA contract has emerged in recent years as an innovative integrated offering in which VA policyholders are given a wider range of investment options as part of their VA contract purchase. In this paper, we adapt the basic structure of the ``MasterKey Investment Protection'' suite offered by MLC \citep{MLCPDS}, which is an example of a hybrid VA offered in Australia. Given the basic hybrid VA structure, we investigate the effect of taxation and variations to the contract structure on optimal policyholder behavior.

In this hybrid contract, if the policyholder decides to withdraw an amount $w(t_k)$ less than the guaranteed withdrawal amount $g(t_k)$ at the withdrawal date $t_k$, the difference $g(t_k) - w(t_k) > 0$ between the amount withdrawn and the guaranteed withdrawal amount is transferred from the VA investment account to the cash fund. In this regard, the cash fund serves as a secondary repository of proceeds from the withdrawal benefit, and its balance then earns interest at a cash rate that the issuer commits to providing. The cash rate $\eta > 0$ is usually benchmarked to the market cash rate and is higher than the risk-free interest rate. However, since the cash fund is separate to the GMWB rider, {\color{black} the balance of the cash fund does not enjoy the same guaranteed downside protection via the benefit base update mechanism. Nonetheless, the cash fund functions as a capital-stable account accruing at a low but positive deterministic rate $\eta$.}\footnote{We emphasize that this contract design is distinct from the lifecycle setting \citep[see, for example,][]{FengJingNg-2025, GaoUlm-2015, Horneff-2009, Horneff-2015, Moenig-2021} in which policyholders have the option to invest their withdrawals in a separate, external investment account or from the setting in which policyholders can purchase an additional option to modify the underlying VA investment composition \citep[see, for example][]{Mahayni-2012}. The cash fund in our setting is seen as an secondary, ``in-house'' investment option that is managed by the issuer of the VA, not the policyholder.} 

{\color{black} To this end, we employ a risk-neutral valuation framework to analyse optimal policyholder behaviour in a VA contract with a GMWB rider, permitting either a static or dynamic withdrawal strategy. In contrast to previous studies analysing VA contracts with GMWB riders in isolation, we investigate policyholder behaviour in hybrid contracts in which a cash fund is offered in tandem with the VA. The risk-neutral valuation framework is standard in the VA literature when the objective is to model a rational policyholder seeking to maximize cash flows from the contract \citep[see, for example,][]{Bacinello-2011, Bauer-2008}, and is particularly well suited to our setting, where the alternative investment option, in the form of a cash fund, is structured within the contract itself rather than held externally}.\footnote{{\color{black} In contrast, a life-cycle or utility approach is more appropriate when the alternative investment is external to the VA contract \citep[see, for example,][]{FengJingNg-2025, GaoUlm-2015, Horneff-2015, Moenig-2021} or when consumption smoothing enters the policyholder's objective function \citep[see, for example,][]{Bruhn-2013}. See also Footnote \ref{fn-RiskNeutralUtility}.}} {\color{black}By doing so, we obtain the behavior and fair fee from a polycholder perspective. The policyholder fair fee is the fee at which a \textit{rational} policyholder is indifferent between entering and not entering the contract; it is a measure of the product's attractiveness to the demand side of the market, irrespective of whether the insurer can profitably supply it at that fee.}\footnote{{\color{black}\citet{AlonsoGarcia-2024} show that, in a GMAB context, if an insurer were to charge the policyholder fair fee, it would earn a positive profit with high probability but face large left-tail losses. In expectation, the insurer will be solvent, and the gap between insurer's fair fee and the policyholder fair fee is itself an informative measure of the distortion introduced by taxation.}} {\color{black} The demand-side perspective adopted in this paper addresses the question of whether the product features of the hybrid VA contract are sufficient to sustain policyholder demand under realistic tax conditions, including early access to retirement funds. The policyholder fair fee is a natural instrument for answering this question.}

In this setting, the guaranteed withdrawal amount $g(t_k)$ is a contractually specified proportion of the benefit base $\{G_t\}_{t\in[0,T]}$, updated through a ratcheting mechanism on contractually specified dates $t_k\in(0,T)$, $k=1,\dots,N$, typically set to coincide with policy anniversary dates. For simplicity, we also assume that withdrawals take place on these dates, the exact timing of which is detailed in the discussions that follow.

Typically, withdrawals and other proceeds from the VA are not tax-exempt. {\color{black} Naturally, taxation introduces a wedge between policyholder and insurers valuation since the fair fee from the policyholder's perspective, defined as the fee that sets the contract  value equal to the initial premium, diverges from the fee that minimises the insurer's liability once taxes are introduced.}\footnote{{\color{black} This occurs because it is the policyholder's post-tax behaviour that governs the state of the contract at every point in time: the withdrawal and surrender boundaries are determined by the policyholder's incentive to maximise post-tax cash flows, and the insurer's liability is a derived quantity that responds to those decisions. As a result, the insurer's liability crosses the initial premium threshold at substantially higher fee rates than the policyholder value does, and in some cases the insurer's fair fee may not exist at all.}} Furthermore, in some taxation regimes, withdrawals are taxed as ordinary income at a rate equal to the policyholder's marginal income tax, whereas interest-bearing investments, such as the cash fund, are taxed on the basis of interest rate income only. Due to the difference in tax treatment of withdrawals from the VA and the cash fund appreciation, a tax-shielding effect may arise from the cash fund, depending on taxation levels, the cash fund rate, and the risk-free rate at which cash flows from the contract are discounted.

As such, the objective of this paper is to analyze optimal policyholder behavior in hybrid VA contracts in the absence and presence of taxation affecting proceeds from the VA.\footnote{For example, in Australia, proceeds from the VA obtained before retirement are subject to tax, whereas those received after retirement are not.} Of particular interest is impact of the rate of appreciation of the cash fund and whether the interest income the cash fund generates is sufficient to influence the policyholder not to withdraw the full guaranteed amount and instead leave some to grow in the cash fund. Furthermore, we also analyze the effect of taxes on the policyholder's strategy and how it interacts with product design features, such as the type of benefit base update scheme (ratchet or no ratchet) and whether or not the cash fund is available. As such, this paper contributes to the literature by analyzing the ratchet mechanism in the context of a GMWB and how it interacts with taxation and other contract features in influencing optimal policyholder withdrawal or surrender. When taking into account taxation, we find through numerical examples that the cash fund plays an influential role with respect to the policyholder's optimal withdrawal strategy. Indeed, in some tax settings, policyholders are unwilling to purchase the contract at all since the policyholder's (risk-neutral) valuation of the contract is less than the initial premium. 

Through a series of numerical experiments, our paper contributes new insights regarding various interactions among taxation, the cash fund, and the ratcheting mechanism in the context of a hybrid VA contract. First, we find that in the absence of taxes, the VA contract becomes more valuable to the policyholder as the cash rate increases. Second, we find that taxation substantially impacts the optimal withdrawal strategy; it is sometimes optimal for the policyholder to not withdraw anything (and leave the guaranteed withdrawal amount in the cash fund) when tax rates are high. This is due to a ``tax-shielding'' effect arising from the difference in the taxation of the ordinary withdrawals and the cash fund. Third, benefit base upgrades in the form of a ratchet are sometimes necessary to ensure the product is worthwhile to the policyholder in the presence of taxation. In addition, when comparing contracts with a return-of-premium and a ratchet specification, we find that the ratchet discourages early surrender due to the enhanced downside risk protection it provides. Finally, the addition of a cash fund discourages active withdrawals with or without taxes; instead, it is optimal for policyholders to leverage the cash fund rate to further grow their proceeds from the contract.

The remainder of the paper is structured as follows. We discuss the mathematical framework for the underlying financial market and the contract specification of the hybrid variable annuity in Section \ref{sec-MathematicalFormulation}. This section also elucidates how the VA investment account value and the benefit base evolve over time and states the policyholder's stochastic optimal control problem, particularly when the policyholder's proceeds from the VA are subject to taxes. Since an analytical solution is not available in our setting, in Section \ref{sec-NumericalSolution}, we discuss the numerical methods used to solve the stochastic optimal control problem. We focus on the numerical approximation of the value of the VA before and after each withdrawal is made by the policyholder. Through a series of numerical illustrations in Section \ref{sec-NumericalIllustrations}, we investigate the impact of contract specifications (that is, the benefit base update scheme and the cash rate) and tax rates on the policyholder's fair fee and optimal withdrawal strategies. We then summarize and conclude the paper in Section \ref{sec-Conclusion}. {\color{black} Appendix \ref{app-SuppFigures} contains additional figures and results supplementing the discussions in Section \ref{sec-NumericalIllustrations} and Appendix \ref{appendix-ComputationalMethods} details the computational algorithms employed in this paper.}

\section{Mathematical Formulation}
\label{sec-MathematicalFormulation} 

\subsection{Notation and VA Assumptions}
\label{sec-Notation}

For the VA contract, we assume that withdrawals, fee charges, and other adjustments occur at discrete points in time. Let $\{t_0,t_1,\dots,t_N\}$ be a partition of the interval $[0,T]$, where $t_0=0$ is the inception of the contract and $t_N=T$ is the contract maturity. We assume that the partition is uniformly spaced, with $\Delta t = t_{k}-t_{k-1}$ for all $k=1,\dots,N$. In particular, the subset $\{t_1,\dots,t_N\}$ are the contractually specified times at which the policyholder can withdraw funds under the VA contract and at which the ratcheting mechanism is applied to adjust the benefit base. 

At time $t = t_0$ (start of the contract), the policyholder pays an initial premium $P_0$, which is then invested into the VA investment account. At each $t=t_k$, $k=1,\dots,N-1$, we assume the following sequence of events:
\begin{enumerate}

    \item At the instant prior to any withdrawals, indicated by $t_k^-$, the guarantee fee $\varphi$, is charged proportional to the VA investment account value $X(t_k^-)$. The fee $\varphi$ is expressed in annual terms and is assumed to cover the cost of financing the guarantee and any associated investment or administrative fees. 
	
    
    
    
    \item The policyholder then makes a withdrawal $w(t_k)$ at time $t_k$. The policyholder may choose not to make a withdrawal at this time, in which case we have $w(t_k) = 0$. The maximum possible withdrawal is the maximum between the post-fee investment account value $X(t_k^-)(1-\varphi \Delta t)$ and the guaranteed withdrawal amount $g(t_k)$. That is,$$0 \leq w(t_k) \leq \max\{X(t_k^-)(1-\varphi \Delta t), g(t_k)\}, \qquad k=1,2,\dots,N.$$ {\color{black} If the policyholder withdraws an amount less than $g(t_k)$, then the difference $g(t_k) - w(t_k)$ is transferred to the cash fund included in the contract.}
    
    \item The VA investment account value and cash fund balance are updated and the benefit base is adjusted (in the event of an excess withdrawal) at the instant after the withdrawal, denoted by $t_k^+$. We assume that the cash fund appreciates at a continuously compounded rate $\eta>0$ per annum.
    
    

    
\end{enumerate}

{\color{black} In this paper where we consider a ratcheting mechanism, the benefit base is an additional state variable which serves as the basis for the guaranteed withdrawal amount.} In this case, the guaranteed withdrawal amount is a function of the current value of the benefit base and the benefit base may increase depending on the performance of the financial market. \citet{Bauer-2008} refer to this type of benefit base as the \textit{ratchet} benefit base. 

In traded VA contracts, issuers allow the policyholder to transfer the ownership of the contract to a beneficiary or to their estate in the event of the policyholder's death during the life of the contract.\footnote{{\color{black}See Page 114 of \cite{MLCPDS} when no Spouse Benefit option is selected.}} As such, we do not consider and model mortality\footnote{A similar assumption is adopted by \citet{Dai-2008} and \citet{AzimzadehForsyth-2015}, among others.}  risk in this paper {\color{black}allowing us to isolate the effects of the cash fund and ratcheting mechanism on the GMWB structure}.\footnote{{\color{black}Bundling life and death benefits within GMxB riders is not innocuous. The literature has established that bundling introduces valuation interdependencies affecting fair fees and policyholder behaviour \citep{Milevsky-2001, Bacinello-2011, Bauer-2023, mackay2017risk,AlonsoGarcia-2025}.}}

\subsection{Financial Market}

Let $(\Omega,\calF,\Q)$ be a complete probability space representing an arbitrage-free financial  market. We assume the financial market consists of a risk-free asset and a risky asset whose price processes are denoted by $S_0$ and $S_1$, respectively. The risk-free asset is assumed to compound continuously at a constant rate of $r>0$ per annum.\footnote{The risk-free rate can also be assumed to be a deterministic function of time or a stochastic process \citep[see, for example,][]{Gudkov-2019, Molent-2020, Shevchenko-2017}.} The processes $S_0$ and $S_1$ are assumed to evolve according to
\begin{align}
\label{eqn-PriceProcess-RiskFree}
\dif S_0(t) & = r S_0(t) \dif t, && S_0(0) = 1,\\
\label{eqn-PriceProcess-Risky}
\dif S_1(t) & = r S_1(t) \dif t + \sigma S_1(t) \dif B(t), && S_1(0) > 0.
\end{align}
Here, $\sigma>0$ is the (constant) volatility coefficient and $B$ is a standard $\Q$-Brownian motion. It is understood that $\Q$ is a probability measure under which $S_1/S_0$ is a $\Q$-martingale.

In traded VA contracts, the policyholder may choose from a menu of investment options into which their initial premium $P_0$ is invested.\footnote{See, for example, the \citet{MLCPDS} product disclosure statement, pp. 15-92.} The investment choices vary mainly in terms of their exposure to risky assets relative to ``risk-free'' instruments.\footnote{A similar assumption was made by \citet{Moenig-2021}, emphasizing that the inclusion of relatively risk-free instruments into the investment choice serves as downside protection and limits the investment's equity exposure.} Suppose the policyholder chooses the investment option which invests a proportion $\varrho\in[0,1]$ of the initial premium into the risky asset $S_1$ and the remainder into the risk-free asset $S_0$. Then from \eqref{eqn-PriceProcess-RiskFree} and \eqref{eqn-PriceProcess-Risky}, the corresponding portfolio value $\tilde{S}$ has dynamics under $\Q$ given by
\begin{equation}
\label{eqn-PriceProcess-Portfolio}
\dif\tilde{S}(t) = r\tilde{S}(t)\dif t + \varrho \sigma \tilde{S}(t) \dif B(t), \qquad \tilde{S}(0) = P_0.
\end{equation}
Standard calculations for a geometric Brownian motion imply that the return on $\tilde{S}$ over the period $[s,t]$ is
\begin{equation}
\label{eqn-PortfolioReturn}
\frac{\tilde{S}(t)}{\tilde{S}(s)} = \exp\Bigg\{\bigg(r - \frac{(\varrho\sigma)^2}{2}\bigg)(t-s) + \varrho \sigma (B(t)-B(s))\Bigg\}.
\end{equation}



\begin{rem}
The portfolio $\tilde{S}$ is different from the VA investment account $X$. We refer to $\tilde{S}$ for the dynamics of $X$, but the latter is affected by guarantee fees and other administrative fees charged by the insurer. 
\end{rem}





\subsection{Update Equations for the State Variables}
\label{sec-VAMechanics-NoDeath}

This section provides the update equations for the VA investment account value $X$, the benefit base $G$, and the cash account balance following the sequence of events outlined in Section \ref{sec-Notation}. 

Over the period $[t_{k-1}^+, t_k^-]$, the VA investment account $X$ grows according to the returns realized by the underlying portfolio $\tilde{S}$. Thus, we have
\begin{align}
    \begin{split}
        \label{eqn-Update-InvestmentAccount-Minus}
        X(t_k^-) 
            & = X(t_{k-1}^+)\frac{\tilde{S}(t_{k}^-)}{\tilde{S}(t_{k-1}^+)} \\
            & = X(t_{k-1}^+)\exp\Bigg\{\bigg(r - \frac{(\varrho\sigma)^2}{2}\bigg)(t_k - t_{k-1}) + \varrho \sigma (B(t_k) - B(t_{k-1})) \Bigg\},
    \end{split}
\end{align}
where $B(t_k) - B(t_{k-1})\sim N(0,t_k - t_{k-1})$ is the increment of $B$ over the time period $[t_{k-1},t_k]$. In other words, $X$ evolves according to the SDE 
\begin{equation}
	\label{eqn-InvestmentAccountSDE}
	\dif X(s) = rX(s) + \varrho \sigma X(s) \dif B(s), \qquad s\in(t_{k-1}^+, t_k^-]
\end{equation} 
with initial value $X(t_{k-1}^+)$. Note that the benefit base does not change between two consecutive withdrawal dates, thus 
\begin{equation}
\label{eqn-Update-BenefitBase-Minus}
G(t_k^-) = G(t_{k-1}^+).
\end{equation}

Prior to the withdrawal event at time $t_k$, we deduct the relevant fees from the VA investment account and calculate the guaranteed withdrawal amount $g(t_k)$. Under a ratchet mechanism, the amount $g(t_k)$ is a proportion $\beta\in(0,1)$ of the maximum between the \textit{post-fee} investment account value $X^{PF}(t_k^-)$, defined as 
\begin{equation}
\label{eqn-Update-InvestmentAccount-PostFee}
X^{PF}(t_k^-) := X(t_k^-)(1 - \varphi \Delta t),
\end{equation}
and the prevailing benefit base $G(t_k^-)$. That is,
\begin{equation}
\label{eqn-GuaranteedWithdrawalAmount}
g(t_k) = \beta \hat{G}(t_k^-)
\end{equation}
where $\hat{G}(t_k^-) := \max\{X^{PF}(t_k^-), G(t_k^-)\}$ denotes the ratchet mechanism applied to the post-fee investment account value and the prevailing benefit base.

\begin{rem}
    \label{rem-mathfrak-notation}
    In the numerical solution discussed in Section \ref{sec-NumericalSolution}, it is convenient to express relevant quantities in terms of the pre-event, pre-fee investment account balance $X(t_k^-)$ and the pre-event benefit base $G(t_k^-)$, which are our state variables. Define the functions $\mathfrak{f}(x)$ and $\mathfrak{g}(x,\gamma)$ as \[\mathfrak{f}(x) := x(1 - \varphi \Delta t) \quad \text{and} \quad \mathfrak{g}(x,\gamma) := \beta \max \{\mathfrak{f}(x), \gamma\}.\] We then have $X^{PF}(t_k^-) = \mathfrak{f}(X(t_k^-))$ and $g(t_k) = \mathfrak{g}(X(t_k^-), G(t_k^-))$.
\end{rem}

At this point, the policyholder withdraws an amount $w(t_k) \in \calW(t_k)$, where $\calW(t_k)$, representing the set of admissible withdrawals at time $t_k$, is given by $$\calW(t_k) := \{w: 0 \leq w \leq \max\{X^{PF}(t_k^-), g(t_k)\}\}.$$ A withdrawal is said to be an \textit{excess withdrawal} if $w(t_k) > g(t_k)$. Excess withdrawals can only occur when $X^{PF}(t_k^-) > g(t_k)$; otherwise, a withdrawal $w(t_k)>g(t_k)$ is not admissible. If the policyholder makes an excess withdrawal, then the benefit base will also decrease in addition to the decrease in the investment account value (these adjustments will be stated formally below). The policyholder is said to \textit{surrender the contract} if $X^{PF}(t_k^-) > g(t_k)$ and $w(t_k) = X^{PF}(t_k^-)$.


\begin{rem}
Intuitively, surrender no longer makes sense if $X^{PF}(t_k^-) \leq g(t_k)$, since the withdrawal is capped at $g(t_k)$. 
\end{rem}

\begin{dfn}
A withdrawal strategy $\vm{w}:=(w(t_1),\dots,w(t_{N-1}))$ is said to be admissible if, for each $k=1,2,\dots,N-1$, $w(t_k) \in \calW(t_k).$ Denote by $\calW$ the set of all admissible withdrawal strategies.
\end{dfn}

After the withdrawal, the investment account value is reduced by the \textit{guaranteed} withdrawal amount (including any excess withdrawal amounts). Thus,
\begin{equation}
\label{eqn-Update-InvestmentAccount-PostWithdrawal}
X(t_k^+) = \max\{0, X^{PF}(t_k^-) - g(t_k) - (w(t_k)-g(t_k))^+\}.
\end{equation}

If $w(t_k)<g(t_k)$, then the amount $g(t_k) - w(t_k)$ is transferred from the investment account to the cash fund and will appreciate over time according to the cash fund rate. 
{\color{black} At any time $t_k$, the balance of the cash fund before the withdrawal at time $t_k$ is represented as \[\mathscr{c}(t_k^-) = e^{\eta \Delta t} \mathscr{c}(t_{k-1}^+),\] reflecting the appreciation of the cash fund at the rate $\eta$. We note that, at the start of the contract, $\mathscr{c}(t_0^-) = \mathscr{c}(t_0^+) = 0$. Once a withdrawal is made at time $t_k$, the cash fund is updated as \[\mathscr{c}(t_k^+) = \mathscr{c}(t_k^-) + (g(t_k) - w(t_k))^+.\] We emphasize that the policyholder cannot withdraw from the cash fund and can only access it upon the expiry of the contract.}\footnote{\color{black} The MLC product disclosure statement is silent on the explicit treatment of the cash fund upon surrender, but it mentions that upon the conclusion of the income protection for any reason (including the cancellation by the policyholder), the investment balance remains in the chosen investment choice \citep[p. 115]{MLCPDS}.  We therefore infer that upon surrender, the cash fund balance is retained within the superannuation wrapper and will continue to earn interest at the cash rate until the contract matures (and thus behaves similar to a term deposit).}

The benefit base is updated depending on whether an excess withdrawal is made and on the value of the post-fee, post-withdrawal investment account relative to $\hat{G}(t_k^-) = \max\{X^{PF}(t_k^-), G(t_k^-)\}$, which is the ratcheted benefit base \textit{prior} to any downward adjustments due to excess withdrawals. Specifically, we have:
\begin{itemize}
    \item If $w(t_k) \leq g(t_k)$, then 
        \begin{equation}
        \label{eqn-Update-BenefitBase-NoExcess}
        G(t_k^+) = \hat{G}(t_k^-).
        \end{equation}
    \item If $w(t_k) > g(t_k)$, then
        \begin{align}
        \begin{split}
        \label{eqn-Update-BenefitBase-Excess}
        G(t_k^+) & = \hat{G}(t_k^-)\left(1 - \frac{w(t_k) - g(t_k)}{X^{PF}(t_k^-) - g(t_k)}\right).        
        \end{split}
        \end{align}
\end{itemize}
In the case of an excess withdrawal, the benefit base is adjusted on a proportional basis, where the proportion is the ratio between the excess withdrawal amount and the post-fee investment account after the reduction of the guaranteed withdrawal amount.\footnote{In a more general setting where withdrawals are permitted anytime and ratcheting occurs only during the policy anniversary dates, the downward adjustment to the benefit base in the case of an excess withdrawal may be made on a dollar-for-dollar basis provided the prevailing investment account value at the time of withdrawal is greater than the benefit base set at the beginning of the year. We do not consider this scenario in the current paper.} Indeed, if the policyholder withdraws the full amount remaining in the investment account (or surrenders the contract), then the benefit base reduces to 0, which further implies that the guaranteed withdrawal amount is zero after surrendering the contract.\footnote{If $w(t_k) = X^{PF}(t_k^-) > g(t_k)$ (i.e. full surrender), we get $G(t_k^+) = \hat{G}(t_k^-)(1 - 1) = 0$, which implies $g(t_j) = 0$ for $k < j \leq N$.]} However, the cash fund will continue to accrue interest until the expiry of the contract. 

\begin{rem}
\label{rem-UpdateEquation}
In the formulation of the optimal control problems below, it is convenient to denote the update equations yielding $X(t_k^+)$ and $G(t_k^+)$ by $h^X(X(t_k^-),G(t_k^-),w(t_k))$ and $h^G(X(t_k^-),G(t_k^-),w(t_k))$, respectively, where
\begin{equation}
    \label{eqn-Update-PostWithdrawalInvestmentAccount}
    h^X(x,\gamma,w) = \max\left\{0, \frakf(x) - \frakg(x,\gamma) - (w - \frakg(x,\gamma))^+\right\},
\end{equation}
and
\begin{equation}
    \label{eqn-Update-PostWithdrawalGuaranteeBase}
    h^G(x,\gamma,w) = 
    \begin{cases}
        \dfrac{1}{\beta} \frakg(x,\gamma) & \text{if $w \leq \frakg(x,\gamma)$},\\
        \dfrac{1}{\beta} \frakg(x,\gamma) \left(1 - \dfrac{w - \frakg(x,\gamma)}{\frakf(x) - \frakg(x,\gamma)}\right) & \text{if $w > \frakg(x,\gamma)$}. 
    \end{cases}
\end{equation}
This notation emphasizes that the updated investment account value and benefit base are functions of the withdrawal amount, the pre-withdrawal investment account value, and the pre-withdrawal benefit base. 
\end{rem}

\subsection{{\color{black} Tax Treatment of Cash Flows from the Contract}}

In the following we present the policyholder's total discounted cash flows when withdrawals and the interest gains on the cash fund are taxed. {\color{black} We note that the tax rates we consider are not agent-specific in the sense that generates non-linearity in \citet{Moenig-2016}: they are statutory parameters applied uniformly to cash flows from the contract, irrespective of the policyholder's external portfolio.}

{\color{black} Our taxation model closely follows Australian taxation rules, given that the hybrid VA contract design studied in this paper is a product offerred in the Australian market. Under Australian superannuation rules, pension payments are taxed at the policyholder's marginal rate\footnote{{\color{black}The marginal tax rates for Australian residents in 2024--25 are as follows \citep{ATOmarginaltax}: nil on income up to \$18,200; 16c per \$1 over \$18,200 (up to \$45,000); \$4,288 plus 30c per \$1 over \$45,000 (up to \$135,000); \$31,288 plus 37c per \$1 over \$135,000 (up to \$190,000); and \$51,638 plus 45c per \$1 over \$190,000.}} if accessed before the preservation age of 60 (with a 15\% offset for early access),\footnote{{\color{black} With the tax offset of 15\% upon early access, the marginal tax rates become 0\%, 1\%, 15\%, 22\% and 30\%, respectively.}} and are tax-free thereafter; lump sum withdrawals face a maximum rate of 20\% before preservation age and are likewise tax-free after.\footnote{See Page 8 of \cite{MLCPDS} based on the rules provided by \cite{ATOsupertax}.} Since our stylised framework does not distinguish between payment types, we study tax rates from 0\% to 20\%. While most policyholders would ordinarily wait until 60 to begin withdrawals, and hence receive tax-free payments throughout, evidence from the COVID-19 pandemic points to a growing incidence of early access \citep{ATOearlyaccess}, motivating our analysis of the taxable case.}

{\color{black} Generally, the premium paid to enter the VA contract, which is a form of a contribution to superannuation, may contain a tax-free component (arising from after-tax member contributions) and a taxable component (arising from concessional contributions and earnings) \citep{ATOtaxvnotax}. In this paper, we treat the initial premium $P_0$ is a post-tax amount; that is, an amount that arises entirely from the taxable component. This assumption represents the reality of higher income earners, who contribute above the concessional cap and are the typical customer base for this type of retirement income product.}\footnote{\color{black} The tax treatment of the initial premium is consistent with the tax-tax-exempt (TTE) model for retirement income products in Australia; in economies where the exempt-exempt-tax (EET) model is prevalent, additional taxes such as capital gains tax may become relevant. However, the EET taxation regime is outside the scope of this paper.}


{\color{black} To simplify, we only consider a single marginal tax rate $\theta$ that is applicable when the policyholder's age $a(t_k)$ at time $t_k$ is below the preservation age $a_p$. Define $\bar{\theta}(t_k)$ by
\[\bar{\theta}(t_k) = \theta \vm{1}_{\{a(t_k) < a_p\}}
	\begin{cases}
		\theta & \text{if $a(t_k) < a_p$} \\
		0 & \text{if $a(t_k) \geq a_p$}.
	\end{cases}\]
Here, $a(t_k) := a_b + t_k$ where $a_b$ is the policyholder's age when they purchased the VA. 
In this case, the post-tax value of a withdrawal at time $t_k$ is given by \[\text{post-tax withdrawal} = (1 - \bar{\theta}(t_k)) w(t_k), \qquad k = 1,\dots,N.\] Moreover, the taxation of the cash fund combines two aspects of Australian taxation rules:
\begin{itemize}
	\item Since the amount accumulated in the cash fund is only accessible to the policyholder at the maturity of the contract, we treat the cash fund as a term deposit. Under Australian taxation rules, the total interest earned on the cash fund over the life of the VA contract is declared when the contract matures at time $t_N$. The total interest earned as of the maturity of the contract is then taxed at the rate $\theta$ \citep{ATOinterest}.
	
	\item However, the cash fund is offered as part of the VA contract, which generally is a superannuation product. As such, if the contract matures after the preservation age, then the interest accrued over the life of the contract will be tax-exempt.
\end{itemize}
Under this taxation scheme, a contribution of $(g(t_k) - w(t_k))^+$ into the cash fund made at time $t_k$ has a post-tax value at $t_N$ given by \[\underbrace{(1 - \theta \vm{1}_{\{a_b + t_N < a_p \}}) (e^{\eta(t_N - t_k)} - 1) (g(t_k) - w(t_k))^+}_{\text{post-tax interest income on CF}} + (g(t_k) - w(t_k))^+,\] for $k=1,\dots,N-1$. Note that this expression can be written as \[\left[(1 - \theta \vm{1}_{\{a_b + t_N < a_p \}}) e^{\eta(t_N - t_k)} + \theta \vm{1}_{\{a_b + t_N < a_p \}}\right] (g(t_k) - w(t_k))^+.\] In the absence of taxes ($\theta = 0$) or when the VA contract matures after the preservation age ($a_b + t_N \geq a_p$), the above expression reduces to $e^{\eta(t_N - t_k)}(g(t_k) - w(t_k))^+$, which is the accumulated value at time $t_N$ of a contribution of $(g(t_k) - w(t_k))^+$ into the cash fund made at time $t_k$.


To properly account for the accrual and taxation of interest over the life of the VA, we denote by $\mathscr{C}(t_k; t_N)$ the post-interest tax accumulated value (at time $t_N$) of the cash fund after the withdrawal made at time $t_k$. This quantity is given by \[\mathscr{C}(t_k; t_N) = \sum_{j=1}^k \left[(1 - \theta \vm{1}_{\{a_b + t_N < a_p \}})e^{\eta(t_N - t_j)} + \theta \vm{1}_{\{a_b + t_N < a_p \}} \right] (g(t_j) - w(t_j))^+, \qquad k = 1, \dots, N - 1.\] We set $\mathscr{C}(t_N; t_N) = \mathscr{C}(t_{N-1}; t_N)$ since no cash fund injections occur at time $t_N$. In the absence of taxes or in a tax-free setting, we have $\mathscr{C}(t_N; t_N) = \mathscr{c}(t_N^-)$.
}

Therefore, given an admissible withdrawal strategy $\vm{w}\in\calW$ and the corresponding state variable trajectories $\vm{X}:=(X(t_1^-),\dots,X(t_N^-))$ and $\vm{G}:=(G(t_1^-),\dots,G(t_N^-))$ achieved under the control $\vm{w}$, the discounted value at time $t=0$ of all cash flows from the VA contract can be represented as
\begin{align}
	\begin{split}
		\label{eqn-DiscountedCashFlowsTaxed}
		H_0(\vm{X}, \vm{G}, \vm{w})
		& = \sum_{k=1}^{N-1} e^{-r t_k} (1 - {\color{black} \bar{\theta}(t_k)}) w(t_k) + e^{-r t_N} \vm{1}_{CF} {\color{black} \mathscr{C}(t_N; t_N)}\\
		& \qquad + e^{-r t_N} (1 - {\color{black} \bar{\theta}(t_N)}) \max \{X^{PF}(t_N^-), g(t_N)\},
	\end{split}
\end{align}
where $\vm{1}_{CF}$ is an indicator function that is equal to 1 when the VA contract includes a cash fund (and is equal to 0 otherwise). The first term can be written as follows
\begin{align*}
	& e^{-r t_N}  \sum_{k=1}^{N-1}  \vm{1}_{CF} \cdot \Big( [(1 - {\color{black} \theta \vm{1}_{\{a_b + t_N < a_p\}}}) e^{\eta (t_N - t_k)} + {\color{black} \theta \vm{1}_{\{a_b + t_N < a_p\}}}] (g(t_k) - w(t_k))^+ \Big) \\
	& \qquad = e^{-r t_N}  \sum_{k=1}^{N-1} \vm{1}_{CF} e^{-r t_k} e^{r t_k} [(1 - {\color{black} \theta \vm{1}_{\{a_b + t_N < a_p\}}}) e^{\eta (t_N - t_k)} + {\color{black} \theta \vm{1}_{\{a_b + t_N < a_p\}}}] (g(t_k) - w(t_k))^+ \\
	& \qquad = \sum_{k=1}^{N-1} \vm{1}_{CF} e^{-r t_k}  e^{-r (t_N - t_k)} [(1 - {\color{black} \theta \vm{1}_{\{a_b + t_N < a_p\}}}) e^{\eta (t_N - t_k)} + {\color{black} \theta \vm{1}_{\{a_b + t_N < a_p\}}}] (g(t_k) - w(t_k))^+ \\
	& \qquad = \sum_{k=1}^{N-1} \vm{1}_{CF} e^{-r t_k} \left[(1 - {\color{black} \theta \vm{1}_{\{a_b + t_N < a_p\}}}) e^{-(r - \eta) (t_N - t_k)} + {\color{black} \theta \vm{1}_{\{a_b + t_N < a_p\}}} e^{-r (t_N - t_k)}\right] (g(t_k) - w(t_k))^+.
\end{align*}

The total discounted cash flows can therefore be written in terms of a ``running cost'' and ``terminal cost'' term (as is usual in optimal control problems) as 
\begin{align}		
	\label{eqn-TotalDiscountedCF_Tax}
	H_0(\vm{X}, \vm{G}, \vm{w}) = \sum_{k=1}^{N-1} e^{-r t_k} C(t_k, X(t_k^-), G(t_k^-), w(t_k)) + e^{-r t_N} D(X(t_N^-), G(t_N^-)),
\end{align}
where
\begin{align}
    \begin{split}
        \label{eqn-TotalDiscountedCF-Running_Tax}
	& C(t_k, X(t_k^-), G(t_k^-), w(t_k)) \\
	& \qquad := (1 - {\color{black}\bar{\theta}(t_k)}) w(t_k) + \vm{1}_{CF} \left[(1 - {\color{black} \theta \vm{1}_{\{a_b + t_N < a_p\}}}) e^{-(r - \eta) (t_N - t_k)} + {\color{black} \theta \vm{1}_{\{a_b + t_N < a_p\}}} e^{-r (t_N - t_k)}\right] (g(t_k) - w(t_k))^+
    \end{split}
\end{align}
and
\begin{equation}
    \label{eqn-TotalDiscountedCF-Terminal_Tax}
    D(X(t_N^-), G(t_N^-)) := (1 - {\color{black} \bar{\theta}(t_N)}) \max \{X^{PF}(t_N^-), g(t_N)\}.
\end{equation}
From \eqref{eqn-Update-InvestmentAccount-PostFee} and \eqref{eqn-GuaranteedWithdrawalAmount}, we note that the right-hand sides of $C(t_k, X(t_k^-), G(t_k^-), w(t_k))$ and $D(X(t_N^-), G(t_N^-))$ are indeed functions of $X(t_k^-)$ and $G(t_k^-)$.

\begin{rem}
	If $\theta = 0$ (no tax), and $\vm{1}_{CF} = 1$ (policyholder has a cash fund), then we also obtain the discounted cash flows written in terms of a ``running cost'' and ``terminal cost'' term as
	\begin{equation}
		\label{eqn-TotalDiscountedCF}
		H_0(\vm{X},\vm{G},\vm{w}) = \sum_{k=1}^{N-1} e^{-r t_k}  C(t_k, X(t_k^-), G(t_k^-), w(t_k)) + e^{-r t_N} D(X(t_N^-), G(t_N^-)),
	\end{equation}
	where
	\begin{align}
		\label{eqn-TotalDiscountedCF-Running}
		C(t_k, X(t_k^-), G(t_k^-), w(t_k)) & := w(t_k) + e^{-(r-\eta)(t_N-t_k)} (g(t_k)-w(t_k))^+, \\
		\label{eqn-TotalDiscountedCF-Terminal}
		D(X(t_N^-), G(t_N^-)) & := \max\{X^{PF}(t_N^-), g(t_N)\}.
	\end{align}
\end{rem}

\subsection{Policyholder Optimization Problem}
\label{sec-PolicyholderOptimizationProblem}

We denote by $J_t(x,\gamma;\vm{w})$ the price of the VA contract at time $t$ given the withdrawal strategy $\vm{w}$, $X(t) = x$, and $G(t) = \gamma$. At $t=0$, the quantity $$J_0(x,\gamma;\vm{w}) := \E^\Q[H_0(\vm{X}, \vm{G}, \vm{w})],$$ yields the contract fair price under the withdrawal strategy $\vm{w}$. The above expectation is dependent on the guarantee fee $\varphi$. Consequently, the value of $\varphi > 0$ such that $P_0 = J_0(P_0,P_0;\vm{w})$ (noting that the right-hand side of the equation is dependent on the parameter $\varphi$) is the \textbf{fair guarantee fee} corresponding to the strategy $\vm{w}$. The fair guarantee fee is interpreted as the fee at which the policyholder receives the same after-tax (risk-neutral expected discounted) cash flows under the strategy $\vm{w}$ as the amount they paid to enter the contract. 

If the policyholder adopts a dynamic strategy, we assume that they choose their withdrawal strategy to maximize the present value of cash flows from the contract. That is, the policyholder seeks the optimal withdrawal strategy $\vm{w}^*\in\calW$ such that
\begin{equation}
\label{eqn-PHOptimization-t0}
J_0(x,\gamma;\vm{w}^*) = \sup_{\vm{w}\in\calW} \E^\Q[H_0(\vm{X},\vm{G},\vm{w})] =: V_0(x,\gamma).
\end{equation}
It is understood that the state variable trajectories $\vm{X}$ and $\vm{G}$ appearing in the expectation above are those corresponding to strategy $\vm{w}$ with initial values $X(t_0) = x$ and $G(t_0) = \gamma$. Of interest in the succeeding numerical analysis is the \textit{fair guarantee fee under the optimal withdrawal strategy}, the guarantee fee $\varphi^* \in [0,\infty)$ such that \[P_0 = V_0(P_0, P_0).\] Under contract specifications or tax rates where $V_0(P_0, P_0) < P_0$ for any assumed value of the guarantee fee $\varphi$, there exists no fair guarantee fee and the contract is not viable for the policyholder. In this case, the policyholder is better off investing their initial capital $P_0$ in a different investment opportunity outside the VA contract. 

More generally, we investigate the optimal control problem at each time $t_k^-$ prior to any withdrawals. To this end, the value function $V_{t_k}(x,\gamma)$ is given by
\begin{align}
\begin{split}
\label{eqn-PHOptimization-tk}
V_{t_k}(x,\gamma) 
    & = \sup_{\substack{w(t_j)\in\calW(t_j)\\ j=k,k+1,\dots,N-1}}\E^\Q\Bigg[\sum_{j=k}^{N-1} e^{-r(t_j-t_k)}C(t_j, X(t_j^-), G(t_j^-), w(t_j))\\
    & \qquad \qquad + e^{-r(t_N-t_k)}D(X(t_N^-), G(t_N^-)) \bigg| X(t_k^-) = x, G(t_k^-) = \gamma\Bigg].
\end{split}
\end{align}
This problem may be solved through backward induction via the Bellman equation
\begin{align}
\begin{split}
\label{eqn-PHOptimization-BellmanEquation}
V_{t_k}(x,\gamma) 
    & = \sup_{w(t_k)\in\calW(t_k)}\Bigg\{C(t_k, x,\gamma,w(t_k))\\
    & \qquad \qquad + \E^\Q_{w(t_k)}\left[e^{-r\Delta t} V_{t_{k+1}}(X(t_{k+1}^-), G(t_{k+1}^-)) | X(t_k^-) = x, G(t_k^-) = \gamma\right]\Bigg\},
\end{split}
\end{align}
with terminal condition $$V_{t_N}(x,\gamma) = D(x,\gamma).$$ 
The subscript ``$w(t_k)$'' on the expectation operator emphasizes that the expectation is taken with respect to the transition probability function of the state variables $(X(t),G(t))$ over the interval $[t_k^-, t_{k+1}^-]$ with initial state $(x,\gamma)$ at time $t_k^-$, if the amount $w(t_k)$ is withdrawn at time $t_k$.

The backward induction \eqref{eqn-PHOptimization-BellmanEquation} can be solved in two stages, first at time $t_k^+$ then at time $t_k^-$, following \citet{Gudkov-2019, Shevchenko-2017}, for example. Denoting by $V_{t_k^-}(\cdot)$ and $V_{t_k^+}(\cdot)$ the value of the contract before and after the withdrawal event at time $t_k$, respectively, we first compute the expectation
\begin{equation}
\label{eqn-PHOptimization-Stage1}
V_{t_k^+}(x,\gamma) = \E^\Q\left[e^{-r\Delta t} V_{t_{k+1}^-}(X(t_{k+1}^-), \gamma) | X(t_k^+) = x, G(t_k^+) = \gamma\right].
\end{equation}
Observe that we have $G(t_{k+1}^-) = \gamma$ in the expectation since the benefit base does not change on the interval $[t_k^+, t_{k+1}^-]$; see \eqref{eqn-Update-BenefitBase-Minus}. The value function prior to withdrawal is then solved as
\begin{equation}
\label{eqn-PHOptimization-Stage2}
V_{t_k^-}(x,\gamma) = \sup_{w(t_k)\in\calW(t_k)}\left\{C(t_k,x,\gamma,w(t_k)) + V_{t_k^+}(h^X(x,\gamma,w(t_k)), h^G(x,\gamma,w(t_k)))\right\}.
\end{equation}
The quantity $V_{t_k^+}(h^X(x,\gamma,w(t_k)), h^G(x,\gamma,w(t_k)))$ denotes the post-withdrawal value function if the policyholder makes a withdrawal of $w(t_k)$, resulting in $X(t_k^+) = h^X(x,\gamma,w(t_k))$ and $G(t_k^+) = h^G(x,\gamma,w(t_k))$; see Remark \ref{rem-UpdateEquation}. As in the Bellman equation, the two-stage process given by \eqref{eqn-PHOptimization-Stage1} and \eqref{eqn-PHOptimization-Stage2} are implemented for $k=N-1,N-2,\dots,0$ with terminal condition $V_{t_N^-} = D(X(t_N^-), G(t_N^-))$.

\section{Numerical Solution}
\label{sec-NumericalSolution}

\subsection{Set-Up and Notation}

In preparation for the numerical implementation of the backward induction in \eqref{eqn-PHOptimization-BellmanEquation}, we introduce alternative notation intended to clarify the quantities that are involved in the calculations. In particular, we denote by $x$ and $\gamma$ the value of the investment account balance and benefit base, respectively, either in the pre- or post-withdrawal context, whichever case applies.

First, we look at the calculation of the conditional expectation in \eqref{eqn-PHOptimization-Stage1}. For a fixed $k=N-1,N-2,\dots,1,0$ and for any $t\in[t_k^+, t_{k+1}^-]$, define the function $v(t,x,\gamma)$ by
\begin{equation}
    v(t,x,\gamma) := \E^\Q\left[e^{-r(t_{k+1}-t)} V_{t_{k+1}^-}(X(t_{k+1}^-), \gamma) | X(t) = x, G(t) = \gamma\right].
\end{equation}
Recall that, on the time interval $(t,t_{k+1}^-]$, the process $X$ evolves according to the SDE \[\dif X(s) = rX(s)\dif s + \varrho\sigma X(s) \dif B(s), \qquad X(t) = x, \quad s\in(t,t_{k+1}^-]\] {\color{black} and that $G(t)$ remains constant at $G(t_k^+)$ for $t\in[t_k^+, t_{k+1}^-]$.} Then, by the (discounted) Feynman-Kac formula \citep[Theorem 6.4.3]{Shreve-2004}, $v(t,x,\gamma)$ is a solution of the partial differential equation (PDE)
\begin{equation}
    \label{eqn-PHOptimization-PDE}
    \pder[v(t,x,\gamma)]{t} + rx\pder[v(t,x,\gamma)]{x} + \frac{1}{2}(\varrho \sigma x)^2 \pder[^2 v(t,x,\gamma)]{x^2} = rv(t,x,\gamma),
\end{equation}
for $(t,x,\gamma)\in[t_k^+, t_{k+1}^-] \times (0,\infty) \times \Gamma$, with terminal condition $v(t_{k+1},x,\gamma) = V_{t_{k+1}^-}(x,\gamma)$. Since the process goes backward in time, $V_{t_{k+1}^-}(x,\gamma)$ is known. Here, $\gamma$ is a parameter, so the PDE must be solved for each value of $\gamma\in\Gamma$. {\color{black} As such, no boundary conditions in $\gamma$ are necessary.} After obtaining the solution of \eqref{eqn-PHOptimization-PDE}, the value of the conditional expectation is given by $V_{t_k^+}(x,\gamma) = v(t_k,x,\gamma)$.

\begin{rem}
    The PDE \eqref{eqn-PHOptimization-PDE} will depend on the underlying dynamics of $X$, particularly the dynamics of the price processes of the primitive assets comprising the investment portfolio associated to the VA.
\end{rem}

In the numerical implementation, it is more convenient to deal with a PDE that proceeds forward in time. To this end, let $\tau = t_{k+1} - t$ and define $U(\tau,x,\gamma) = v(t_{k+1}-t, x,\gamma)$. Then $U$ is a solution of the PDE
\begin{equation}
    \label{eqn-PHOptimization-PDE2}
    -\pder[U(\tau,x,\gamma)]{\tau} + rx\pder[U(\tau,x,\gamma)]{x} + \frac{1}{2}(\varrho \sigma x)^2 \pder[^2 U(\tau,x,\gamma)]{x^2} = rU(\tau,x,\gamma),
\end{equation}
with \textit{initial} condition $U(0,x,\gamma) = V_{t_{k+1}^-}(x,\gamma)$. {\color{black} This PDE will be solved numerically over the truncated domain $[0, x_{\max}]$ for some suitably chosen $x_{\max}$. In the numerical solution of \eqref{eqn-PHOptimization-PDE2}, we will also assume that $\gamma$ takes values in an interval $[0, \gamma_{\max}]$ for some suitably chosen $\gamma_{\max}$.}\footnote{{\color{black}} The selection of $x_{\max}$ and $\gamma_{\max}$ is discussed in Section \ref{sec-NumericalIllustrations}.}

\subsection{Applying the Jump Condition via Interpolation}
\label{sec-JumpConditionMethod}

Second, we consider the jump condition represented by the maximization problem \eqref{eqn-PHOptimization-Stage2} with respect to the withdrawal strategy $w(t_k)$ at time $t_k$. The following algorithm is adapted from \citet{Gudkov-2019} and \citet{Shevchenko-2017} and uses the notation introduced in Remarks \ref{rem-mathfrak-notation} and \ref{rem-UpdateEquation}. Fix $k \in\{N-1,N-2,\dots,1,0\}$ and suppose that $V_{t_k^+}(x,\gamma)$ has been approximated by a numerical solution of \eqref{eqn-PHOptimization-PDE2}. This approximation is available only on the values of $(x,\gamma)$ on the grid {\color{black} $\mathbb{X} \times \mathbb{G} := \{0, x_1,\dots, x_{N_x}\} \times \{0, \gamma_1, \dots, \gamma_{N_\gamma}\}$ (that is, a discretization of $[0, x_{\max}] \times [0, \gamma_{\max}]$), where $N_x$ and $N_\gamma$ denote the number of grid points and $x_{N_x} = x_{\max}$ and $\gamma_{N_\gamma} = \gamma_{\max}$}. However, for a given $w\in\calW(t_k)$, the values $x' = h^X(x,\gamma,w)$ and $\gamma' = h^G(x,\gamma,w)$ may not be on this grid. Therefore, a two-dimensional interpolation in $(x,y)$ is necessary to approximate the value of $V_{t_k}^+(h^X(x,\gamma,w),h^G(x,\gamma,w))$ from the known values of $V_{t_k}^+(x,\gamma)$.

The interpolation and numerical solution of \eqref{eqn-PHOptimization-Stage2} proceeds as follows for each $(x,\gamma)\in\mathbb{X}\times\mathbb{G}$:
\begin{enumerate}
    \item {\color{black} Construct a grid $\mathbb{W} = \{0, w_1, \dots, w_{N_\gamma}\}$ of possible values of $w$ based on $\calW(t_k) = [0, \max\{\frakf(x), \frakg(x,\gamma)\}]$. The grid is the same size as the grid for $\gamma$ (for convenience) and is constructed with linearly spaced points, ensuring that the guaranteed withdrawal amount $g(t_k)$ is included in the grid.} 

    \item Compute the value of $h^X(x,\gamma,w)$ and $h^G(x,\gamma,w)$. 

    \item Approximate $V_{t_k}^+(h^X(x,\gamma,w),h^G(x,\gamma,w))$ by {\color{black} linear} interpolation from $\{V_{t_k}^+(x,\gamma)\}_{(x,\gamma)\in\mathbb{X}\times\mathbb{G}}$, the approximate solution  the PDE \eqref{eqn-PHOptimization-PDE2}.

    \item Choose the value of $w^*\in\mathbb{W}$ such that 
    \begin{equation}
        \label{eqn-OptimalWithdrawal}
        w^*(t_k) = w^*(t_k,x,\gamma) = \argmax_{w\in\mathbb{W}}\left\{C(t_k,x,\gamma,w(t_k)) + V_{t_k^+}(h^X(x,\gamma,w), h^G(x,\gamma,w))\right\}.
    \end{equation}
    {\color{black} The optimal withdrawal amount is determined via a grid search over $\mathbb{W}$ (similar to the implementation of \citet{Gudkov-2019}) since the objective function is not known to be concave in $w$ a priori.} Consequently, we have \[V_{t_k}^-(x,\gamma) = C(t_k, x, \gamma, w^*(t_k) + V_{t_k}^+(h^X(x,\gamma,w^*(t_k)), h^G(x,\gamma,w^*(t_k)).\]
\end{enumerate}
The approximation $\{V_{t_k}^-(x,\gamma)\}_{(x,\gamma)\in\mathbb{X}\times\mathbb{G}}$ is then used as the terminal condition to solve PDE \eqref{eqn-PHOptimization-PDE2} for $V_{t_{k-1}}^+(x,\gamma)$.

\subsection{Numerical Solution of the PDE via Method of Lines (MOL)}

We first consider the interval $[t_{N-1}^+,t_N^-]$, the time period before the VA matures. We solve PDE \eqref{eqn-PHOptimization-PDE2}, for a fixed $\gamma\in\Gamma\subseteq[0,+\infty)$, for $\tau\in(0,t_N-t_{N-1})=(0,\Delta t)$ and $x\in[0,+\infty)$, with initial condition
\begin{equation}
\label{eqn-PHOptimization-PDE2-InitialCondition}
U(0,x,\gamma) = D(x,\gamma) = \max\{\frakf(x),\frakg(x,\gamma)\}, \qquad x\in[0,+\infty).
\end{equation}
Boundary conditions in $x$ are discussed below:
\begin{itemize}
    \item \textit{Boundary condition along $x=0$:} Recall that
    \begin{align*}
        V_{t_{N-1}^+}(0,\gamma)
            & = \E_\Q\left[e^{-r(t_N-t_{N-1})} V_{t_N^-}(X(t_N^-), \gamma) | X(t_{N-1}^+) = 0, G(t_{N-1}^+) = \gamma\right]\\
            & = \E_\Q\left[e^{-r(t_N-t_{N-1})} D(X(t_N^-), \gamma) | X(t_{N-1}^+) = 0, G(t_{N-1}^+) = \gamma\right].
    \end{align*}
    From \eqref{eqn-Update-InvestmentAccount-Minus}, if $X(t_{N-1}^+) = 0$, then $X(t_N^-) = 0$ a.s. Therefore, \[V_{t_{N-1}^+}(0,\gamma) = e^{-r(t_N - t_{N-1})} D(0,\gamma) = e^{-r(t_N-t_{N-1})}\beta\gamma,\] and so we have the boundary condition
    \begin{equation}
    \label{eqn-PHOptimization-PDE2-BoundaryConditionx0}
    U(\tau,0,\gamma) = e^{-r\tau}\beta\gamma, \qquad \tau\in(0,\Delta t].
    \end{equation}
    In particular, we have $U(0,0,\gamma) = \beta\gamma$, which is consistent with the initial condition.

    \item \textit{Asymptotic boundary condition as $x\to+\infty$:} We adopt the boundary condition
    \begin{equation}
        \label{eqn-PHOptimization-PDE2-BoundaryConditionxinf}
        \pder[^2 U(\tau,x,\gamma)]{x^2} \to 0 \qquad \text{as $x\to\infty$}, \qquad \tau\in(0,\Delta t].
    \end{equation}
    This is justified by a formal differentiation of the Black-Scholes-type PDE and the application of Fichera theory for boundary conditions \citep[see, for example,][Example 1.17, with $\alpha=2$, $q=0$]{Meyer-2015}. This means that for very high values of the underlying investment account value, the value of the contract to the policyholder increases linearly with respect to the value of the investment account. \color{black} Since the PDE \eqref{eqn-PHOptimization-PDE2} will be solved numerically over a truncated domain $[0, x_{\max}]$, the asymptotic boundary condition will be applied as the boundary condition \[\pder[^2 U(\tau, x_{\max}, \gamma)]{x^2} = 0, \qquad \tau \in (0, \Delta \tau].\]
\end{itemize}

Let $\{\tau_0,\tau_1,\dots,\tau_{N_\tau}\}$ be a partition of $[0,\Delta t]$, where $\tau_0 = 0$, $\tau_{N_\tau} = \Delta t$, and $\Delta\tau = \tau_{n}-\tau_{n-1}$ for all $n=1,2,\dots,N_\tau$. Furthermore, let $U_n(x;\gamma) = U(\tau_n,x,\gamma)$ be the solution of \eqref{eqn-PHOptimization-PDE2} at $\tau = \tau_n$, parameterized by $\gamma$. We employ the three-level backward difference approximation for $\partial U/\partial \tau$,
\begin{equation}
    \label{eqn-MOL-FD-Time}
    \pder[U(\tau_n,x,\gamma)]{\tau} \approx 
    \begin{cases}
        \frac{U_n(x;\gamma) - U_{n-1}(x;\gamma)}{\Delta \tau}, & \text{for $n=1,2$}\\
        \frac{3}{2}\frac{U_n(x;\gamma) - U_{n-1}(x;\gamma)}{\Delta \tau} - \frac{1}{2}\dfrac{U_{n-1}(x;\gamma) - U_{n-2}(x;\gamma)}{\Delta \tau}, & \text{for $n=3,4,\dots,N_\tau$}.
    \end{cases}
\end{equation}
This approximation is unconditionally stable \citep[Appendix 2.2]{Meyer-2015}. The time discretization yields the approximation
\begin{equation}
    \label{eqn-MOL-PDEApproximation}
    a(x) U_{n}''(x;\gamma) + b(x) U_n'(x;\gamma) - c_n(x) U_n(x;\gamma) = F_n(x;\gamma)
\end{equation}
of PDE \eqref{eqn-PHOptimization-PDE2} at $\tau = \tau_n$, where
\begin{align*}
    a(x) & = {\color{black} \frac{1}{2}(\varrho \sigma x)^2} \\ 
    b(x) & = rx\\
    c_n(x) & =  \begin{cases}
                r + \frac{1}{\Delta \tau}, & \text{if $n=1,2$}\\
                r + \frac{3}{2\Delta \tau}, & \text{if $n=3,4,\dots,N_\tau$}
                \end{cases}\\
    F_n(x;\gamma) & =  \begin{cases}
                -\frac{1}{\Delta\tau} U_{n-1}(x;\gamma), & \text{if $n=1,2$}\\
                -\frac{4U_{n-1}(x;\gamma) - U_{n-2}(x;\gamma)}{2\Delta\tau}, & \text{if $n=3,4,\dots,N_\tau$}.
                \end{cases}
\end{align*}
The second-order differential equation can then be solved, for example, via the Riccati transform approach (see Appendix \ref{appendix-RiccatiTransform}). 

The same process is then used to numerically solve the equation \eqref{eqn-PHOptimization-PDE2} for $U(\tau, x, \gamma) = v(t_{k+1} - t, x, \gamma)$, for $k=1,\dots,N-1$, given the initial condition $U(0, x, \gamma) = V_{t_{k+1}^-}(x, \gamma)$ (obtained after applying the jump condition as discussed in Section \ref{sec-JumpConditionMethod}) and the same boundary conditions \eqref{eqn-PHOptimization-PDE2-BoundaryConditionx0} and \eqref{eqn-PHOptimization-PDE2-BoundaryConditionxinf}. We recover $V_{t_k^+}(x,\gamma) = v(t_k, x, \gamma)$ by reverting from $\tau$ to $t$.

\section{Numerical Illustrations}
\label{sec-NumericalIllustrations}

\subsection{Set-Up}

We conduct several numerical experiments to illustrate the valuation of the hybrid VA product and the effect of taxes and certain product features, such as the presence of a cash fund, the cash fund appreciation rate, and the type of benefit base update scheme presented in Section \ref{sec-MathematicalFormulation}. {\color{black} The numerical illustrations below consider two cases: (i) a \textit{full taxation case} in which the VA contract commencement and maturity occur before the preservation age; and (ii) a \textit{partial taxation case} where the contract commences before preservation age but matures after the preservation age. These cases reflect the emerging trend where policyholders commence early access on retirement income products before the preservation age and hence cash flows are fully or partially taxed. The case where the policyholder commences all withdrawals after the preservation age is covered by the case where the tax rate is 0, since cash flows occurring after the preservation age are tax-free.} 

In the base case, we assume that the model parameters, initial premium, the guaranteed withdrawal rate are as stated in Table \ref{tab-Parameters}. Specifically, we consider a 10-year product where withdrawals occur once a year at the policy anniversary. For each one-year period, the numerical estimate of the contract value is calculated over an 80-point grid in both the $x$- and $\gamma$-direction and a 50-point grid in the $\tau$-direction.\footnote{All computations were implemented in \texttt{R} on a MacBook Air running on the Apple M1 chip with 8 cores and 16GB RAM. 
{\color{black} A detailed analysis of grid sizes and computation times is provided in Appendix \ref{app-GridSize}.}
} 


\begin{table}[h]
\begin{threeparttable}
    \caption{Default parameter values and mesh sizes}
    \label{tab-Parameters}
    \begin{tabular}{@{}ccccc@{}}
    \toprule
    $P_0 = 100$ & $r = 0.03$ & $\varrho = 0.80$ & $\sigma = 0.20$ & $\beta = 0.10$ \\
    $T = 10$ & $\Delta t = 1$ & $N_x = 80$ & $N_\gamma = 80$ & $N_\tau = 50$ \\ \bottomrule
\end{tabular}
\end{threeparttable}
\end{table}

{\color{black} The computational grid $\mathbb{X} \times \mathbb{G}$ for the numerical solution of \eqref{eqn-PHOptimization-PDE2} is constructed as follows. To determine a suitable value of $x_{\max}$, we perform an exact simulation the geometric Brownian motion \eqref{eqn-InvestmentAccountSDE} with an initial value of $X(0) = P_0$ to obtain values of $X(T)$. Denote the set of simulated values of $X(T)$ by $\mathscr{X}$. We then set $x_{\max}$ equal to $\mathtt{mean}(\mathscr{X}) + 3 \mathtt{stdev}(\mathscr{X})$. The interval $[0, x_{\max}]$ is then split into three regions: $[0, \mathscr{x}_1]$, $[\mathscr{x}_1, \mathscr{x}_2]$, and $[\mathscr{x}_2, x_{\max}]$, where $\mathscr{x}_1$ and $\mathscr{x}_2$ are the 10th and 90th quantile of $\mathscr{X}$. Grid points in each sub-interval are then selected linearly such that the first and third sub-intervals each contain 10\% each of the total number of grid points $N_x$ and the remaining 80\% lie in the second sub-interval. The initial premium value $P_0$ is also included as a grid point (in the second sub-interval). The grid $\mathbb{G} = [0, \gamma_{\max}]$ is constructed in the same manner.}

To illustrate the effect of the cash rate and the tax rate, we allow the values of $\eta$ and $\theta$ to vary over the sets $\{2\%, 3\%, 4\%, 5\%\}$ and $\{0, 2.5\%, 5\%, 10\%, 20\%\}$, respectively. The values for $\eta$ are chosen such that it reflects a cash fund that performs worse than, similar to, or better than the risk-free asset.\footnote{As of 23 May 2025, the Reserve Bank of Australia's cash rate target is 3.85\%, while the spread between the term deposit rates offered by Australian financial institutions and the cash rate target is typically between 0.7\% and 0.8\%. Thus, the values of $\eta$ assumed in our numerical experiments represent a reasonable spread with respect to the risk-free rate $r$.} In this regard, we assume a constant risk-free rate of $r = 3\%$ to best analyze the effect of the spread between risk-free rate and the cash fund rate on the policyholder's valuation of the contract and their subsequent optimal withdrawal strategy. The values for $\theta$ allow us to determine the effect of having no taxes and marginally increasing the tax rate. 

\subsection{Analysis of Policyholder's Fair Fee and Optimal Withdrawal Behavior}

First, we investigate the fair guarantee fee $\varphi^*$ introduced in Section \ref{sec-PolicyholderOptimizationProblem} under different contract specifications and tax rates. To determine the fair guarantee fee, we consider an equally-spaced partition of the interval $[0, 400]$ (in basis points) and value the contract for each $\varphi$ in the partition following the method outlined in Section \ref{sec-NumericalSolution}. We then approximate $\varphi^*$ as the root of the cubic spline interpolant of the equation $P_0 - V_{t_0}(P_0, P_0; \varphi) = 0$ using the pairs of guarantee fee and VA contract values obtained in the previous step. 

\begin{table}[h]
	\caption{Fair guarantee fees $\varphi^*$ (in basis points) under a static and dynamic withdrawal strategy for various marginal tax rates $\theta$, when the benefit base is updated by a ratchet mechanism or not and $\eta=4\%$}
	\label{tab-FairGuaranteeFee-StaticvsDynamic}
	\begin{subtable}{\textwidth}
		\centering
		\begin{tabular}{@{}crrrrrr@{}}
			\toprule
			& \multicolumn{2}{c}{Static} & \multicolumn{4}{c}{Dynamic} \\ \cmidrule(l){4-7} 
			& \multicolumn{2}{c}{} &\multicolumn{2}{c}{No Cash Fund} & \multicolumn{2}{c}{Cash Fund ($\eta = 4\%$)} \\ \cmidrule(l){4-7} 
			Tax Rate &\multicolumn{1}{c}{No Ratchet} & \multicolumn{1}{c}{Ratchet} & \multicolumn{1}{c}{No Ratchet} & \multicolumn{1}{c}{Ratchet} & \multicolumn{1}{c}{No Ratchet} & \multicolumn{1}{c}{Ratchet}  \\ \midrule
			No Tax &54.9559 & 86.6630 & 112.4728 & 126.8871 & 204.6804 & 230.1654 \\
			$\theta = 2.5\%$ & 7.6246 & 33.5875 & 27.0643 & 57.2017 & 154.0414 & 198.3923  \\
			$\theta = 5\%$ &NA & NA &NA & NA & 118.2448 & 172.2789 \\
			$\theta = 10\%$ & NA & NA &  NA & NA & 67.9520 & 127.1135\\ 
			$\theta = 20\%$ & NA & NA &  NA & NA & NA & 48.1429 \\
			\bottomrule
		\end{tabular}
		\caption{Full taxation case}
	\end{subtable} \\
	\begin{subtable}{\textwidth}
		\centering
		\begin{tabular}{@{}crrrrrr@{}}
			\toprule
			& \multicolumn{2}{c}{Static} & \multicolumn{4}{c}{Dynamic} \\ \cmidrule(l){4-7} 
			& \multicolumn{2}{c}{} &\multicolumn{2}{c}{No Cash Fund} & \multicolumn{2}{c}{Cash Fund ($\eta = 4\%$)} \\ \cmidrule(l){4-7} 
			Tax Rate &\multicolumn{1}{c}{No Ratchet} & \multicolumn{1}{c}{Ratchet} & \multicolumn{1}{c}{No Ratchet} & \multicolumn{1}{c}{Ratchet} & \multicolumn{1}{c}{No Ratchet} & \multicolumn{1}{c}{Ratchet}  \\ \midrule
			No Tax & 54.9559 & 86.6630 & 112.4728 & 126.8871 & 204.6804 & 230.1654 \\
			$\theta = 2.5\%$ & 41.2510 & 69.6740 & 61.4828 & 90.6220 & 179.0962 & 222.6583 \\
			$\theta = 5\%$ & 27.8230 & 52.9606 & 39.5713 & 65.5065 & 178.0443 & 222.6208 \\
			$\theta = 10\%$ & 4.4572 & 20.4030 & NA & 17.0322 & 178.0370 & 222.6208 \\
			$\theta = 20\%$ & NA & NA & NA & NA & 178.0370 & 222.6208 \\
			\bottomrule
		\end{tabular}
		\caption{\color{black} Partial taxation case}
	\end{subtable}
\end{table}

\begin{table}[h]
	\caption{Fair guarantee fees $\varphi^*$ (in basis points) with a ratchet under a static and dynamic withdrawal strategy for various marginal tax rates $\theta$ and cash account appreciation rates $\eta$.}
	\label{tab-FairGuaranteeFee-EE80}
	\begin{subtable}{\textwidth}
		\centering
		\begin{tabular}{@{}crrrrr@{}}
			\toprule
			& \multicolumn{1}{c}{} & \multicolumn{4}{c}{Dynamic} \\ \cmidrule(l){3-6} 
			Tax Rate & \multicolumn{1}{c}{Static} & \multicolumn{1}{c}{$\eta = 2\%$} & \multicolumn{1}{c}{$\eta = 3\%$} & \multicolumn{1}{c}{$\eta = 4\%$} & \multicolumn{1}{c}{$\eta = 5\%$} \\ \midrule
			No Tax & 86.6630 & 126.8871 & 126.8871 & 230.1654 & 351.6089 \\
			$\theta = 2.5\%$ & 35.5875 & 58.4112 & 92.3337 & 198.3923 & 319.5487 \\
			$\theta = 5\%$ & NA & 5.1702 & 67.7839 & 172.2789 & 291.3602 \\
			$\theta = 10\%$ & NA & NA & 31.1690 & 127.1135 & 239.0831 \\ 
			$\theta = 20\%$ & NA & NA & NA & 48.1429 & 147.6102 \\
			\bottomrule
		\end{tabular}
		\caption{Full taxation case}
	\end{subtable} \\
	\begin{subtable}{\textwidth}
		\centering
		\begin{tabular}{@{}crrrrr@{}}
			\toprule
			& \multicolumn{1}{c}{} & \multicolumn{4}{c}{Dynamic} \\ \cmidrule(l){3-6} 
			Tax Rate & \multicolumn{1}{c}{Static} & \multicolumn{1}{c}{$\eta = 2\%$} & \multicolumn{1}{c}{$\eta = 3\%$} & \multicolumn{1}{c}{$\eta = 4\%$} & \multicolumn{1}{c}{$\eta = 5\%$} \\ \midrule
			No Tax & 86.6630 & 126.8871 & 126.8871 & 230.1654 & 351.6089 \\
			$\theta = 2.5\%$ & 69.6740 & 90.6220 & 115.9566 & 222.6583 & 344.7695 \\
			$\theta = 5\%$ & 52.9606 & 65.5065 & 115.9611 & 222.6208 & 344.8070 \\
			$\theta = 10\%$ & 20.4030 & 43.5070 & 115.9637 & 222.6208 & 344.8070 \\
			$\theta = 20\%$ & NA & 43.5070 & 115.9637 & 222.6208 & 344.8070 \\
			\bottomrule
		\end{tabular}
		\caption{\color{black} Partial taxation case}
	\end{subtable}
\end{table}

Table \ref{tab-FairGuaranteeFee-StaticvsDynamic} illustrates the impact of taxation and contract features on the fair guarantee fee, whereas Table \ref{tab-FairGuaranteeFee-EE80} shows the impact of the cash rate and taxation on the fair guarantee fee. {\color{black} We write ``NA'' to indicate that no $\varphi^*\geq 0$ exists which satisfies the above equation (that is, when the policyholder valuation is strictly less than $P_0$ for all values of $\varphi$ considered).} In Table \ref{tab-FairGuaranteeFee-StaticvsDynamic}, The effect of the ratchet mechanism for the benefit base update, the presence of a cash fund (in the dynamic case) and its interaction with various tax rates is examined. For the dynamic case, we consider a cash rate of 4\%, since the cash fund is typically managed to outperform a benchmark such as the risk-free rate (see, for example, \citet[pp. 49-50]{MLCPDS}) which in our case is chosen as $r=3\%$ (Table \ref{tab-Parameters}). Note that in the static case there is no cash fund effect as the static withdrawal behavior corresponds to withdrawing exactly the guaranteed amount, meaning that no cash fund injections are done. 

Second, we analyze policyholder's optimal withdrawal behavior under the fair guarantee fee under various contract and taxation specifications. Tables \ref{tab-OptWithSummary-ScenarioAnalysis} and \ref{tab-OptWithSummary-ScenarioAnalysis_etaVAR} present simulation results on optimal withdrawal and surrender behavior {\color{black} under scenarios generated under the risk-neutral dynamics of the underlying investment fund, assuming that the guarantee fee is equal to the policyholder fair guarantee fee $\varphi^*$. These results therefore characterize optimal withdrawal and surrender decisions undertaken by rational policyholders.} Table \ref{tab-OptWithSummary-ScenarioAnalysis} explores the impact of contract design features, specifically the presence or absence of a cash fund and a ratchet mechanism, while Table \ref{tab-OptWithSummary-ScenarioAnalysis_etaVAR}  varies the cash rate $\eta$ and tax rate $\theta$. The surrender rate indicates the proportion of scenarios where the policyholder surrenders the contract, and the average surrender time refers to the mean time until surrender. We write ``NA'' if the scenario does not yield a policyholder fair guarantee fee $\varphi^*$ or, if $\varphi^*$ exists, to denote that surrender was not observed. The average duration measures time spent in the contract, including pre-surrender periods. The final columns report the proportion of withdrawal opportunities ($T - 1$ total) where policyholders engage in each withdrawal type, conditional on not surrendering: ``No Withdrawal'', ``Withdrawal Below Guaranteed'', ``Withdrawal at Guaranteed Amount'', and ``Excess Withdrawal''. {\color{black} Further details on the simulation methodology and how these quantities are calculated are provided in Appendix \ref{app-SimulatedWithdrawals}. The results shown below are based on 10,000 simulations.}

\begin{table}[t!]
	\caption{Simulation analysis of optimal withdrawal behavior and surrender for each contract configuration or scenario considered in Table \ref{tab-FairGuaranteeFee-StaticvsDynamic}.}
	\label{tab-OptWithSummary-ScenarioAnalysis}
	\begin{subtable}{\textwidth}
		\centering
		\resizebox{\columnwidth}{!}{
		\begin{tabular}{@{}cccrrrrrrrr@{}}
			\toprule
			Tax Rate & Cash Fund? & Ratchet? & \multicolumn{1}{c}{$\varphi^*$} & \multicolumn{1}{c}{Surr. Rate} & \multicolumn{1}{c}{\begin{tabular}[c]{@{}c@{}}Avg. \\ Surr. Time\end{tabular}} & \multicolumn{1}{c}{\begin{tabular}[c]{@{}c@{}}Avg. \\ Duration\end{tabular}} & \multicolumn{1}{c}{No With.} & \multicolumn{1}{c}{\begin{tabular}[c]{@{}c@{}}With.\\ Below Guar.\end{tabular}} & \multicolumn{1}{c}{\begin{tabular}[c]{@{}c@{}}With.\\ At Guar.\end{tabular}} & \multicolumn{1}{c}{\begin{tabular}[c]{@{}c@{}}Excess\\ With.\end{tabular}} \\ \midrule
			0.0\% & FALSE & FALSE & 112.4728 & 0.3227 & 2.5357 & 7.6062 & 0.0579 & 0.0000 & 0.5520 & 0.3902 \\
			2.5\% & FALSE & FALSE & 27.0643 & 0.1506 & 3.6108 & 9.0429 & 0.0000 & 0.0000 & 0.7087 & 0.2913 \\
			5.0\% & FALSE & FALSE & NA & NA & NA & NA & NA & NA & NA & NA \\
			10.0\% & FALSE & FALSE & NA & NA & NA & NA & NA & NA & NA & NA \\
			20.0\% & FALSE & FALSE & NA & NA & NA & NA & NA & NA & NA & NA \\
			0.0\% & FALSE & TRUE & 126.8871 & 2e-04 & 8.0000 & 9.9996 & 0.0002 & 0.0000 & 0.6507 & 0.3491 \\
			2.5\% & FALSE & TRUE & 57.2017 & 0.0000 & NA & 10.0000 & 0.0000 & 0.0000 & 0.7437 & 0.2563 \\
			5.0\% & FALSE & TRUE & NA & NA & NA & NA & NA & NA & NA & NA \\
			10.0\% & FALSE & TRUE & NA & NA & NA & NA & NA & NA & NA & NA \\
			20.0\% & FALSE & TRUE & NA & NA & NA & NA & NA & NA & NA & NA \\
			0.0\% & TRUE & FALSE & 204.6804 & 0.2953 & 2.0308 & 7.6467 & 0.8155 & 0.0311 & 0.0000 & 0.1534 \\
			2.5\% & TRUE & FALSE & 154.0414 & 0.1991 & 2.9702 & 8.6060 & 0.8675 & 0.0288 & 0.0000 & 0.1036 \\
			5.0\% & TRUE & FALSE & 118.2448 & 0.1194 & 4.1323 & 9.2994 & 0.8782 & 0.0218 & 0.0000 & 0.1000 \\
			10.0\% & TRUE & FALSE & 67.9520 & 0.0132 & 7.7045 & 9.9697 & 0.9687 & 0.0093 & 0.0000 & 0.0220 \\
			20.0\% & TRUE & FALSE & NA & NA & NA & NA & NA & NA & NA & NA \\
			0.0\% & TRUE & TRUE & 230.1654 & 0.0000 & NA & 10.0000 & 0.8275 & 0.0647 & 1.1e-05 & 0.1079 \\
			2.5\% & TRUE & TRUE & 198.3923 & 0.0000 & NA & 10.0000 & 0.8658 & 0.0406 & 0.0000 & 0.0936 \\
			5.0\% & TRUE & TRUE & 172.2789 & 0.0000 & NA & 10.0000 & 0.8842 & 0.0291 & 1.1e-05 & 0.0867 \\
			10.0\% & TRUE & TRUE & 127.1135 & 0.0000 & NA & 10.0000 & 0.9412 & 0.0196 & 1.1e-05 & 0.0392 \\
			20.0\% & TRUE & TRUE & 48.1429 & 0.0000 & NA & 10.0000 & 0.9710 & 0.0108 & 1.1e-05 & 0.0182 \\\bottomrule
		\end{tabular}
		}
		\caption{Full taxation case}
	\end{subtable} \\
	\begin{subtable}{\textwidth}
		\centering
		\resizebox{\columnwidth}{!}{
		\begin{tabular}{@{}cccrrrrrrrr@{}}
			\toprule
			Tax Rate & Cash Fund? & Ratchet? & \multicolumn{1}{c}{$\varphi^*$} & \multicolumn{1}{c}{Surr. Rate} & \multicolumn{1}{c}{\begin{tabular}[c]{@{}c@{}}Avg. \\ Surr. Time\end{tabular}} & \multicolumn{1}{c}{\begin{tabular}[c]{@{}c@{}}Avg. \\ Duration\end{tabular}} & \multicolumn{1}{c}{No With.} & \multicolumn{1}{c}{\begin{tabular}[c]{@{}c@{}}With.\\ Below Guar.\end{tabular}} & \multicolumn{1}{c}{\begin{tabular}[c]{@{}c@{}}With.\\ At Guar.\end{tabular}} & \multicolumn{1}{c}{\begin{tabular}[c]{@{}c@{}}Excess\\ With.\end{tabular}} \\ \midrule
			0.0\% & FALSE & FALSE & 112.4728 & 0.3227 & 2.5357 & 7.6062 & 0.0579 & 0.0000 & 0.5520 & 0.3902 \\
			2.5\% & FALSE & FALSE & 61.4828 & 0.1028 & 5.6998 & 9.5588 & 1.2E-05 & 0.0000 & 0.7610 & 0.2390 \\
			5.0\% & FALSE & FALSE & 39.5713 & 0.1054 & 5.6435 & 9.5478 & 0.0000 & 0.0000 & 0.7463 & 0.2537 \\
			10.0\% & FALSE & FALSE & NA & NA & NA & NA & NA & NA & NA & NA \\
			20.0\% & FALSE & FALSE & NA & NA & NA & NA & NA & NA & NA & NA \\
			0.0\% & FALSE & TRUE & 126.8871 & 2.0E-04 & 8.0000 & 9.9996 & 0.0002 & 0.0000 & 0.6507 & 0.3491 \\
			2.5\% & FALSE & TRUE & 90.6220 & 0.0000 & NA & 10.0000 & 0.0000 & 0.0000 & 0.7905 & 0.2095 \\
			5.0\% & FALSE & TRUE & 65.5065 & 0.0000 & NA & 10.0000 & 0.0000 & 0.0000 & 0.7905 & 0.2095 \\
			10.0\% & FALSE & TRUE & 17.0322 & 0.0000 & NA & 10.0000 & 0.0000 & 0.0000 & 0.7762 & 0.2238 \\
			20.0\% & FALSE & TRUE & NA & NA & NA & NA & NA & NA & NA & NA \\
			0.0\% & TRUE & FALSE & 204.6804 & 0.2953 & 2.0308 & 7.6467 & 0.8155 & 0.0311 & 0.0000 & 0.1534 \\
			2.5\% & TRUE & FALSE & 179.0962 & 0.1162 & 3.7214 & 9.2767 & 0.8426 & 0.0331 & 0.0000 & 0.1244 \\
			5.0\% & TRUE & FALSE & 178.0443 & 0.0840 & 5.5409 & 9.6290 & 0.8352 & 0.0338 & 0.0000 & 0.1309 \\
			10.0\% & TRUE & FALSE & 178.0370 & 0.0838 & 5.5396 & 9.6280 & 0.8352 & 0.0339 & 0.0000 & 0.1309 \\
			20.0\% & TRUE & FALSE & 178.0370 & 0.0841 & 5.5415 & 9.6295 & 0.8352 & 0.0339 & 0.0000 & 0.1309 \\
			0.0\% & TRUE & TRUE & 230.1654 & 0.0000 & NA & 10.0000 & 0.8275 & 0.0647 & 1.1E-05 & 0.1079 \\
			2.5\% & TRUE & TRUE & 222.6583 & 0.0000 & NA & 10.0000 & 0.8778 & 0.0381 & 0.0000 & 0.0841 \\
			5.0\% & TRUE & TRUE & 222.6208 & 0.0000 & NA & 10.0000 & 0.8778 & 0.0381 & 0.0000 & 0.0841 \\
			10.0\% & TRUE & TRUE & 222.6208 & 0.0000 & NA & 10.0000 & 0.8778 & 0.0381 & 0.0000 & 0.0841 \\
			20.0\% & TRUE & TRUE & 222.6208 & 0.0000 & NA & 10.0000 & 0.8778 & 0.0381 & 0.0000 & 0.0841 \\
			\bottomrule
		\end{tabular}
		}
		\caption{\color{black} Partial taxation case}
	\end{subtable}
\end{table}

\begin{table}[h]
	\caption{Simulation analysis of optimal withdrawal behavior and surrender for each case considered in Table \ref{tab-FairGuaranteeFee-EE80}.}
	\label{tab-OptWithSummary-ScenarioAnalysis_etaVAR}
	\begin{subtable}{\textwidth}
		\centering
		\resizebox{\columnwidth}{!}{
			\begin{tabular}{@{}ccrrrrrrrr@{}}
				\toprule
				Cash Rate & Tax Rate & \multicolumn{1}{c}{$\varphi^*$} & Surr. Rate & \begin{tabular}[c]{@{}c@{}}Avg.\\ Surr. Time\end{tabular} & \multicolumn{1}{c}{\begin{tabular}[c]{@{}c@{}}Avg. \\ Duration\end{tabular}} & \multicolumn{1}{c}{No With.} & \multicolumn{1}{c}{\begin{tabular}[c]{@{}c@{}}With. \\ Below Guar.\end{tabular}} & \multicolumn{1}{c}{\begin{tabular}[c]{@{}c@{}}With. \\ At Guar.\end{tabular}} & \multicolumn{1}{c}{\begin{tabular}[c]{@{}c@{}}Excess\\ With.\end{tabular}} \\ \midrule
				2.0\% & 0.0\% & 126.8871 & 2e-04 & 8.0000 & 9.9996 & 0.0002 & 0.0000 & 0.6507 & 0.3491 \\
				3.0\% & 0.0\% & 126.8871 & 0.0000 & NA & 10.0000 & 0.4727 & 0.2744 & 0.0531 & 0.1993 \\
				4.0\% & 0.0\% & 230.1654 & 0.0000 & NA & 10.0000 & 0.8275 & 0.0647 & 1.1e-05 & 0.1079 \\
				5.0\% & 0.0\% & 351.6089 & 0.0000 & NA & 10.0000 & 0.8283 & 0.0734 & 0.0000 & 0.0983 \\
				2.0\% & 2.5\% & 58.4112 & 0.0000 & NA & 10.0000 & 0.1333 & 0.0333 & 0.6228 & 0.2105 \\
				3.0\% & 2.5\% & 92.3337 & 0.0000 & NA & 10.0000 & 0.8186 & 0.0488 & 0.0000 & 0.1326 \\
				4.0\% & 2.5\% & 198.3923 & 0.0000 & NA & 10.0000 & 0.8658 & 0.0406 & 0.0000 & 0.0936 \\
				5.0\% & 2.5\% & 319.5487 & 0.0000 & NA & 10.0000 & 0.8666 & 0.0597 & 0.0000 & 0.0737 \\
				2.0\% & 5.0\% & 5.1702 & 0.0000 & NA & 10.0000 & 0.2998 & 0.0343 & 0.5366 & 0.1293 \\
				3.0\% & 5.0\% & 67.7839 & 0.0000 & NA & 10.0000 & 0.8697 & 0.0277 & 1.1e-05 & 0.1025 \\
				4.0\% & 5.0\% & 172.2789 & 0.0000 & NA & 10.0000 & 0.8842 & 0.0291 & 1.1e-05 & 0.0867 \\
				5.0\% & 5.0\% & 291.3602 & 0.0000 & NA & 10.0000 & 0.9115 & 0.0301 & 0.0000 & 0.0584 \\
				2.0\% & 10.0\% & NA & NA & NA & NA & NA & NA & NA & NA \\
				3.0\% & 10.0\% & 31.1690 & 0.0000 & NA & 10.0000 & 0.9428 & 0.0209 & 0.0000 & 0.0363 \\
				4.0\% & 10.0\% & 127.1135 & 0.0000 & NA & 10.0000 & 0.9412 & 0.0196 & 1.1e-05 & 0.0392 \\
				5.0\% & 10.0\% & 239.0831 & 0.0000 & NA & 10.0000 & 0.9386 & 0.0146 & 0.0000 & 0.0468 \\
				2.0\% & 20.0\% & NA & NA & NA & NA & NA & NA & NA & NA \\
				3.0\% & 20.0\% & NA & NA & NA & NA & NA & NA & NA & NA \\
				4.0\% & 20.0\% & 48.1429 & 0.0000 & NA & 10.0000 & 0.9710 & 0.0108 & 1.1e-05 & 0.0182 \\
				5.0\% & 20.0\% & 147.6102 & 0.0000 & NA & 10.0000 & 0.9734 & 0.0101 & 0.0000 & 0.0165 \\ \bottomrule
			\end{tabular}
			}
		\caption{Full taxation case}
	\end{subtable} \\
	\begin{subtable}{\textwidth}
		\centering
		\resizebox{\columnwidth}{!}{
			\begin{tabular}{@{}ccrrrrrrrr@{}}
				\toprule
				Cash Rate & Tax Rate & \multicolumn{1}{c}{$\varphi^*$} & Surr. Rate & \begin{tabular}[c]{@{}c@{}}Avg.\\ Surr. Time\end{tabular} & \multicolumn{1}{c}{\begin{tabular}[c]{@{}c@{}}Avg. \\ Duration\end{tabular}} & \multicolumn{1}{c}{No With.} & \multicolumn{1}{c}{\begin{tabular}[c]{@{}c@{}}With. \\ Below Guar.\end{tabular}} & \multicolumn{1}{c}{\begin{tabular}[c]{@{}c@{}}With. \\ At Guar.\end{tabular}} & \multicolumn{1}{c}{\begin{tabular}[c]{@{}c@{}}Excess\\ With.\end{tabular}} \\ \midrule
				2.0\% & 0.0\% & 126.8871 & 2.0e-04 & 8.0000 & 9.9996 & 0.0002 & 0.0000 & 0.6507 & 0.3491 \\
				3.0\% & 0.0\% & 126.8871 & 0.0000 & NA & 10.0000 & 0.4727 & 0.2744 & 0.0531 & 0.1993 \\
				4.0\% & 0.0\% & 230.1654 & 0.0000 & NA & 10.0000 & 0.8275 & 0.0647 & 1.1e-05 & 0.1079 \\
				5.0\% & 0.0\% & 351.6089 & 0.0000 & NA & 10.0000 & 0.8283 & 0.0734 & 0.0000 & 0.0983 \\
				2.0\% & 2.5\% & 90.6220 & 0.0000 & NA & 10.0000 & 0.0000 & 0.0000 & 0.7905 & 0.2095 \\
				3.0\% & 2.5\% & 115.9566 & 0.0000 & NA & 10.0000 & 0.6295 & 0.1721 & 0.0558 & 0.1423 \\
				4.0\% & 2.5\% & 222.6583 & 0.0000 & NA & 10.0000 & 0.8778 & 0.0381 & 0.0000 & 0.0841 \\
				5.0\% & 2.5\% & 344.7695 & 0.0000 & NA & 10.0000 & 0.8930 & 0.0331 & 0.0000 & 0.0739 \\
				2.0\% & 5.0\% & 65.5065 & 0.0000 & NA & 10.0000 & 0.0000 & 0.0000 & 0.7905 & 0.2095 \\
				3.0\% & 5.0\% & 115.9611 & 0.0000 & NA & 10.0000 & 0.6488 & 0.1537 & 0.0558 & 0.1415 \\
				4.0\% & 5.0\% & 222.6208 & 0.0000 & NA & 10.0000 & 0.8778 & 0.0381 & 0.0000 & 0.0841 \\
				5.0\% & 5.0\% & 344.8070 & 0.0000 & NA & 10.0000 & 0.8930 & 0.0330 & 1.1e-05 & 0.0739 \\
				2.0\% & 10.0\% & 43.5070 & 0.0000 & NA & 10.0000 & 0.4444 & 0.0000 & 0.3368 & 0.2188 \\
				3.0\% & 10.0\% & 115.9637 & 0.0000 & NA & 10.0000 & 0.5997 & 0.2008 & 0.0565 & 0.1428 \\
				4.0\% & 10.0\% & 222.6208 & 0.0000 & NA & 10.0000 & 0.8778 & 0.0381 & 0.0000 & 0.0841 \\
				5.0\% & 10.0\% & 344.8070 & 0.0000 & NA & 10.0000 & 0.8930 & 0.0330 & 1.1e-05 & 0.0739 \\
				2.0\% & 20.0\% & 43.5070 & 0.0000 & NA & 10.0000 & 0.4444 & 0.0000 & 0.3368 & 0.2188 \\
				3.0\% & 20.0\% & 115.9637 & 0.0000 & NA & 10.0000 & 0.6419 & 0.1602 & 0.0556 & 0.1420 \\
				4.0\% & 20.0\% & 222.6208 & 0.0000 & NA & 10.0000 & 0.8778 & 0.0381 & 0.0000 & 0.0841 \\
				5.0\% & 20.0\% & 344.8070 & 0.0000 & NA & 10.0000 & 0.8930 & 0.0330 & 1.1e-05 & 0.0739 \\ \bottomrule
			\end{tabular}
		}
		\caption{{\color{black} Partial taxation case}}
	\end{subtable}
\end{table}

To better understand the impact of tax rates, cash rates, and other contract features on fair fee dynamics, we also examine the optimal withdrawal strategy profiles $w^*(t_k) = w^*(t_k, x, \gamma)$ over values of $(x,\gamma)$ in the computational grid $\mathbb{X} \times \mathbb{G}$, as these ultimately affect the fair fees in our policyholder based valuation. Since the guaranteed withdrawal amount $g(t_k) = \mathfrak{g}(X(t_k^-), G(t_k^-))$ and the maximum withdrawal amount $\overline{w}(t_k) := \max\{\mathfrak{f}(X(t_k^-)), \mathfrak{g}(X(t_k^-), G(t_k^-))\}$ are state-dependent, the optimal withdrawal strategy $w^*(t_k)$ (presented in \eqref{eqn-OptimalWithdrawal}) and its relative magnitude are best understood as a proportion of either the guaranteed withdrawal amount or the maximum allowable withdrawal.

\begin{figure}[h!]
	\begin{subfigure}[b]{0.3\textwidth}
		\includegraphics[width = \textwidth]{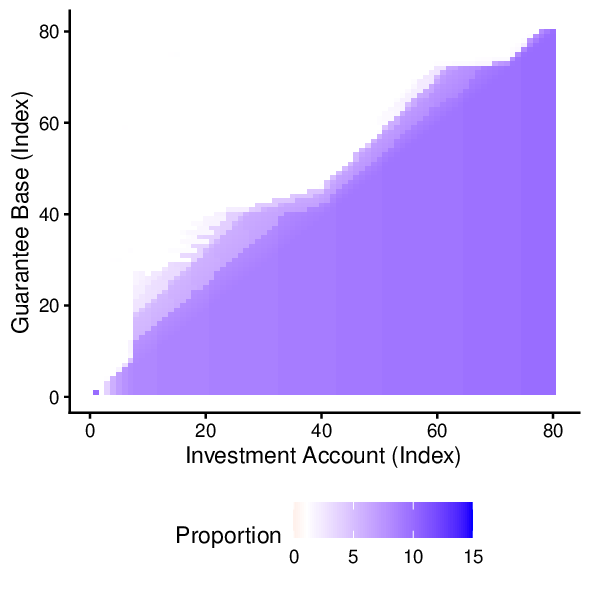}
		\caption{$\theta=0$, $\eta = 2\%$}
	\end{subfigure}
	\hfill
	\begin{subfigure}[b]{0.3\textwidth}
		\includegraphics[width = \textwidth]{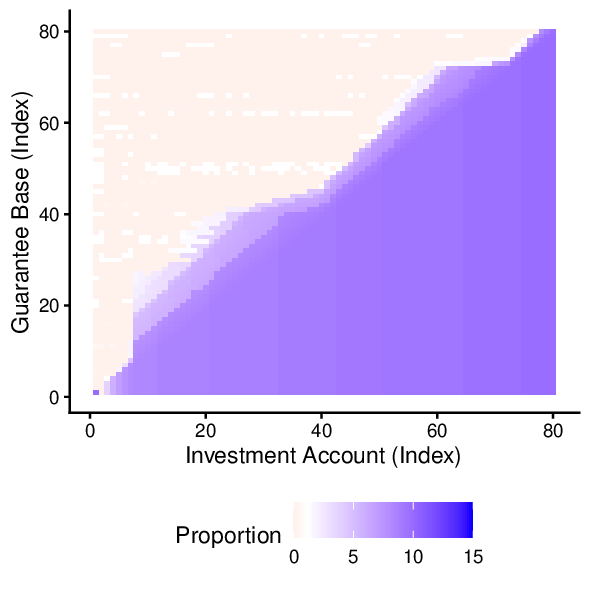}
		\caption{$\theta=0$, $\eta = 3\%$}
	\end{subfigure}
	\hfill
	\begin{subfigure}[b]{0.3\textwidth}
		\includegraphics[width = \textwidth]{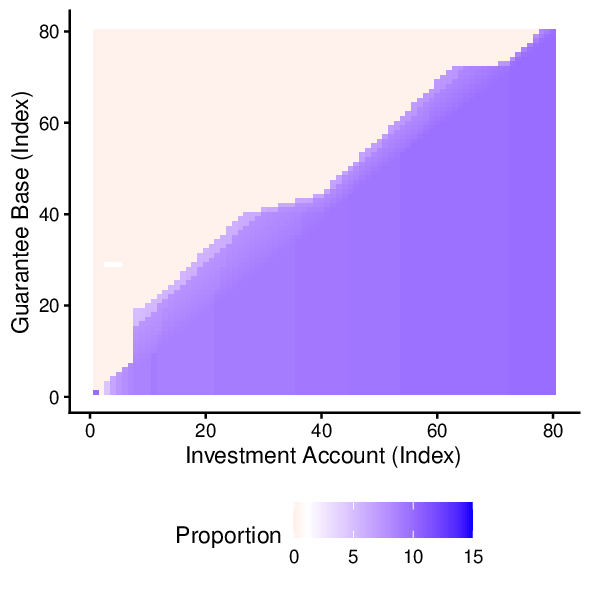}
		\caption{$\theta=0$, $\eta = 4\%$}
	\end{subfigure}

	\begin{subfigure}[b]{0.3\textwidth}
		\includegraphics[width = \textwidth]{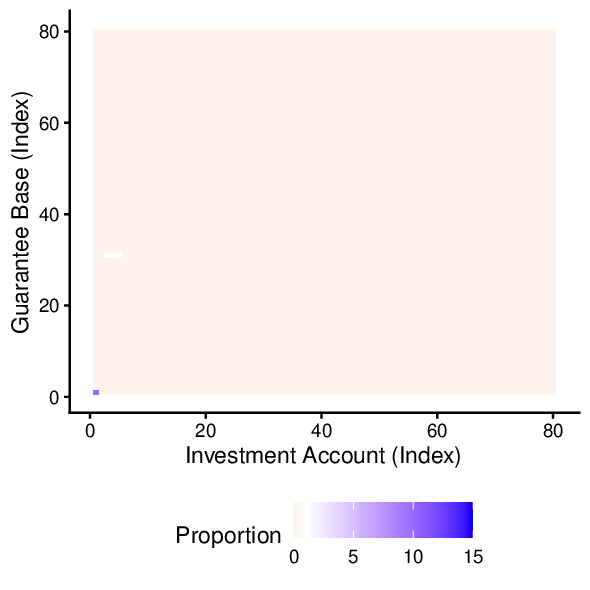}
		\caption{$\theta=10\%$, $\eta = 2\%$}
	\end{subfigure}
	\hfill
	\begin{subfigure}[b]{0.3\textwidth}
		\includegraphics[width = \textwidth]{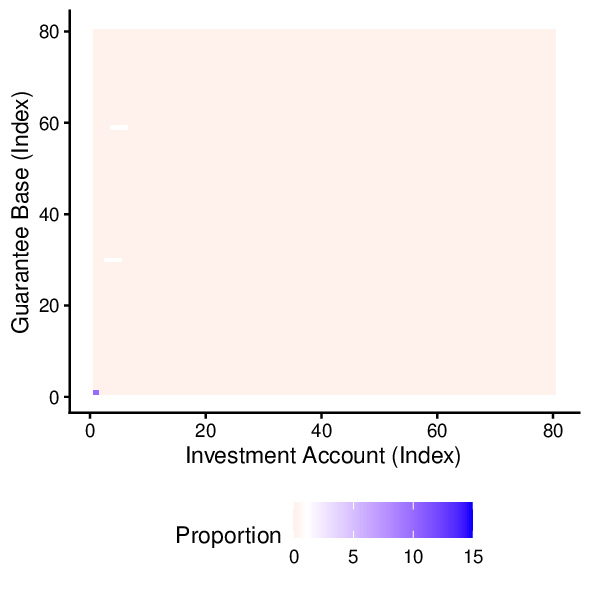}
		\caption{$\theta=10\%$, $\eta = 3\%$}
	\end{subfigure}
	\hfill
	\begin{subfigure}[b]{0.3\textwidth}
		\includegraphics[width = \textwidth]{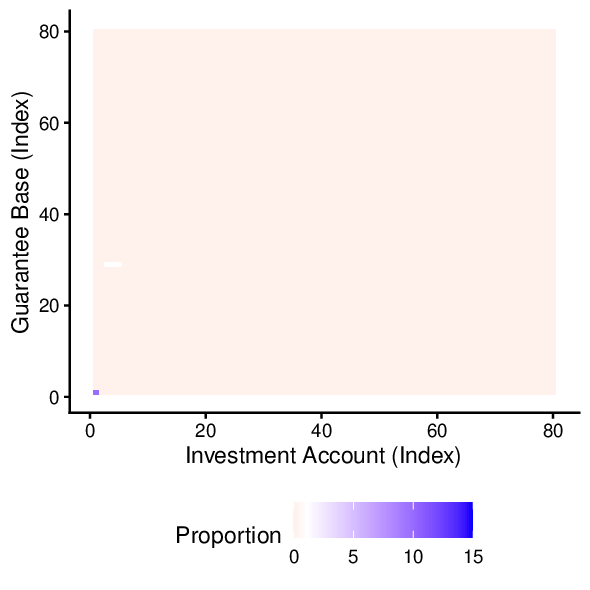}
		\caption{$\theta=10\%$, $\eta = 4\%$}
	\end{subfigure}
	\caption{{\color{black} (Partial taxation case)} The optimal withdrawal strategy for cash rates $\eta \in \{2\%, 3\%, 4\%\}$ expressed in terms of $\mathscr{w}^*_{\rm guar}(4,x,\gamma)$ when $\theta = 0$ (1st row) and $\theta=10\%$ (2nd row).}
	\label{fig:OptWithdraw_NoTaxTax_eta_t4_guar}
\end{figure}

\begin{figure}[h!]
	\begin{subfigure}[b]{0.3\textwidth}
		\includegraphics[width = \textwidth]{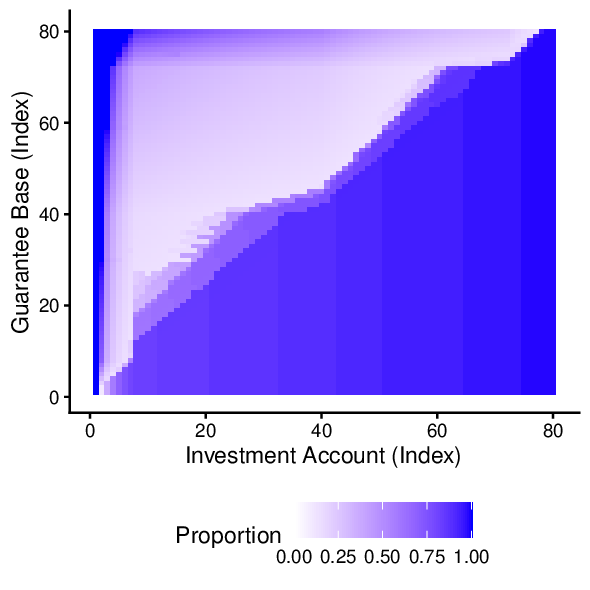}
		\caption{$\theta=0$, $\eta = 2\%$}
	\end{subfigure}
	\hfill
	\begin{subfigure}[b]{0.3\textwidth}
		\includegraphics[width = \textwidth]{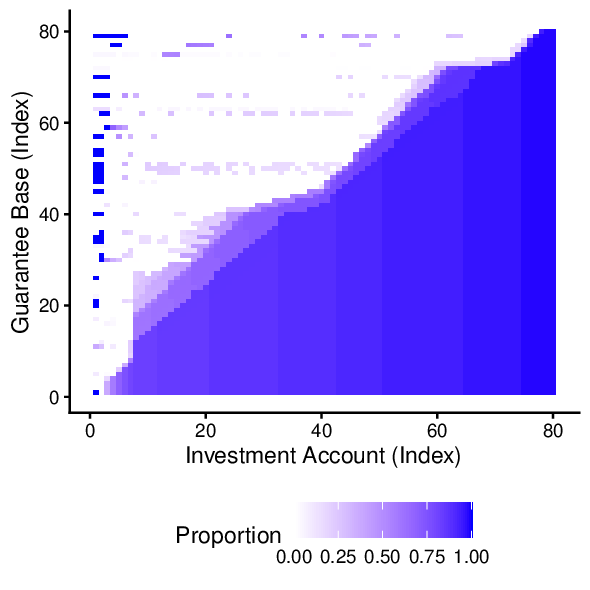}
		\caption{$\theta=0$, $\eta = 3\%$}
	\end{subfigure}
	\hfill
	\begin{subfigure}[b]{0.3\textwidth}
		\includegraphics[width = \textwidth]{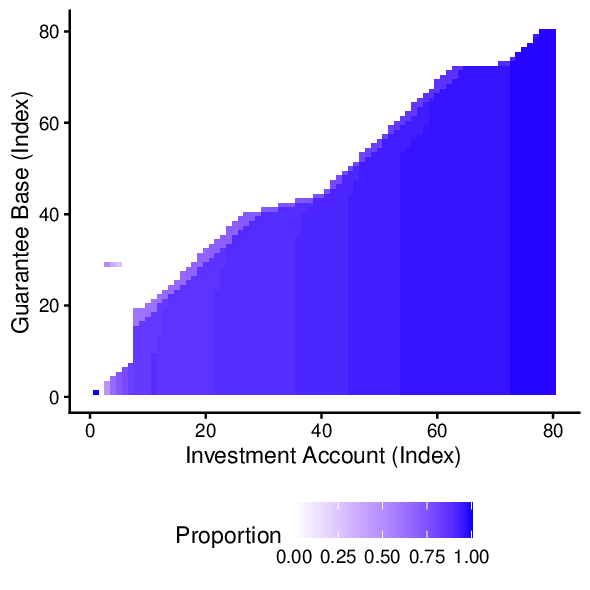}
		\caption{$\theta=0$, $\eta = 4\%$}
	\end{subfigure}
	\begin{subfigure}[b]{0.3\textwidth}
		\includegraphics[width = \textwidth]{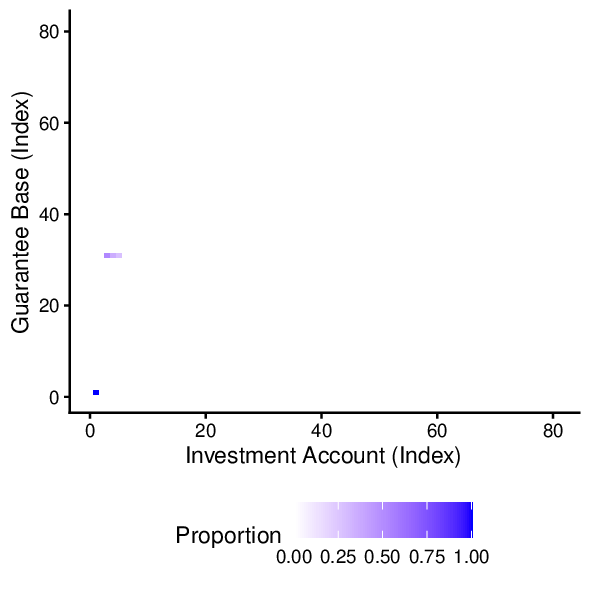}
		\caption{$\theta=10\%$, $\eta = 2\%$}
	\end{subfigure}
	\hfill
	\begin{subfigure}[b]{0.3\textwidth}
		\includegraphics[width = \textwidth]{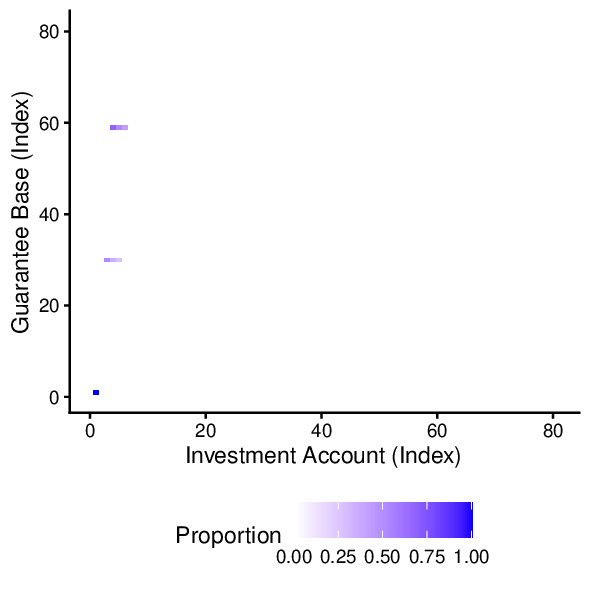}
		\caption{$\theta=10\%$, $\eta = 3\%$}
	\end{subfigure}
	\hfill
	\begin{subfigure}[b]{0.3\textwidth}
		\includegraphics[width = \textwidth]{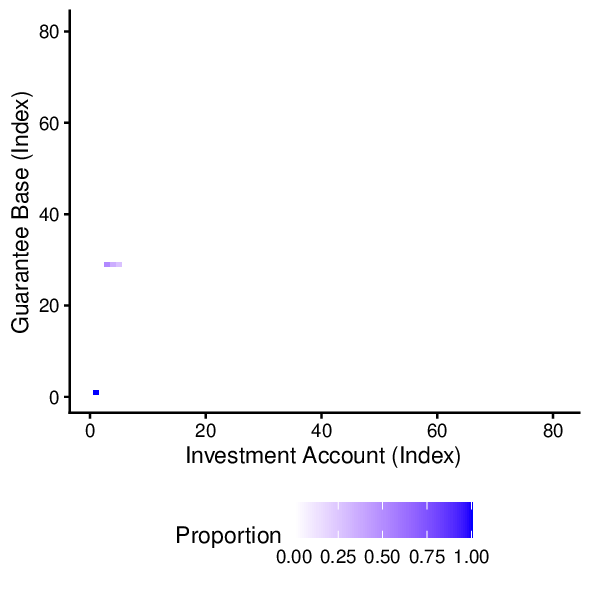}
		\caption{$\theta=10\%$, $\eta = 4\%$}
	\end{subfigure}
	\caption{{\color{black} (Partial taxation case)} The optimal withdrawal strategy for cash rates $\eta \in \{2\%, 3\%, 4\%\}$ expressed in terms of $\mathscr{w}^*_{\max}(4,x,\gamma)$ when $\theta = 0$ (1st row) and $\theta=10\%$ (2nd row).}
	\label{fig:OptWithdraw_NoTaxTax_eta_t4_max}
\end{figure}

To this end, let us define the proportion of the optimal withdrawal amount with respect to the guaranteed amount (resp. maximum possible withdrawal) as $\mathscr{w}^*_{\rm guar}$ (resp.  $\mathscr{w}^*_{\max}$). Mathematically, these are defined as follows:
\begin{equation}\label{eq:wguar_wmax}
	\mathscr{w}^*_{\rm guar}(t_k, x, \gamma) := \frac{w^*(t_k, x, \gamma)}{g(t_k)}, \qquad
	\mathscr{w}^*_{\max}(t_k, x, \gamma) := \frac{w^*(t_k, x,\gamma)}{\overline{w}(t_k, x,\gamma)},
\end{equation}
for a given value of $x = X(t_k^-)$ and $\gamma = G(t_k^-)$. This notation facilitates the visualizations of the optimal withdrawal strategy that will be presented below. Obviously, if $\mathscr{w}^*_{\rm guar}(t_k, x,\gamma) = 1$, then the policyholder withdraws exactly the guaranteed amount. If no withdrawal is made, then both $\mathscr{w}^*_{\rm guar}(t_k, x,\gamma)$ and $\mathscr{w}^*_{\max}(t_k, x,\gamma)$ are 0. Alternatively, if both $\mathscr{w}^*_{\rm guar}(t_k, x,\gamma)$ and $\mathscr{w}^*_{\max}(t_k, x,\gamma)$ are equal to 1, then the policyholder withdraws exactly the guaranteed amount that in turn coincides with the maximum account value. This scenario is a result of the investment account being sufficiently low and should not be mistaken with full surrender (see, for example, the case when $\eta=2\%$ and $\theta = 0\%$ in Figures \ref{fig:OptWithdraw_NoTaxTax_eta_t4_guar} and \ref{fig:OptWithdraw_NoTaxTax_eta_t4_max}). If $\mathscr{w}^*_{\rm guar}(t_k, x,\gamma) > 1$, then the policyholder makes an excess withdrawal which represents a total surrender only if $\mathscr{w}^*_{\max}(t_k, x,\gamma)$ is also equal to 1. Additional figures illustrating optimal withdrawal strategies, mentioned in footnotes in the succeeding discussion, are placed in Appendix \ref{app-SuppWithdrawalFigures}.

Tables \ref{tab-FairGuaranteeFee-StaticvsDynamic} and \ref{tab-FairGuaranteeFee-EE80} confirm that, as expected, fair fees under the dynamic case are substantially higher than those under the static case. This difference arises from the additional flexibility policyholders are given in deciding the amount they wish to withdraw from the contract. Furthermore, the presence of a ratchet increases fair fees, as the benefit base and, consequently, guaranteed withdrawals grow over time. In contrast, under the no-ratchet scenario, the guarantee feature is locked at inception with no opportunity for growth. Including a cash fund, even without taxation, increases product attractiveness as it allows for withdrawals to grow further until maturity at a riskless cash account appreciation rate $\eta>r$ (Table \ref{tab-FairGuaranteeFee-StaticvsDynamic}). 

Without taxation, adding a ratchet increases fees by approximately 10\% to 60\% (Table \ref{tab-FairGuaranteeFee-StaticvsDynamic}). However, in the presence of taxes, fees increase significantly by at least 25\% in the dynamic case and up to four times in the static case. Ratchets have a greater impact on fees when taxes are present due to the implied rational behavior of policyholders. Assuming no cash fund, Table \ref{tab-OptWithSummary-ScenarioAnalysis} shows that policyholders make excess withdrawals slightly more often when no ratchet is present. However, these excess withdrawals result in effective surrenders in about 32\% of cases when $\theta = 0$ and in 15\% of cases when all cash flows are taxed ($\theta = 2.5\%$) (Panel A) and 10\% under the partial taxation case (Panel B). When ratchets are present, surrenders disappear altogether. The combination of taxes and ratchets leads to more \textit{static}-like behavior, allowing the guarantee base to grow without penalties, which yields greater payments and proportionally higher fair fees. With a cash fund, similar reasoning applies. In this case, however, behavior primarily splits between no withdrawals (implying cash fund injections at the ratcheted guarantee level) and excess withdrawals, rather than between withdrawals at the guarantee level and excess withdrawals. Furthermore, the cash fund essentially forces a minimum withdrawal equal to the guarantee. This discourages the underlying fund from growing excessively and thus avoiding overly aggressive ratcheting, which typically happens when the policyholder chooses not to make a withdrawal. {\color{black} Moreover, routing the guaranteed benefit through the cash fund vehicle ensures that the product is suitable within a superannuation retirement environment, where retirement income streams must exceed the minimum annual withdrawal prescribed by the Australian Taxation Office \citep{ATOminwith}.}\footnote{Indeed, it may be optimal for a policyholder to not make any withdrawals as part of a bang-bang strategy (i.e. zero withdrawal, withdrawal of the contractual amount, or full surrender of the contract); see, for example, \citet{AzimzadehForsyth-2015}.}


Considering a ratchet, in isolation, does not generally yield positive fees more often. Indeed, without a cash fund, it increases the fair fees due to the non-decreasing guaranteed benefit base, but still yields no fees for $\theta > 20\%$ in the partial taxation case and for $\theta > 2.5\%$ when fully liable for taxation. Jointly considering a cash fund and a ratchet, however, yields positive fair fees for all studied $\theta$. This is primarily due to allowing a cash fund, but the effect is greater if both product features are combined. Indeed, even if the presence of the cash fund optimally leads to policyholders doing cash fund injections and appreciate at $\eta$, the injections will increase over time since the guarantee level $g(t_k)$ increases with the guarantee base. This highlights a key finding: when taxes are considered, the cash fund and ratchet feature become essential for product feasibility. 

{\color{black} It is important to highlight the effect of partial versus full taxation. Tables \ref{tab-FairGuaranteeFee-StaticvsDynamic} and \ref{tab-FairGuaranteeFee-EE80} show that the overall effect of taxation, be it through ratcheting, cash fund inclusion, or both, is more muted when proceeds are only partially taxed, as expected. Indeed, in the partial taxation case, the proceeds of the cash fund will not be taxed since these are paid after maturity, which occurs after preservation age. Taxation hence only affects active withdrawals made between inception and age 60, that is, during the first 5 years of the contract. Without the cash fund, Tables \ref{tab-OptWithSummary-ScenarioAnalysis} and \ref{tab-OptWithSummary-ScenarioAnalysis_etaVAR} show that policyholders withdraw at or above the guarantee, and hence partial versus full taxation simply increases product feasibility as $\theta$ decreases. However, once the cash fund is introduced and $\eta>r$, active withdrawals are mostly avoided as capital injections to the cash fund are preferred; since the associated cash flows are no longer taxed, the fee does not decrease further with $\theta$. Minor differences in the partial taxation case are still present between $\theta=0$ and $\theta>0$ simply because excess withdrawals still marginally occur, and when made before preservation age, these are taxed.}	
	
Policyholder behavior and guarantee fair fees are sensitive to the relationship between the cash fund appreciation rate $\eta$, the taxation level $\theta$ and the risk-free rate $r$. Assuming the cash fund does not outperform the risk-free rate ($\eta = 2\% < r$), the optimal dynamic withdrawal strategy suggests that policyholders avoid injecting funds into the cash account. This is evident in Figures \ref{fig:OptWithdraw_NoTaxTax_eta_t4_guar} and \ref{fig:OptWithdraw_NoTaxTax_eta_t4_max} and Table \ref{tab-OptWithSummary-ScenarioAnalysis_etaVAR}, where no cash fund deposits occur when there are no taxes. 
In addition, the valuation of the product remains stable in some scenarios despite differences in strategy. When there are no taxes, the fair guarantee fee does not decline further as $\eta$ decreases below $r$. For example, while the optimal withdrawal behavior differs between $\eta = 2\%$ and $\eta = 3\%$ (both without taxation), the product's valuation remains unchanged, resulting in an identical fair guarantee fee. This occurs because, when $\eta < r$, withdrawing less than the guaranteed amount is suboptimal, and the cash fund is not utilized. 
Unlike the static case, the dynamic framework allows for withdrawals exceeding the guaranteed amount, which explains the greater fee in the dynamic scenario compared to the static case even if $\eta<r$. 

However, when taxation is introduced, the role of the cash fund becomes more significant, likely due to its ``tax shielding'' effect, even when its performance is lower than the risk-free rate (compare $\eta=2\%$ and $3\%$ in Table \ref{tab-FairGuaranteeFee-EE80} with the no cash fund case in Table \ref{tab-FairGuaranteeFee-StaticvsDynamic}). The attractiveness of the cash fund increases substantially when $\eta>r$, even in a tax-free environment as evidenced by the optimal withdrawal behavior that exhibits a {\color{black} bang-bang-like} strategy whereby no withdrawal (that is, injection into cash fund or complete surrender) is optimal (Figure \ref{fig:OptWithdraw_NoTaxTax_eta_t4_guar}). In this case a cash fund nearly doubles the fair guarantee fee. This suggests that policyholders are willing to forgo regular income in favor of making cash fund injections to obtain a higher risk-free return even without the ``tax-shielding'' feature.

In summary, incorporating ratcheting features enhances product feasibility, especially in high-tax environments. More importantly, integrating a secondary investment opportunity, such as a cash fund with preferential tax treatment, where only 
gains from interest are taxed at maturity {if maturity occurs before preservation age, further improves the product's attractiveness. This contrasts with the tax treatment of withdrawals, which are taxed as ordinary income before preservation age. As a result, the product functions similarly to a risk-free savings account when taxes are imposed, which undermines its intended role as a supplemental retirement income product through regular withdrawals. This likely explains why the product is more effectively marketed as an income source after the preservation age, as reflected in Australia, since cash flows are no longer taxed after preservation age. However, even in a tax-free context, allowing the option to withdraw less than the guarantee and let part of your income appreciate at cash account appreciation rate $\eta$, undermines its goal as a retirement income product since no withdrawals become more prevalent compared to the no cash fund case. Indeed, Table \ref{tab-OptWithSummary-ScenarioAnalysis} shows in the ratchet case that in over 90\% of cases policyholders will not withdraw anything of strictly below the guarantee and inject in the cash fund compared to 0\% in absence of a cash fund. 

Figure \ref{fig:OptWithdraw_NoTaxTax_t1t4t9_max} shows the optimal withdrawal strategies expressed in terms of $\mathscr{w}^*_{\rm max}$ at the beginning, midway and prior to maturity when $\eta=4\%>r$ without and with tax (first row vs. second row) in a partial taxation context. Recall that if $\mathscr{w}^*_{\rm guar}>1$ and $\mathscr{w}^*_{\rm max}=1$ this implies a surrender.\footnote{Figure \ref{fig:OptWithdraw_NoTaxTax_t1t4t9_guar} shows the corresponding figure for $\mathscr{w}^*_{\rm guar}$, {\color{black} supplementing Figure \ref{fig:OptWithdraw_NoTaxTax_t1t4t9_max} which shows $\mathscr{w}^*_{\max}$}.} Early in the contract and near maturity, policyholders exhibit a {\color{black} bang-bang-like} strategy: no withdrawals initially and primarily excess withdrawals or surrender close to maturity. At the midpoint ($t = 4$, the year before the policyholder reaches the preservation age), behavior becomes more nuanced, with decisions more contingent on state variables, taxation, and product features. For this reason, subsequent figures will focus on $t = 4$ to facilitate detailed analysis. 

\begin{figure}
	\begin{subfigure}[b]{0.3\textwidth}
		\includegraphics[width = \textwidth]{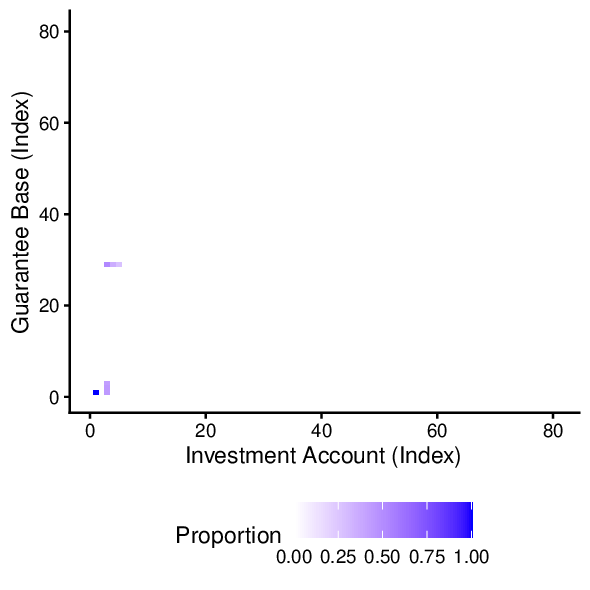}
		\caption{$\theta=0$, $t=1$ }
	\end{subfigure}
	\hfill
	\begin{subfigure}[b]{0.3\textwidth}
		\includegraphics[width = \textwidth]{Figures/WPFigPA/t4_cash4_tax0_max.eps}
		\caption{$\theta=0$, $t=4$}
	\end{subfigure}
	\hfill
	\begin{subfigure}[b]{0.3\textwidth}
		\includegraphics[width = \textwidth]{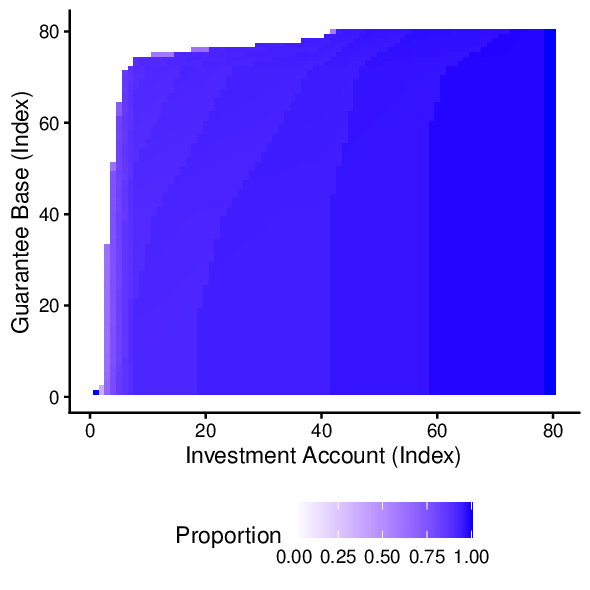}
		\caption{$\theta=0$, $t=9$}
	\end{subfigure}
	\begin{subfigure}[b]{0.3\textwidth}
		\includegraphics[width = \textwidth]{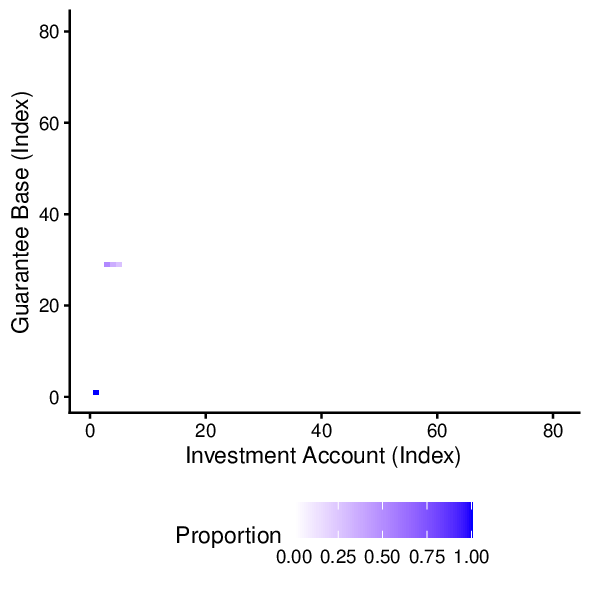}
		\caption{$\theta=10\%$, $t=1$}
	\end{subfigure}
	\hfill
	\begin{subfigure}[b]{0.3\textwidth}
		\includegraphics[width = \textwidth]{Figures/WPFigPA/t4_cash4_tax10_max.eps}
		\caption{$\theta=10\%$, $t=4$}
	\end{subfigure}
	\hfill
	\begin{subfigure}[b]{0.3\textwidth}
		\includegraphics[width = \textwidth]{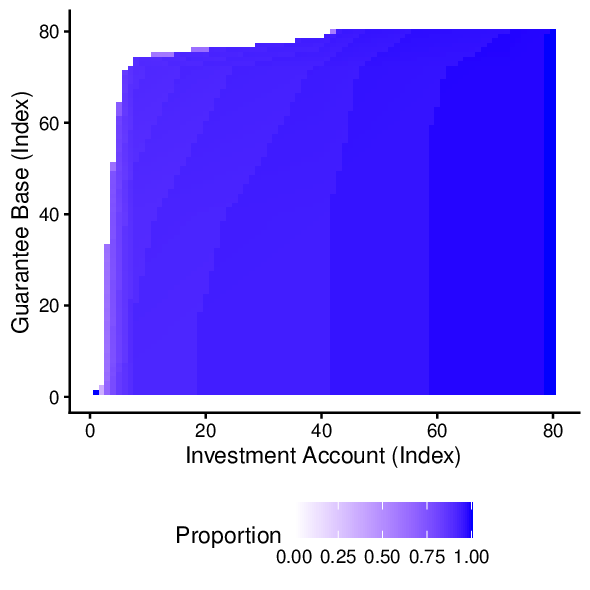}
		\caption{$\theta=10\%$, $t=9$}
	\end{subfigure}
	\caption{{\color{black} (Partial taxation case)} The optimal withdrawal strategy at $t \in \{1, 4, 9\}$ expressed in terms of  $\mathscr{w}^*_{\max}(t,x,\gamma)$ for $\eta = 4\%$ when $\theta = 0$ (1st row) and $\theta=10\%$ (2nd row).}
	\label{fig:OptWithdraw_NoTaxTax_t1t4t9_max}
\end{figure}

It is important to note that although the optimal withdrawal profiles may suggest that surrender will occur, this might not happen in practice, as the underlying asset may never reach the relevant value thresholds. For example, while dark blue regions at $t = 9$ may indicate surrender behavior in scenarios with jointly a high underlying value and a high guarantee base, such combinations are unlikely to materialize late in the contract given that the underlying generally decreases over time as a result of regular withdrawals; see \eqref{eqn-Update-InvestmentAccount-PostWithdrawal}. In fact, when a ratchet is present, our simulation studies indicate that surrenders do not occur precisely for this reason.

	

\begin{figure}
	\begin{subfigure}[b]{0.3\textwidth}
		\includegraphics[width = \textwidth]{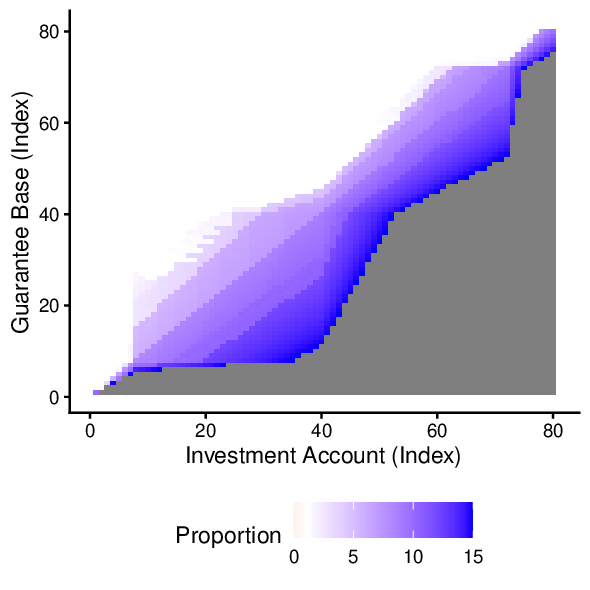}
		\caption{no CF, no ratchet}
	\end{subfigure}
	\qquad
	\begin{subfigure}[b]{0.3\textwidth}
		\includegraphics[width = \textwidth]{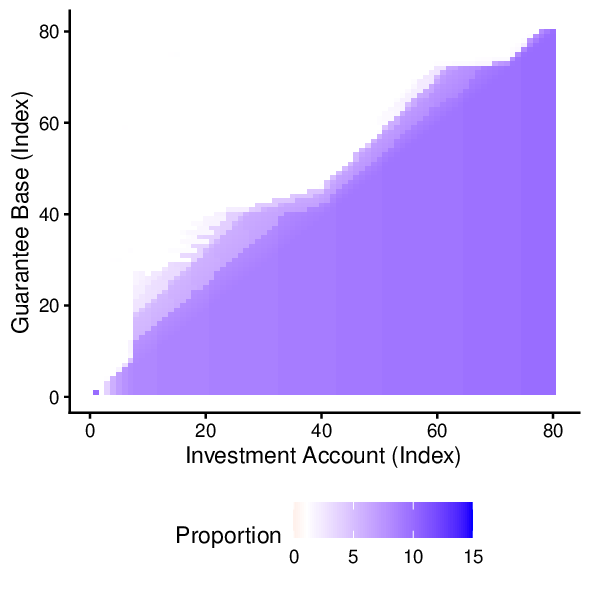}
		\caption{no CF, with ratchet}
	\end{subfigure}
	
	\begin{subfigure}[b]{0.3\textwidth}
		\includegraphics[width = \textwidth]{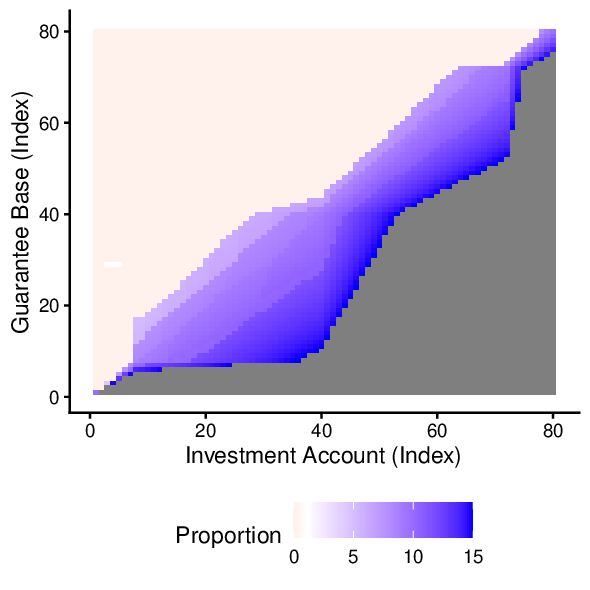}
		\caption{CF, no ratchet}
	\end{subfigure}
	\qquad
	\begin{subfigure}[b]{0.3\textwidth}
		\includegraphics[width = \textwidth]{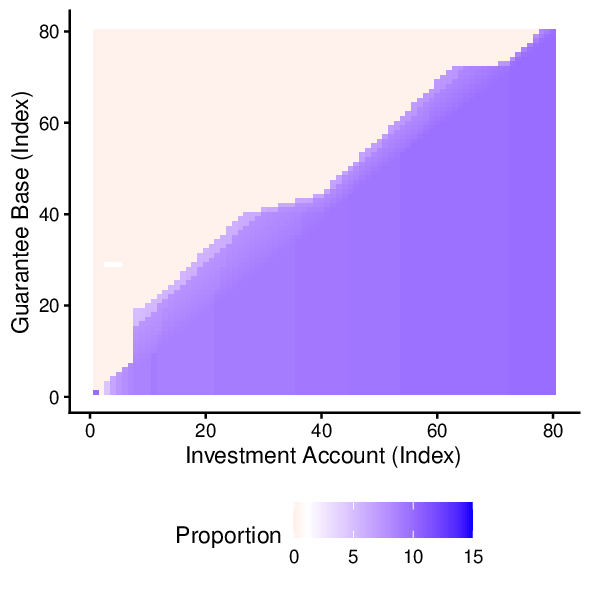}
		\caption{CF, with ratchet}
	\end{subfigure}
	\caption{{\color{black} (Partial taxation case)} The optimal withdrawal strategy for various contract specifications (cash fund vs. no cash fund, ratchet vs. no ratchet/return-to-premium) expressed in terms of $\mathscr{w}^*_{\rm guar}(t,x,\gamma)$ for $\theta = 0$ and $t = 4$. (Note: Regions in dark gray indicate withdrawals where $\mathscr{w}^*_{\rm guar}(t,x,\gamma) > 15$. An upper bound of 15 is used uniformly across all plots of $\mathscr{w}^*_{\rm guar}(t,x,\gamma) > 15$ to facilitate comparison.)}
	\label{fig:OptWithdraw_NoTax_ContractSpec_t4_guar}
\end{figure}

\subsection{Cash Funds Discourage Active Withdrawals}

In the absence of a cash fund, policyholders typically withdraw at least the guaranteed amount. Figure \ref{fig:OptWithdraw_NoTax_ContractSpec_t4_guar} and Table \ref{tab-OptWithSummary-ScenarioAnalysis} confirm this: without taxation\footnote{Figure \ref{fig:OptWithdraw_NoTax_ContractSpec_t4_max} shows $\mathscr{w}^*_{\max}$ in the non-taxed case, {\color{black} supplementing Figure \ref{fig:OptWithdraw_NoTax_ContractSpec_t4_guar} which shows $\mathscr{w}^*_{\rm guar}$}.}, over half of all withdrawals equal the guaranteed amount $g(t_k)$. With taxes\footnote{Figure \ref{fig:OptWithdraw_Tax_ContractSpec_t4} shows $\mathscr{w}^*_{\rm guar}$  and $\mathscr{w}^*_{\max}$ in the taxed case for the partial taxation case and Figure \ref{fig:OptWithdraw_Tax_ContractSpec_t4_FullTax} for the full taxation case.}, this share rises to approximately 75\%, highlighting a shift toward more \textit{static}-like behavior. This behavioral shift also results in markedly lower surrender rates, increasing the average contract duration by at least one year.

When a cash fund is available, policyholders exhibit a {\color{black} \textit{bang-bang}-like} strategy: they either withdraw nothing or surrender the contract entirely. Across all tax settings, between 85\% and 91\% of policyholders either refrain from withdrawals or withdraw below $g(t_k)$, instead injecting funds into the cash account. Only 1-20\% surrender, depending on $\theta$ (higher $\theta$ increasing average contract duration).

Higher $\theta$ values intensify this trend, as many policyholders prefer to defer income, avoid tax, and maximize tax-deferred growth in the appreciating cash fund (Figure~\ref{fig:OptWithdraw_Tax_t4}) specially under the partial taxation environment. Notably, even without taxation, the mere availability of a cash fund shifts behavior toward non-withdrawal, illustrating its design influence beyond tax incentives.

Table \ref{tab-OptWithSummary-ScenarioAnalysis_etaVAR} further highlights how this effect depends on the cash rate $\eta$. When $\eta < r$, the behavior approximates the no-cash-fund setting, especially when $\theta = 0$, with minimal cash injections (Figures \ref{fig:OptWithdraw_NoTaxTax_eta_t4_guar} and \ref{fig:OptWithdraw_NoTaxTax_eta_t4_max}). Under taxation, modest capital injections occur, yet over two-thirds of policyholders still withdraw at least $g(t_k)$.

When $\eta \geq r$, behavior shifts markedly. Table~\ref{tab-OptWithSummary-ScenarioAnalysis_etaVAR} shows that 74--93\% of policyholders withdraw less than the guaranteed amount, depending on $\theta$. A high $\eta$ disincentivizes active withdrawals (Figure~\ref{fig:OptWithdraw_NoTaxTax_eta_t4_guar}), undermining the product's suitability as a retirement income product that supplements first-pillar state-based pensions. Rather than providing a stream of retirement income, the product transforms into a savings mechanism: guaranteed (ratcheted) withdrawals are redirected into the cash fund appreciating at $\eta$ that can only be accessed at maturity.

Nevertheless, policyholder value increases with $\eta$, leading to higher fair guarantee fees. Policyholders benefit from both ratcheting, which raises the guaranteed withdrawal base, and from tax-deferred growth in the cash fund above the risk-free rate.

\subsection{Ratcheting Eliminates Surrender Incentives}

Even when a cash fund is available (for example, at a cash rate of $\eta = 4\%$), policyholders tend to surrender the contract if it lacks a ratchet component. In the assumed financial environment and in the absence of taxes, the inability to benefit from investment gains through increases in the guaranteed withdrawal amount (due to a static benefit base) prompts some policyholders to fully surrender the contract. This allows them to realize gains from a rising investment account rather than remain locked into a contract with limited upside. However, increasing the tax rate reduces surrender likelihood in these scenarios, as policyholders face substantial tax liabilities upon withdrawing their entire account balance. 

Introducing a ratchet feature changes this interaction significantly as highlighted in Figure \ref{fig:OptWithdraw_NoTax_ContractSpec_t4_guar}. Indeed, without ratchet the proportion of withdrawals is greater (dark blue) compared to the ratchet case (light blue). Adding a ratchet enhances the attractiveness of remaining in the contract by allowing the guarantee base to rise with market performance, thereby eliminating the rational incentive to surrender. In such settings, approximately one-third of policyholders make excess withdrawals without surrendering, reducing overall lapsation. Table~\ref{tab-OptWithSummary-ScenarioAnalysis} shows that the presence of a ratchet suppresses surrender behavior even when a cash account is also present.

Indeed, managing lapse risk is a significant challenge, given its implications for insurer solvency \citep{Loisel-2011, Moodys-2013}. A range of surrender mitigation mechanisms has been proposed, including state-dependent fees levied  only if the account value falls below \citep{BernardHardyMacKay-2014} or above \citep{FengJingNg-2025} a specified threshold, time-dependent fees that decline over time \citep{Moenig-2018, Bernard-2019}, and two-account structures that impose charges only on a secondary, risk-free guarantee account \citep{AlonsoGarcia-2025}.

Our results suggest that a relatively simple structural feature, the ratchet, can eliminate rational surrender behavior effectively, even in the presence of a cash fund. Without taxation, adding a ratchet increases the fair fee by only about 10\%, yet surrender rates fall from roughly one-third to zero. With taxation, the same surrender reduction is observed, although the fee impact becomes more substantial: the required fair fee increases by between 40\% and 200\%  when no cash fund is present and increases by between 23\% and 50\% when a cash fund is included. {\color{black} The effect of the cash fund inclusion is greater in the full taxation than the partial taxation case, as expected.}

\begin{landscape}
	\begin{figure}
		\begin{subfigure}[b]{0.3\textwidth}
			\includegraphics[width = \textwidth]{Figures/WPFigPA/t4_cash4_tax0_guar.eps}
			\caption{$\mathscr{w}^*_{\rm guar}(4,x,\gamma)$ for $\theta = 0$}
		\end{subfigure}
		\hfill
		\begin{subfigure}[b]{0.3\textwidth}
			\includegraphics[width = \textwidth]{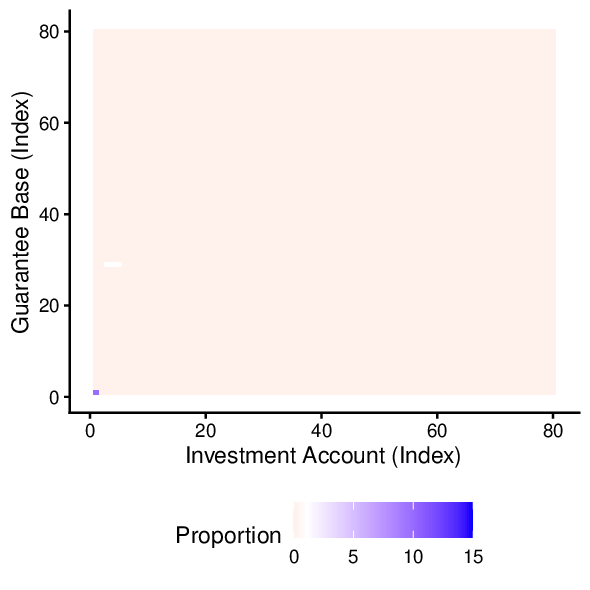}
			\caption{$\mathscr{w}^*_{\rm guar}(4,x,\gamma)$ for $\theta = 2.5\%$}
		\end{subfigure}
		\hfill
		\begin{subfigure}[b]{0.3\textwidth}
			\includegraphics[width = \textwidth]{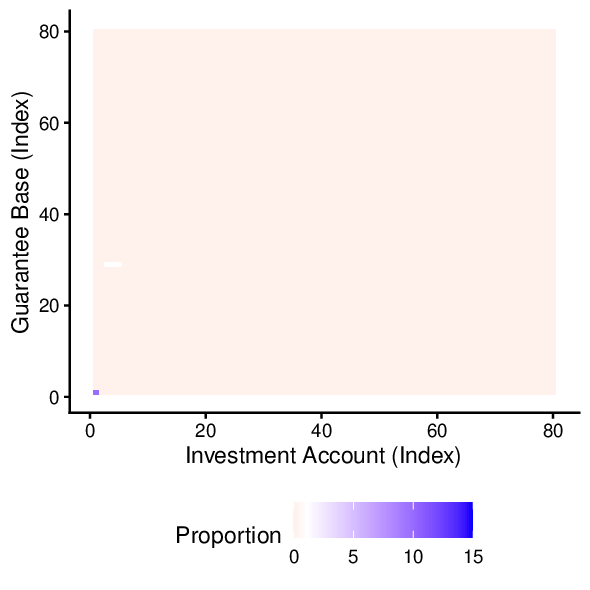}
			\caption{$\mathscr{w}^*_{\rm guar}(4,x,\gamma)$ for $\theta = 5\%$}
		\end{subfigure}
		\hfill
		\begin{subfigure}[b]{0.3\textwidth}
			\includegraphics[width = \textwidth]{Figures/WPFigPA/t4_cash4_tax10_guar.eps}
			\caption{$\mathscr{w}^*_{\rm guar}(4,x,\gamma)$ for $\theta = 10\%$}
		\end{subfigure}
		
		\begin{subfigure}[b]{0.3\textwidth}
			\includegraphics[width = \textwidth]{Figures/WPFigPA/t4_cash4_tax0_max.eps}
			\caption{$\mathscr{w}^*_{\max}(4,x,\gamma)$ for $\theta = 0$}
		\end{subfigure}
		\hfill
		\begin{subfigure}[b]{0.3\textwidth}
			\includegraphics[width = \textwidth]{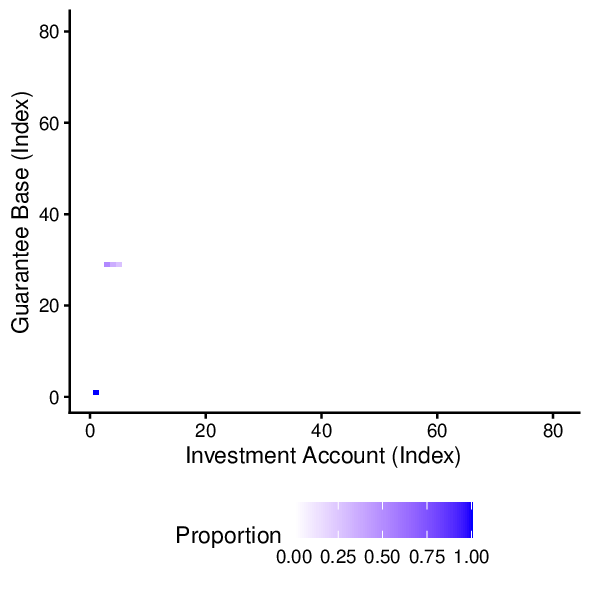}
			\caption{$\mathscr{w}^*_{\max}(4,x,\gamma)$ for $\theta = 2.5\%$}
		\end{subfigure}
		\hfill
		\begin{subfigure}[b]{0.3\textwidth}
			\includegraphics[width = \textwidth]{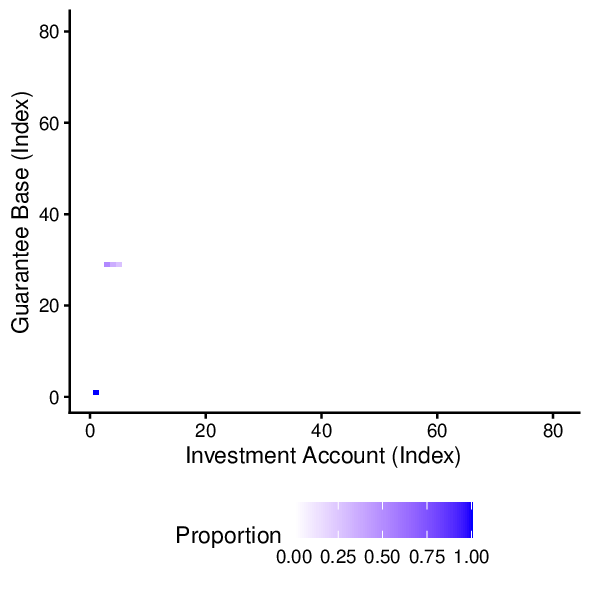}
			\caption{$\mathscr{w}^*_{\max}(4,x,\gamma)$ for $\theta = 5\%$}
		\end{subfigure}
		\hfill
		\begin{subfigure}[b]{0.3\textwidth}
			\includegraphics[width = \textwidth]{Figures/WPFigPA/t4_cash4_tax10_max.eps}
			\caption{$\mathscr{w}^*_{\max}(4,x,\gamma)$ for $\theta = 10\%$}
		\end{subfigure}
		\caption{{\color{black} (Partial taxation case)} The optimal withdrawal strategy for various tax rates $\theta \in \{0, 2.5\%, 5\%, 10\%\}$ expressed in terms of $\mathscr{w}^*_{\rm guar}(4,x,\gamma)$ and $\mathscr{w}^*_{\max}(4,x,\gamma)$ for $\eta = 4\%$.}
		\label{fig:OptWithdraw_Tax_t4}
	\end{figure}
\end{landscape}


\section{Conclusion}
\label{sec-Conclusion}

This paper investigates the pricing and analysis of optimal policyholder behavior in a VA contract with a GMWB rider that includes a supplementary cash fund maintained by the provider. In many jurisdictions, ordinary withdrawals/income and interest income are taxed differently, so this paper investigates how optimal policyholder behavior changes given the interaction of tax rates and the cash fund {\color{black} appreciation} rate. Furthermore, we assume that the benefit base of the VA contract evolves using a ratchet mechanism, reflecting contract specifications in traded VAs. The analysis is conducted through a risk-neutral valuation approach in which the policyholder seeks to maximize the net present value of their cash flows from the VA contract.

Our analysis and numerical experiments show that taxation, the cash fund, and the ratchet mechanism substantially influence policyholders' valuation of the contract, optimal withdrawal strategy, and surrender behavior. When either taxation is present or absent, the policyholder's fair guarantee fee increases as the cash fund appreciation rate increases. However, taxation significantly impacts the optimal policyholder withdrawal strategy since, when withdrawals are taxed, it is sometimes optimal for the policyholder to withdraw nothing. This then transfers the guaranteed withdrawal amount to the cash fund, where it will grow at the cash fund rate and will only be taxed for its interest income {\color{black} if the contract matures before preservation age}. This unveils a tax-shielding effect stemming from the difference in tax treatment of withdrawals and interest income. Furthermore, since taxation generally reduces the policyholder's valuation of the contract, benefit base update schemes such as the ratchet mechanism enhance policyholder participation in these contracts. Since the ratchet provides better downside risk protection compared to a return-of-premium, the ratchet mechanism tends to discourage strategic surrender in policyholders, especially in the presence of taxation.

This paper can also be extended in several directions. First, one can consider a more complex asset class as a supplementary investment. Doing so may involve other tax considerations, such as capital gains tax and other asset class-specific taxation schemes \citep{AlonsoGarcia-2024}. Alternatively, more complex hybrid contract designs may allow {\color{black} policyholders} to dictate the investment portfolio composition, which then enters the optimization problem \citep{FengJingNg-2025, Horneff-2015, Mahayni-2012, Moenig-2021}. Second, the analysis can be extended to allow policyholders to surrender in between event dates. Methodologically, this induces an American-type option valuation in between event dates, leading to a free-boundary PDE problem. Such problems can also be solved using the method of lines. Third, one can also include other VA riders in the contract design. Additional riders may interact with the specification of the GMWB rider to influence policyholder behavior; an important example is the inclusion of a GMDB, which emphasizes the need to model mortality risk \citep[see, for example,][]{Bauer-2023}. {\color{black} Finally, the insurer's perspective of the richer hybrid contract architecture analyzed in this paper can be investigated, determining whether a similar taxation wedge exists between the valuation of the insurer and the policyholder. Such an analysis potentially builds on the work of \citet{AlonsoGarcia-2024} for VAs with GMAB riders.} 

{
\setstretch{1.15}
\bibliographystyle{hapalike}
\bibliography{references}





}

\newpage

\appendix

\section{Supplementary Figures and Results}
\label{app-SuppFigures}

\subsection{Optimal Withdrawal Strategies}
\label{app-SuppWithdrawalFigures}

This document contains additional figures illustrating the optimal withdrawal strategies for various configurations of the cash rate and the tax rate. These supplement some of the discussions in Section 4.2 of the main paper.

\begin{figure}[h!]
	\begin{subfigure}[b]{0.3\textwidth}
		\includegraphics[width = \textwidth]{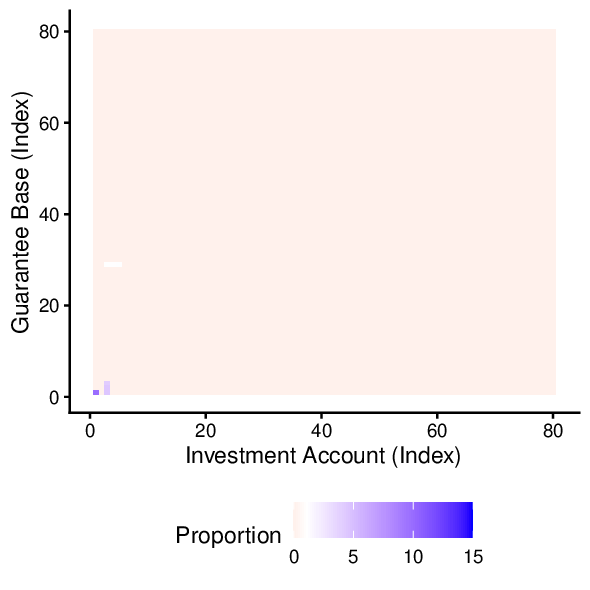}
		\caption{$\theta=0$, $t=1$ }
	\end{subfigure}
	\hfill
	\begin{subfigure}[b]{0.3\textwidth}
		\includegraphics[width = \textwidth]{Figures/WPFigPA/t4_cash4_tax0_guar.eps}
		\caption{$\theta=0$, $t=4$}
	\end{subfigure}
	\hfill
	\begin{subfigure}[b]{0.3\textwidth}
		\includegraphics[width = \textwidth]{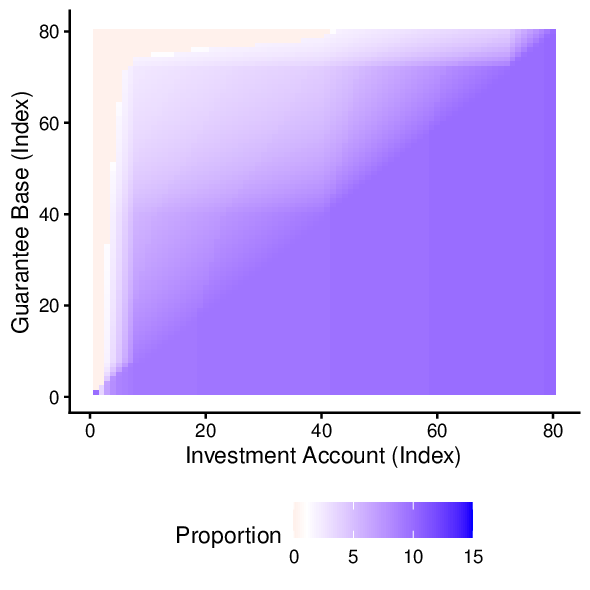}
		\caption{$\theta=0$, $t=9$}
	\end{subfigure}
	\begin{subfigure}[b]{0.3\textwidth}
		\includegraphics[width = \textwidth]{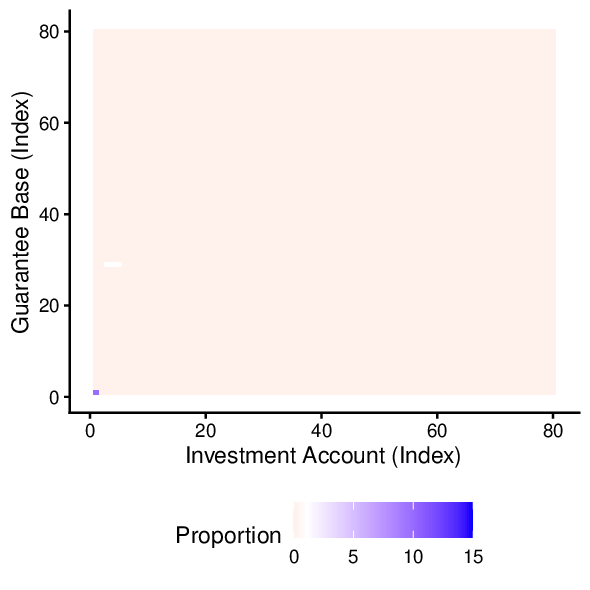}
		\caption{$\theta=10\%$, $t=1$}
	\end{subfigure}
	\hfill
	\begin{subfigure}[b]{0.3\textwidth}
		\includegraphics[width = \textwidth]{Figures/WPFigPA/t4_cash4_tax10_guar.eps}
		\caption{$\theta=10\%$, $t=4$}
	\end{subfigure}
	\hfill
	\begin{subfigure}[b]{0.3\textwidth}
		\includegraphics[width = \textwidth]{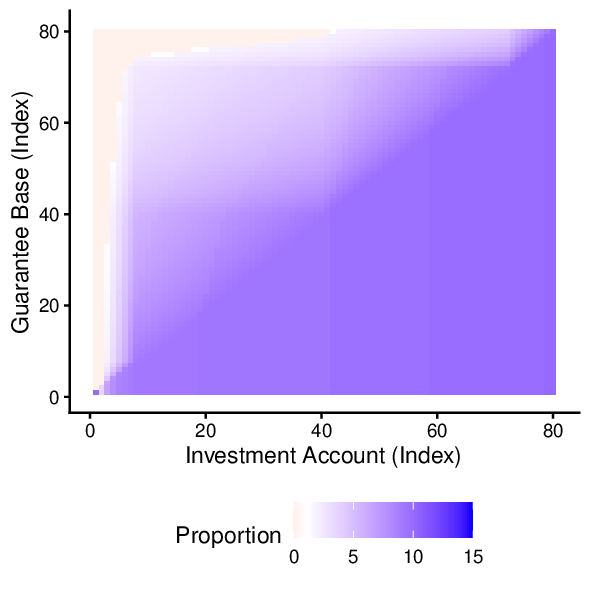}
		\caption{$\theta=10\%$, $t=9$}
	\end{subfigure}
	\caption{{\color{black} (Partial taxation case)} The optimal withdrawal strategy at $t \in \{1, 4, 9\}$ expressed in terms of  $\mathscr{w}^*_{\rm guar}(t,x,\gamma)$ for $\eta = 4\%$ when $\theta = 0$ (1st row) and $\theta=10\%$ (2nd row).}
	\label{fig:OptWithdraw_NoTaxTax_t1t4t9_guar}
\end{figure}

\begin{figure}
	\begin{subfigure}[b]{0.3\textwidth}
		\includegraphics[width = \textwidth]{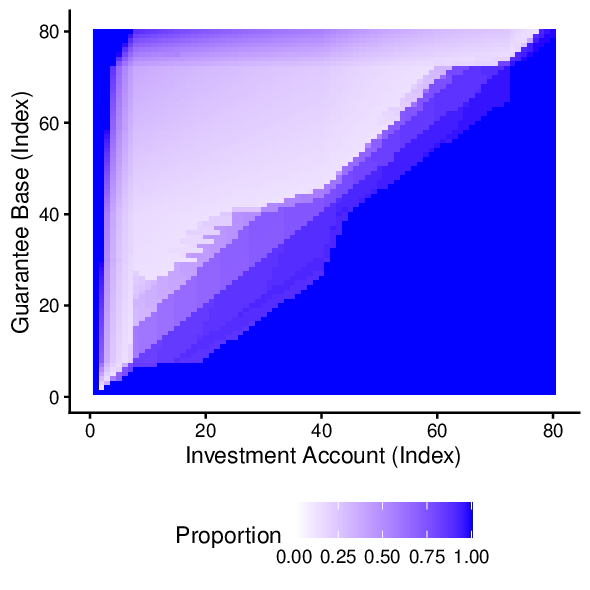}
		\caption{no CF, no ratchet}
	\end{subfigure}
	\qquad
	\begin{subfigure}[b]{0.3\textwidth}
		\includegraphics[width = \textwidth]{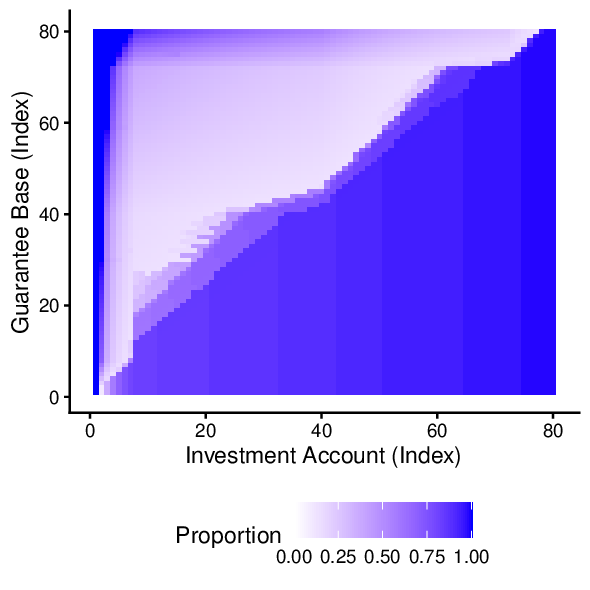}
		\caption{no CF, with ratchet}
	\end{subfigure}
	
	\begin{subfigure}[b]{0.3\textwidth}
		\includegraphics[width = \textwidth]{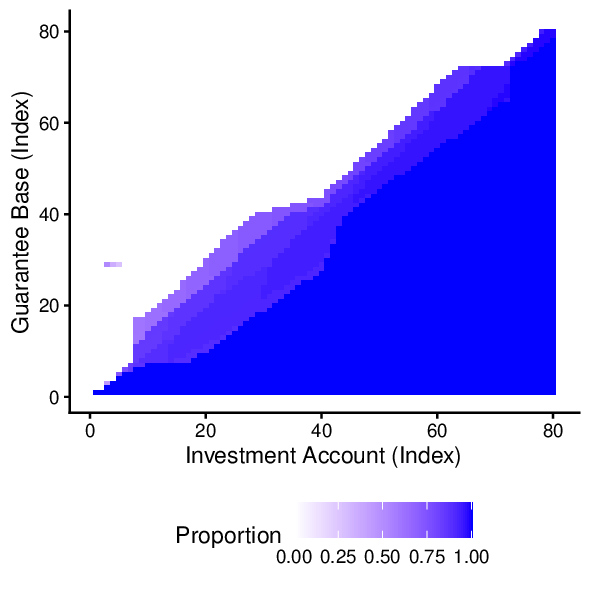}
		\caption{CF, no ratchet}
	\end{subfigure}
	\qquad
	\begin{subfigure}[b]{0.3\textwidth}
		\includegraphics[width = \textwidth]{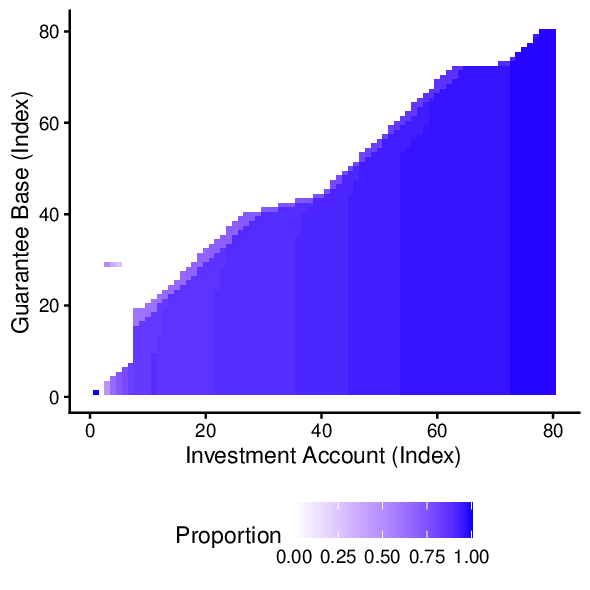}
		\caption{CF, with ratchet}
	\end{subfigure}
	\caption{{\color{black} (Partial taxation case)} The optimal withdrawal strategy for various contract specifications (cash fund vs. no cash fund, ratchet vs. no ratchet/return-to-premium) expressed in terms of $\mathscr{w}^*_{\rm max}(t,x,\gamma)$ for $\theta = 0$ and $t = 4$.}
	\label{fig:OptWithdraw_NoTax_ContractSpec_t4_max}
\end{figure}

	

\begin{landscape}
	\begin{figure}
		\begin{subfigure}[b]{0.3\textwidth}
			\includegraphics[width = \textwidth]{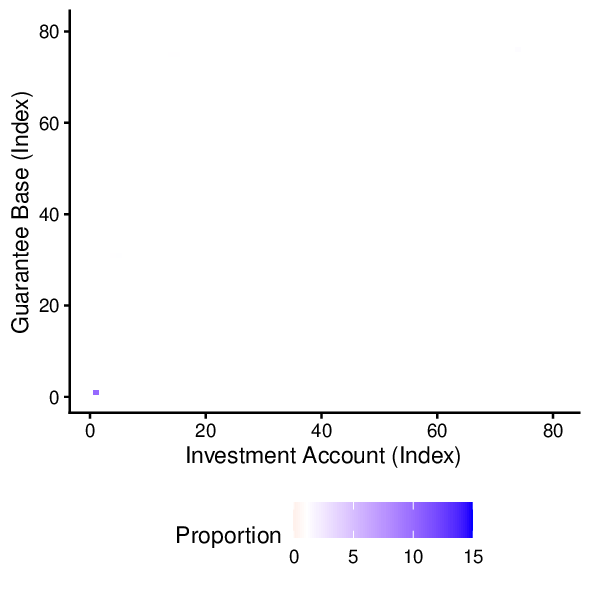}
			\caption{$\mathscr{w}^*_{\rm guar}(4,x,\gamma)$, no CF, no ratchet}
		\end{subfigure}
		\hfill
		\begin{subfigure}[b]{0.3\textwidth}
			\includegraphics[width = \textwidth]{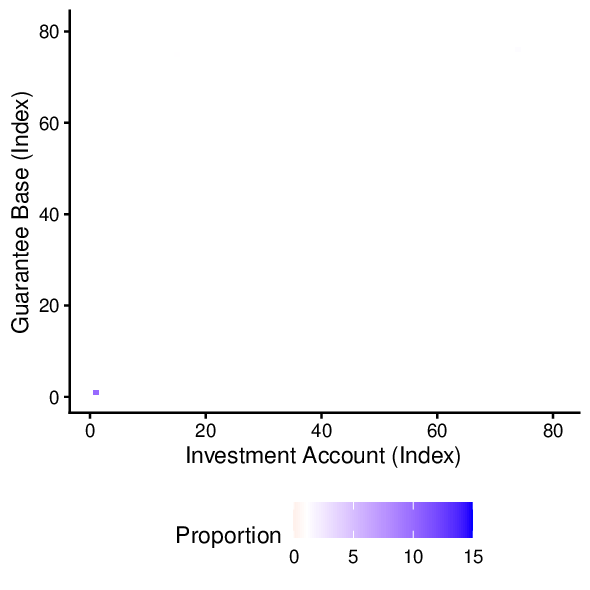}
			\caption{$\mathscr{w}^*_{\rm guar}(4,x,\gamma)$, no CF, with ratchet}
		\end{subfigure}
		\hfill
		\begin{subfigure}[b]{0.3\textwidth}
			\includegraphics[width = \textwidth]{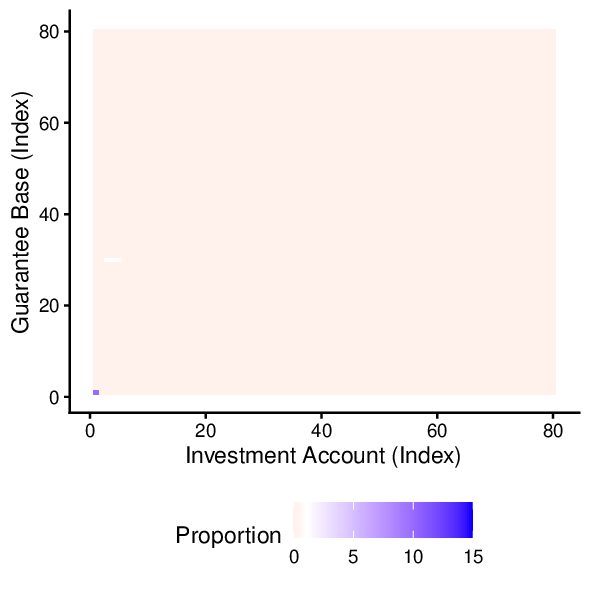}
			\caption{$\mathscr{w}^*_{\rm guar}(4,x,\gamma)$, CF, no ratchet}
		\end{subfigure}
		\hfill
		\begin{subfigure}[b]{0.3\textwidth}
			\includegraphics[width = \textwidth]{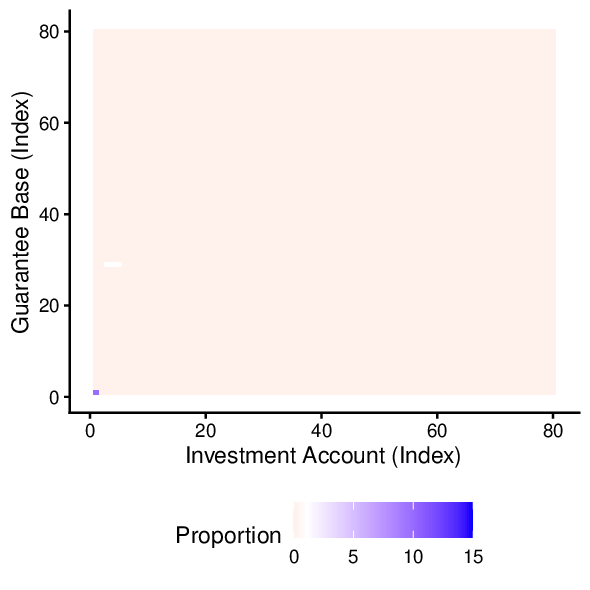}
			\caption{$\mathscr{w}^*_{\rm guar}(4,x,\gamma)$, CF, with ratchet}
		\end{subfigure}
		
		\begin{subfigure}[b]{0.3\textwidth}
			\includegraphics[width = \textwidth]{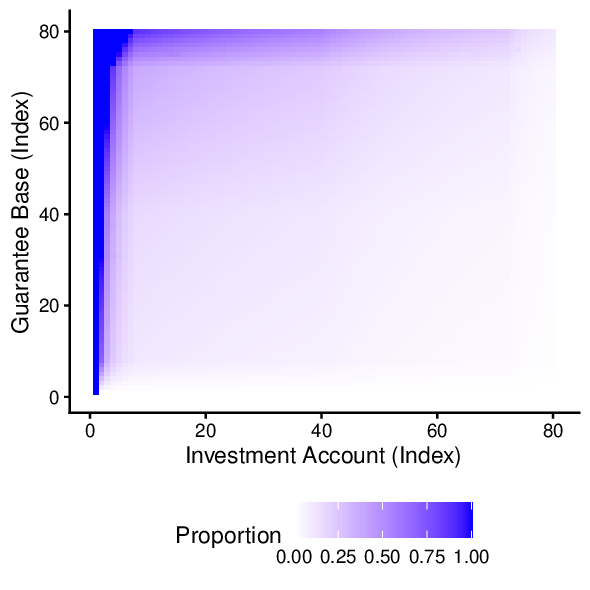}
			\caption{$\mathscr{w}^*_{\rm max}(4,x,\gamma)$, no CF, no ratchet}
		\end{subfigure}
		\hfill
		\begin{subfigure}[b]{0.3\textwidth}
			\includegraphics[width = \textwidth]{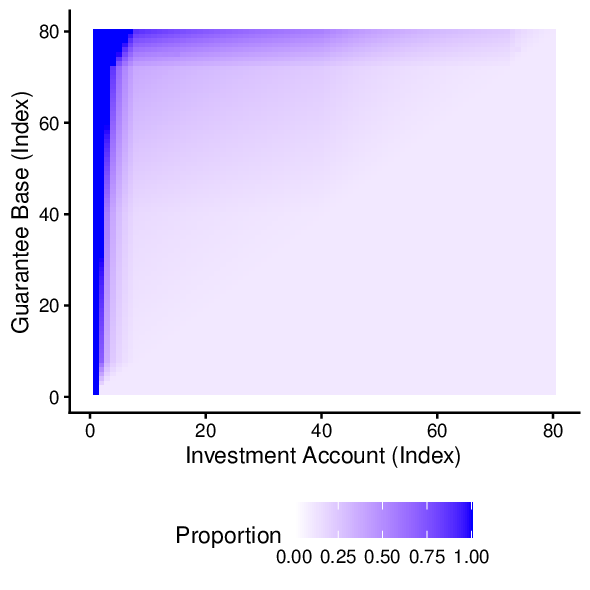}
			\caption{$\mathscr{w}^*_{\rm max}(4,x,\gamma)$, no CF, with ratchet}
		\end{subfigure}
		\hfill
		\begin{subfigure}[b]{0.3\textwidth}
			\includegraphics[width = \textwidth]{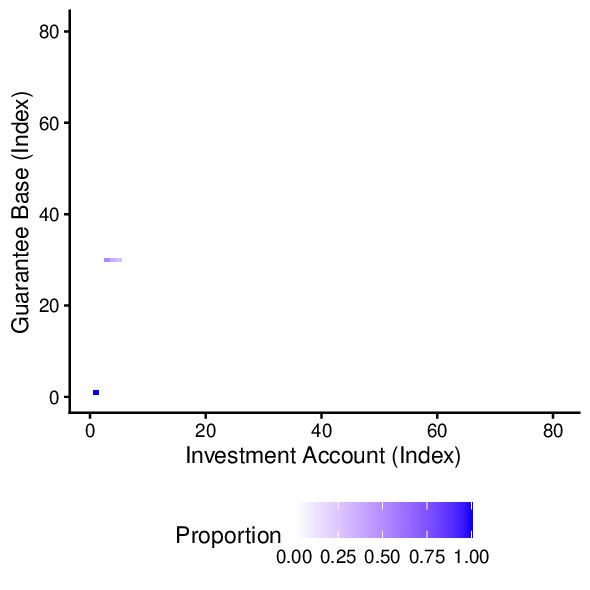}
			\caption{$\mathscr{w}^*_{\rm max}(4,x,\gamma)$, CF, no ratchet}
		\end{subfigure}
		\hfill
		\begin{subfigure}[b]{0.3\textwidth}
			\includegraphics[width = \textwidth]{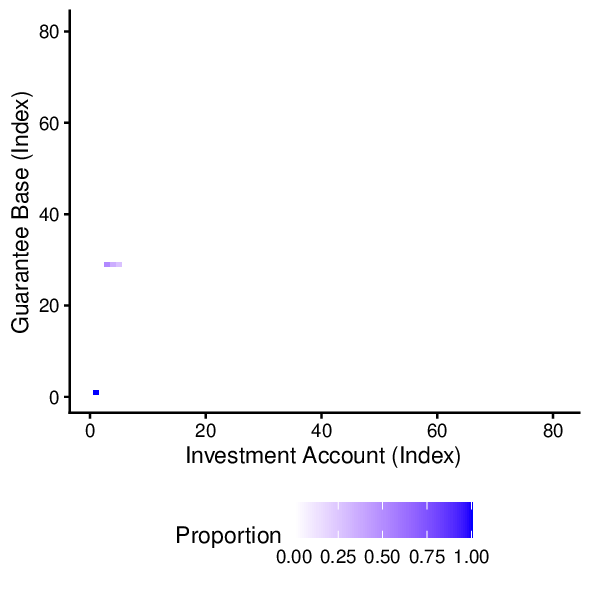}
			\caption{$\mathscr{w}^*_{\rm max}(4,x,\gamma)$, CF, with ratchet}
		\end{subfigure}		
		\caption{{\color{black} (Partial taxation case)} The optimal withdrawal strategy for various contract specifications (cash fund vs. no cash fund, ratchet vs. no ratchet/return-to-premium) expressed in terms of $\mathscr{w}^*_{\rm guar}(t,x,\gamma)$ and $\mathscr{w}^*_{\max}(t,x,\gamma)$ for $\theta = 5\%$ and $t = 4$.}
		\label{fig:OptWithdraw_Tax_ContractSpec_t4}
	\end{figure}
\end{landscape}

\begin{landscape}
	\begin{figure}
		\begin{subfigure}[b]{0.3\textwidth}
			\includegraphics[width = \textwidth]{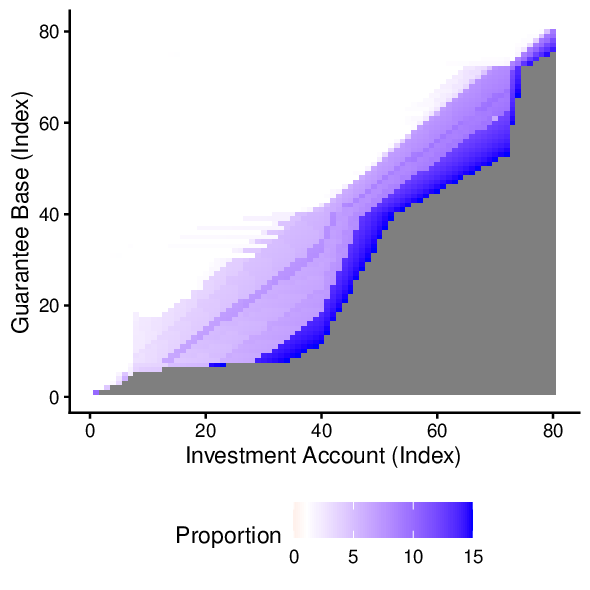}
			\caption{$\mathscr{w}^*_{\rm guar}(4,x,\gamma)$, no CF, no ratchet}
		\end{subfigure}
		\hfill
		\begin{subfigure}[b]{0.3\textwidth}
			\includegraphics[width = \textwidth]{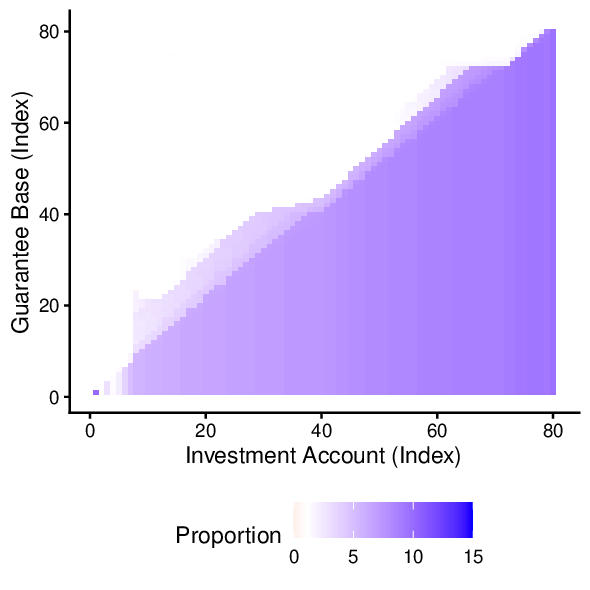}
			\caption{$\mathscr{w}^*_{\rm guar}(4,x,\gamma)$, no CF, with ratchet}
		\end{subfigure}
		\hfill
		\begin{subfigure}[b]{0.3\textwidth}
			\includegraphics[width = \textwidth]{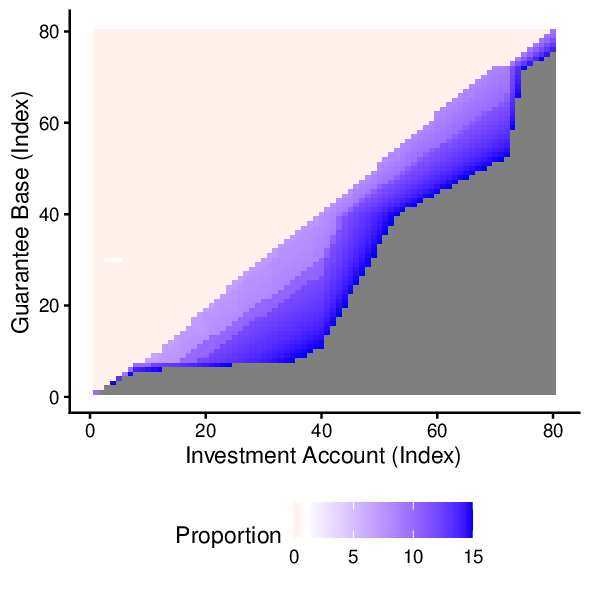}
			\caption{$\mathscr{w}^*_{\rm guar}(4,x,\gamma)$, CF, no ratchet}
		\end{subfigure}
		\hfill
		\begin{subfigure}[b]{0.3\textwidth}
			\includegraphics[width = \textwidth]{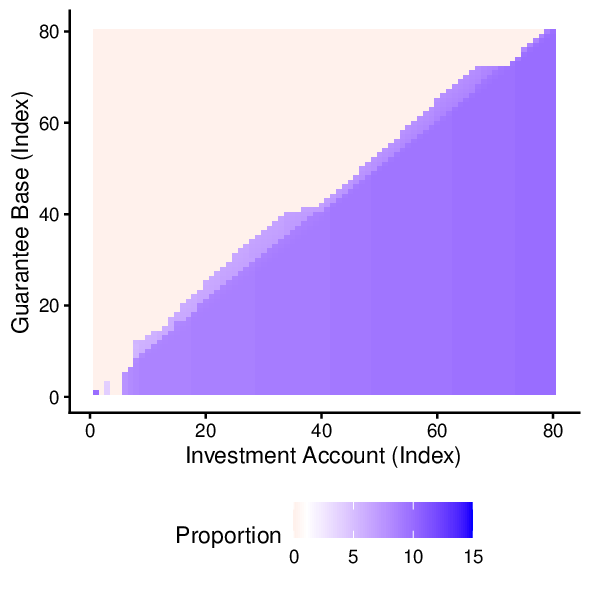}
			\caption{$\mathscr{w}^*_{\rm guar}(4,x,\gamma)$, CF, with ratchet}
		\end{subfigure}
		
		\begin{subfigure}[b]{0.3\textwidth}
			\includegraphics[width = \textwidth]{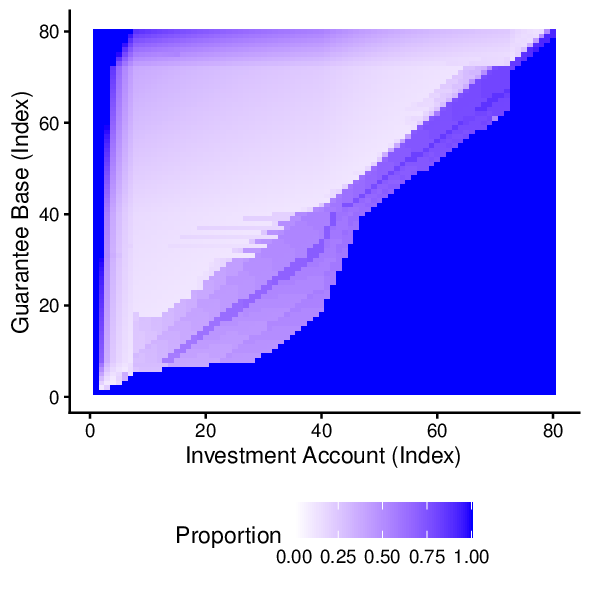}
			\caption{$\mathscr{w}^*_{\rm max}(4,x,\gamma)$, no CF, no ratchet}
		\end{subfigure}
		\hfill
		\begin{subfigure}[b]{0.3\textwidth}
			\includegraphics[width = \textwidth]{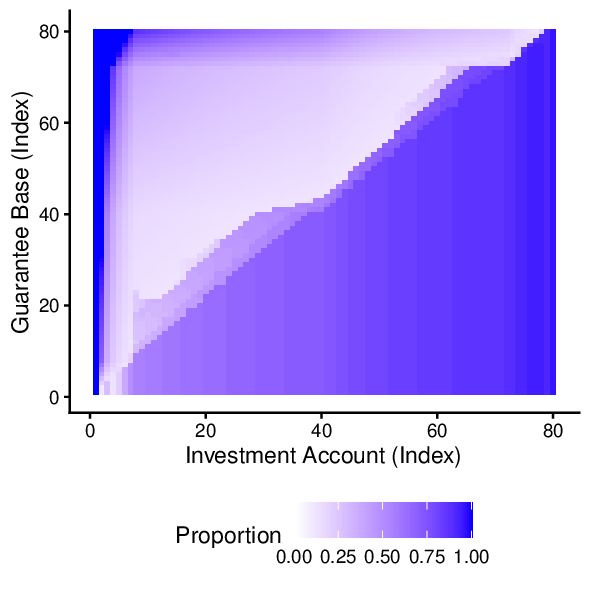}
			\caption{$\mathscr{w}^*_{\rm max}(4,x,\gamma)$, no CF, with ratchet}
		\end{subfigure}
		\hfill
		\begin{subfigure}[b]{0.3\textwidth}
			\includegraphics[width = \textwidth]{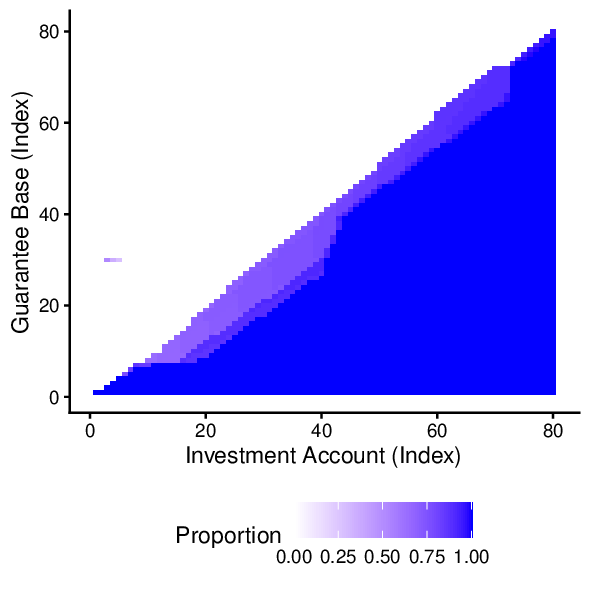}
			\caption{$\mathscr{w}^*_{\rm max}(4,x,\gamma)$, CF, no ratchet}
		\end{subfigure}
		\hfill
		\begin{subfigure}[b]{0.3\textwidth}
			\includegraphics[width = \textwidth]{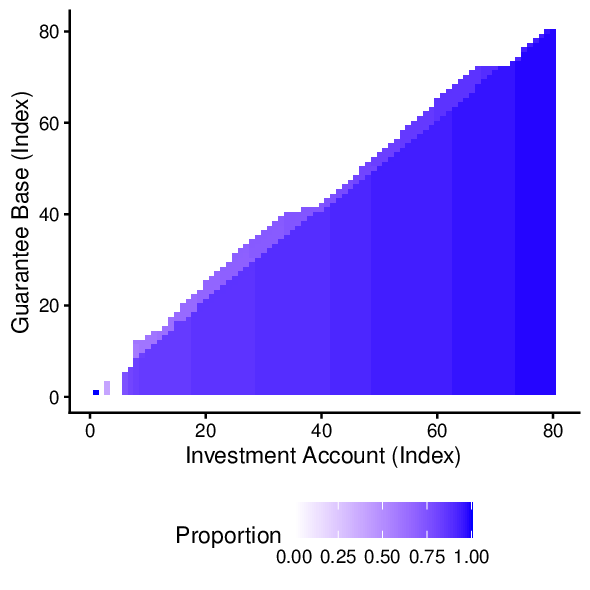}
			\caption{$\mathscr{w}^*_{\rm max}(10,x,\gamma)$, CF, with ratchet}
		\end{subfigure}		
		\caption{{\color{black} (Full taxation case)} The optimal withdrawal strategy for various contract specifications (cash fund vs. no cash fund, ratchet vs. no ratchet/return-to-premium) expressed in terms of $\mathscr{w}^*_{\rm guar}(t,x,\gamma)$ and $\mathscr{w}^*_{\max}(t,x,\gamma)$ for $\theta = 2.5\%$ and $t = 4$.}
		\label{fig:OptWithdraw_Tax_ContractSpec_t4_FullTax}
	\end{figure}
\end{landscape}

\clearpage 
\newpage

{\color{black}
\subsection{Sensitivity to Grid Sizes}
\label{app-GridSize}

Table \ref{tab-GridSize} shows the valuation of the VA value at the fair guarantee fee under both static and dynamic policyholder behavior and the computation time (in CPU time in seconds) required with respect to grid sizes $N_x$ and $N_\gamma$, assuming $N_x = N_\gamma$ in all calculations. Note that the same grid sizes are used for $w$ when implementing the grid search at each time step to determine the optimal withdrawal amount $w^*(t_k)$. The computation time includes the time required to calculate the fair guarantee fee and the calculation of the VA value at time $t = 0$ when the investment account value and the guarantee benefit base are both equal to $P_0 = 100$ (that is, $V_{t_0}(P_0, P_0; \varphi^*)$), which should be equal to $P_0$. In all calculations, we assume a tax rate of 0\% and that the guarantee benefit base evolves according to a ratchet mecahnism; in the dynamic case, we assume further that the cash fund is included in the contract and that the cash rate is 4\%. The relative error is computed as $|V_{t_0}(P_0, P_0; \varphi^*) - P_0| / P_0$.

\begin{table}[h]
\caption{Computation time and valuation of the VA contract at the fair guarantee fee under static and dynamic policyholder strategies with respect to grid sizes $N_x$ and $N_\gamma$. In all implementations, we assume $N_x = N_\gamma$.}
\label{tab-GridSize}
	\begin{tabular}{@{}rrrrlrrr@{}}
	\toprule
	\multicolumn{1}{c}{} & \multicolumn{3}{c}{Static} &  & \multicolumn{3}{c}{Dynamic} \\ \cmidrule(lr){2-4} \cmidrule(l){6-8} 
	\multicolumn{1}{c}{Grid Size} & \multicolumn{1}{c}{CPU Time (s)} & \multicolumn{1}{c}{VA Value} & \multicolumn{1}{c}{Rel. Error} &  & \multicolumn{1}{c}{CPU Time (s)} & \multicolumn{1}{c}{VA Value} & \multicolumn{1}{c}{Rel. Error} \\ \midrule
	40 & 32.89 & 99.9999 & 7.13E-07 &  & 232.10 & 100.0153 & 1.53E-04 \\
	60 & 69.99 & 100.0002 & 2.31E-06 &  & 734.21 & 100.0090 & 9.00E-05 \\
	80 & 135.68 & 100.0000 & 1.31E-07 &  & 1687.81 & 99.9863 & 1.37E-04 \\
	100 & 232.54 & 99.9972 & 2.77E-05 &  & 3266.04 & 99.9997 & 2.87E-06 \\
	150 & 644.37 & 99.9916 & 8.38E-05 &  & 14252.69 & 100.0055 & 5.52E-05 \\
	200 & 1387.09 & 99.9931 & 6.94E-05 &  & 34776.56 & 100.0044 & 4.37E-05 \\ \bottomrule
	\end{tabular}%
\end{table}

Notably, the relative error, is not monotonically decreasing with respect to the grid size. However, the relative error for the static case is minimized when $N_x = N_\gamma = 80$. For comparison purposes, we use the same grid size for the dynamic case. As such, based on these results, we employ in the main analysis an 80-by-80 grid (i.e. $N_x = N_\gamma$) since it is sufficiently granular to allow a detailed examination of the optimal withdrawal behavior without being too prohibitive in terms of the required computational time. 
}

{\color{black}
\section{Details on Computational Methodology}
\label{appendix-ComputationalMethods}

\subsection{Riccati Transform Approach}
\label{appendix-RiccatiTransform}

The Riccati transform approach to solve the second-order ODE \eqref{eqn-MOL-PDEApproximation} over the truncated domain $x \in [0, x_{\max}]$ for each $n=1,\dots,N_\tau$ is as follows. Our discussion here closely follows \citet{Meyer-2015}. 

First, for fixed $\gamma \in [0, \gamma_{\max}]$, we let $Y_n(x;\gamma) := U'_n(x;\gamma)$, so the second-order ODE \eqref{eqn-MOL-PDEApproximation} can be expressed as a system of first-order ODEs of the form
\begin{equation}
	\label{eqn-MOL-PDEApproximation-FirstOrderODESystem}
	\begin{cases}
		U_n'(x; \gamma) = \mathsf{a}(x; \gamma) U_n(x;\gamma) + \mathsf{b}(x; \gamma) Y_n(x; \gamma) + \mathsf{f}_n(x; \gamma) \\
		Y_n'(x; \gamma) = \mathsf{c}(x; \gamma) U_n(x;\gamma) + \mathsf{d}(x; \gamma) Y_n(x; \gamma) + \mathsf{g}_n(x; \gamma),
	\end{cases}
\end{equation}
where the functions $\mathsf{a}(x;\gamma), \dots, \mathsf{g}_n(x; \gamma)$ are functions of $x \in [0, x_{\max}]$, possibly parameterized by $\gamma$, and possibly dependent on $n$. These functions are determined using the coefficients of equation \eqref{eqn-MOL-PDEApproximation}. Equation \eqref{eqn-PHOptimization-PDE2-BoundaryConditionx0} implies the (left) boundary condition \[U_n(0; \gamma) = e^{-r \tau_n} \beta \gamma\] and the condition \eqref{eqn-PHOptimization-PDE2-BoundaryConditionxinf} applied to $Y_n(x; \gamma)$ at $x = x_{\max}$ yields the (right) boundary condition 
\begin{equation}
	\label{eqn-MOL-PDEApproximation-FirstOrderODESystem-RightBoundary}
	0 = \mathsf{c}(x_{\max}; \gamma) U_n(x_{\max}; \gamma) + \mathsf{d}(x_{\max}; \gamma) Y_n(x_{\max}; \gamma) + \mathsf{g}(x_{\max}; \gamma).
\end{equation}

We then assume that $U_n(x;\gamma)$ and $Y_n(x; \gamma)$ are related via the Riccati transformation
\begin{equation}
	\label{eqn-MOL-RiccatiTransform}
	U_n(x;\gamma) = \mathsf{R}_n(x; \gamma) Y_n(x; \gamma) + \mathsf{W}_n(x; \gamma)
\end{equation}
for some functions $\mathsf{R}_n$ and $\mathsf{W}_n$ of $x \in [0, x_{\max}]$ parameterized by $\gamma$. Indeed, using \eqref{eqn-MOL-RiccatiTransform}, the system of ODEs \eqref{eqn-MOL-PDEApproximation-FirstOrderODESystem} can be written as
\begin{equation}
	\label{eqn-MOL-RiccatiSystem}
	\begin{cases}
		\mathsf{R}_n'(x;\gamma) = \mathsf{b}(x;\gamma) + [\mathsf{a}(x;\gamma) - \mathsf{d}(x;\gamma)] \mathsf{R}_n(x;\gamma) - \mathsf{c}(x;\gamma) (\mathsf{R}_n(x;\gamma))^2, & \quad \mathsf{R}_n(0; \gamma) = 0 \\
		\mathsf{W}_n'(x;\gamma) = [\mathsf{a}(x;\gamma) - \mathsf{c}(x;\gamma) \mathsf{R}_n(x;\gamma)] - \mathsf{R}_n(x;\gamma) \mathsf{g}_n (x;\gamma) + \mathsf{f}_n(x;\gamma), & \quad \mathsf{W}_n(0; \gamma) = e^{-r \tau_n} \beta \gamma.
	\end{cases}
\end{equation}
This system of first-order ODEs has a unique local solution pair $(\mathsf{R}_n, \mathsf{W}_n)$ given the continuity of the coefficients $\mathsf{a}(x;\gamma), \dots, \mathsf{g}_n(x; \gamma)$ for $x \in (0, x_{\max})$ \citep{Meyer-1997, Meyer-2015}.

The functions $\mathsf{R}_n(x; \gamma)$ and $\mathsf{W}_n(x; \gamma)$ can then be solved numerically from $x = 0$ to $x = x_{\max}$ using any method for numerical integration over the grid $\mathbb{X} = \{0, x_1, \dots, x_{N_x}\}$, where $x_{N_x} = x_{\max}$. In this paper, we used the trapezoidal rule, which discussed in depth in the context of Riccati systems of equations in \citet[Section 3.3]{Meyer-2015}. This step is known as the \textit{forward sweep}.

Once $\mathsf{R}_n(x_{\max}; \gamma)$ and $\mathsf{W}_n(x_{\max}; \gamma)$ are known, we can use \eqref{eqn-MOL-RiccatiTransform} to determine the value of $Y_n(x_{\max}; \gamma) = y^*$ which satisfies right boundary condition \eqref{eqn-MOL-PDEApproximation-FirstOrderODESystem-RightBoundary}. Moreover, using \eqref{eqn-MOL-RiccatiTransform} in the second equation of the system \eqref{eqn-MOL-PDEApproximation-FirstOrderODESystem} yields the linear ODE \[Y_n'(x;\gamma) = [\mathsf{c}(x; \gamma) \mathsf{R}_n(x; \gamma) + \mathsf{d}(x; \gamma)] Y_n(x; \gamma) + \mathsf{c}(x; \gamma) \mathsf{W}_n(x; \gamma) + \mathsf{g}_n(x; \gamma), \qquad Y_n(x_{\max}; \gamma) = y^*.\] This ODE can then be solved numerically backwards from $x = x_{\max}$ to $x = 0$; this process is referred to as the \textit{backward sweep}. Once $Y_n(x; \gamma)$ has been solved for all $x \in \mathbb{X}$, we can then compute for $U_n(x; \gamma)$ over the grid $\mathbb{X}$.

The process discussed above is then repeated for all possible values of $\gamma$ over a grid $\mathbb{G} := \{0, \gamma_1, \dots, \gamma_{N_\gamma}\}$, where $\gamma_{N_\gamma} = \gamma_{\max}$ to obtain $U_n(x; \gamma)$ for all $(x, \gamma) \in \mathbb{X} \times \mathbb{G}$. The process is then iterated over all $n = 1,\dots,N_\tau$ to obtain a numerical approximation of the solution $U(\tau, x, \gamma)$ of the PDE \eqref{eqn-PHOptimization-PDE2} on $(0, t_k - t_{k-1}) \times [0, x_{\max}] \times [0, \gamma_{\max}]$.

\subsection{Simulation of Optimal Policyholder Withdrawal Behavior}
\label{app-SimulatedWithdrawals}

This appendix discusses the simulation procedure used to generate simulated optimal policyholder withdrawal behavior which have been summarized in Tables \ref{tab-OptWithSummary-ScenarioAnalysis} and \ref{tab-OptWithSummary-ScenarioAnalysis_etaVAR}. This procedure requires as inputs the parameter values and the grid sizes (as specified in Table \ref{tab-Parameters}), the grids $\mathbb{X}$ and $\mathbb{G}$ for the state variables, the optimal withdrawal profiles $\{w^*(t_k, x, \gamma)\}_{k=1,\dots,10; x \in \mathbb{X}; \gamma \in \mathbb{G}}$, and the withdrawal ratio profiles $\{\mathscr{w}_{\rm guar}^*(t_k, x, \gamma)\}_{k=1,\dots,10; x \in \mathbb{X}; \gamma \in \mathbb{G}}$ and $\{\mathscr{w}_{\rm max}^*(t_k, x, \gamma)\}_{k=1,\dots,10; x \in \mathbb{X}; \gamma \in \mathbb{G}}$ obtained through the numerical method discussed in Section \ref{sec-NumericalSolution}.

The process discussed below outlines how we obtain \textit{one simulated path} of optimal withdrawal strategies. This process can be replicated to obtain as many simulated paths as desired. The results shown in Tables \ref{tab-OptWithSummary-ScenarioAnalysis} and \ref{tab-OptWithSummary-ScenarioAnalysis_etaVAR} are based on 10,000 simulations.

Denote by $\tilde{X}(t_k^-)$ and $\tilde{G}(t_k^-)$ the simulated values of the VA investment account value and the guarantee benefit base immediately before the $k$th withdrawal made at time $t_k$. We set \[\tilde{X}(t_0^-) = \tilde{G}(t_0^-) = \tilde{X}(t_0^+) = \tilde{G}(t_0^+) = P_0.\]

\medskip

\noindent \textbf{For $k = 1,\dots,T$:}
\begin{enumerate}
	\item Simulate $\tilde{X}(t_k^-)$ using the risk-neutral dynamics of the investment account value described by \eqref{eqn-Update-InvestmentAccount-Minus}.\footnote{Antithetic variables were used in the (Monte Carlo) simulation of the Brownian motion increments from the standard normal distribution.} Set $\tilde{G}(t_k^-) = \tilde{G}(t_{k-1}^+)$ and calculate the guaranteed withdrawal amount $g(t_k)$ using \eqref{eqn-GuaranteedWithdrawalAmount} given $\tilde{X}(t_k^-)$ and $\tilde{G}(t_k^-)$.
	
	\item Using two-dimensional linear interpolation, interpolate the (simulated) optimal withdrawal strategy $\tilde{w}^*(t_k, \tilde{X}(t_k^-), \tilde{G}(t_k^-))$ from the optimal withdrawal profile grid $\{w^*(t_k, x, \gamma)\}_{x \in \mathbb{X}; \gamma \in \mathbb{G}}$. Likewise, interpolate the (simulated) optimal withdrawal ratios $\tilde{\mathscr{w}}_{\rm guar}^*(t_k, \tilde{X}(t_k^-), \tilde{G}(t_k^-))$ and $\tilde{\mathscr{w}}_{\rm max}^*(t_k, \tilde{X}(t_k^-), \tilde{G}(t_k^-))$ from the grids $\{\mathscr{w}_{\rm guar}^*(t_k, x, \gamma)\}_{x \in \mathbb{X}; \gamma \in \mathbb{G}}$ and $\{\mathscr{w}_{\rm max}^*(t_k, x, \gamma)\}_{x \in \mathbb{X}; \gamma \in \mathbb{G}}$, respectively. \textit{For simplicity, we drop the dependence on $\tilde{X}(t_k^-)$ and $\tilde{G}(t_k^-)$ in the notation for the interpolated optimal withdrawal amount $\tilde{w}^*(t_k)$ and the withdrawal ratios $\tilde{\mathscr{w}}^*_{\rm guar}(t_k)$ and $\tilde{\mathscr{w}}^*_{\rm max}(t_k)$.}
	
	\item Classify the optimal withdrawal behavior at time $t_k$:
	\begin{itemize}
		\item If $\tilde{w}^*(t_k) = 0$ and $\tilde{\mathscr{w}}^*_{\rm max}(t_k) = 0$, then \texttt{no withdrawal}.
		\item If $\tilde{\mathscr{w}}^*_{\rm guar}(t_k) < 1$ and $\tilde{\mathscr{w}}^*_{\rm max}(t_k) > 0$, then \texttt{withdrawal below guaranteed amount}.
		\item If $\tilde{\mathscr{w}}^*_{\rm guar}(t_k) = 1$, then \texttt{withdrawal at guaranteed amount}.
		\item If $\tilde{\mathscr{w}}^*_{\rm guar}(t_k) > 1$ and $\tilde{\mathscr{w}}^*_{\rm max}(t_k) < 1$, then \texttt{excess withdrawal}.
		\item If $\tilde{\mathscr{w}}^*_{\rm guar}(t_k) > 1$ and $\tilde{\mathscr{w}}^*_{\rm max}(t_k) = 1$, then \texttt{surrender}.
	\end{itemize}

	\item Calculate the post-withdrawal investment account value $\tilde{X}(t_k^+)$ and guarantee benefit base $\tilde{G}(t_k^+)$ using equations \eqref{eqn-Update-PostWithdrawalInvestmentAccount} and \eqref{eqn-Update-PostWithdrawalGuaranteeBase}, respectively, given $w = \tilde{w}^*(t_k)$.
	
	If the optimal withdrawal at time $t_k$ is \texttt{surrender}, then $\tilde{X}(t_k^+)$ and $\tilde{G}(t_k^+)$ both become zero and the for-loop in $k$ terminates. Otherwise, we move to the next time step.
\end{enumerate}

The \textit{surrender rate} is obtained by dividing the number of optimal withdrawal strategy paths that end in surrender by the total number of simulations. The average number of time steps before surrender calculated among the paths that end in surrender gives the \textit{average surrender time}. The \textit{average duration} is obtained by averaging the time spent in the contract ($T$ if the simulated strategy does not end in surrender or the corresponding surrender time if it does) across all simulated strategy paths.
}

\end{document}